\documentclass[reprint,nofootinbib,amsmath,amssymb,aps,eqsecnum,showpacs,twocolumn,superscriptaddress,prb,amsart]{revtex4-2}
\usepackage{graphicx}
\usepackage{multirow}
\usepackage{color}
\usepackage{bm}
\usepackage{xcolor}
\usepackage{times}
\usepackage{amsmath,bm,amsfonts}
\usepackage{dcolumn}
\usepackage{graphicx}
\usepackage{latexsym}
\usepackage{hhline}
\usepackage{braket}
\usepackage{bbold}

\def\t#1{\textrm{#1}}
\def\ket#1{|#1\rangle }

\usepackage{graphicx}
\usepackage{dcolumn}
\usepackage{bm}
\usepackage{multirow}

\usepackage{centernot}

\usepackage[normalem]{ulem}

\usepackage{amsmath}
\usepackage{amsthm}
\usepackage{amstext}
\usepackage{amssymb}
\usepackage{mathrsfs}
\usepackage{amsfonts}
\usepackage{amsbsy} 
\usepackage{tensor}
\usepackage{physics}

\usepackage[all,matrix,cmtip]{xy}

\usepackage{csquotes}
\MakeOuterQuote{"}

\usepackage{color}
\definecolor{niceBlue}{RGB}{20,10,237}

\usepackage[colorlinks,citecolor=niceBlue,linkcolor=niceBlue,urlcolor=niceBlue,hyperindex]{hyperref}

\definecolor{brn}{rgb}{.8,.4,.0}
\definecolor{redo}{rgb}{1,.5,.0}
\definecolor{ddgrn}{rgb}{0,0.4,0}
\definecolor{dgrn}{rgb}{0,0.55,0}
\definecolor{dbl}{rgb}{0,0,0.5}

\usepackage[bbgreekl]{mathbbol}

\newcommand{\Z}{\mathbb{Z}}

\newcommand{\ZN}[1]{\Z_N^{(#1)}}

\renewcommand{\t}[1]{\widetilde{#1}} 
\newcommand{\h}[1]{\hat{#1}} 

\newcommand{\ii}{\hspace{1pt}\mathrm{i}\hspace{1pt}}
\newcommand{\ee}{\hspace{1pt}\mathrm{e}}

\newcommand{\da}{\dagger}

\newcommand{\Rf}[1]{Ref.~\onlinecite{#1}}

\newcommand{\pp}{\partial}

\newcommand{\etc}{{\it etc}}

\newcommand{\bpm}{\begin{pmatrix}}
\newcommand{\epm}{\end{pmatrix}}
\newcommand{\bmm}{\begin{matrix}}
\newcommand{\emm}{\end{matrix}}

\newcommand{\cV}{ {\cal V} }

\usepackage{euscript}

\newcommand\eX         {\EuScript{X}}

\newcommand\eZ         {\EuScript{Z}}

\newcommand{\del}{\delta} 
 
\newcommand{\eps}{\epsilon} 
 
\newcommand{\ga}{\gamma} 
\newcommand{\Ga}{\Gamma}

\newcommand{\La}{\Lambda} 
\newcommand{\om}{\omega}

\newcommand{\txti}[1]{\textit{#1}}
\renewcommand{\txt}[1]{\text{#1}}

\newcommand{\hstar}{\mathop{*}} 

\newcommand{\nn}{\nonumber \\}

\definecolor{darkcyan}{rgb}{0.0, 0.55, 0.55}
\newcommand{\lcm}[2]{{\rm lcm}\left(#1,#2 \right)}
\renewcommand{\gcd}[2]{{\rm gcd}\left(#1,#2 \right)}

\begin{document}
\title{Aspects of $\mathbb{Z}_N$ rank-2 gauge theory in $(2+1)D$: \\ construction schemes, holonomies, and sublattice one-form symmetries}
\author{Yun-Tak Oh}
\affiliation{Division of Display and Semiconductor Physics, Korea University, Sejong 30019, Korea}
\author{Salvatore D. Pace}
\affiliation{Department of Physics, Massachusetts Institute of Technology, Cambridge, Massachusetts 02139, USA}
\author{Jung Hoon Han}
\affiliation{Department of Physics, Sungkyunkwan University, Suwon 16419, Korea}
\author{Yizhi You}
\email[Electronic address:$~~$]{y.you@northeastern.com}
\affiliation{Department of Physics, Northeastern University, 360 Huntington Ave, Boston, MA 02115, USA}
\author{Hyun-Yong Lee}
\email[Electronic address:$~~$]{hyunyong@korea.ac.kr }
\affiliation{Division of Display and Semiconductor Physics, Korea University, Sejong 30019, Korea}
\affiliation{Department of Applied Physics, Graduate School, Korea University, Sejong 30019, Korea}
\affiliation{Interdisciplinary Program in E·ICT-Culture-Sports Convergence, Korea University, Sejong 30019, Korea}

\begin{abstract} 
Rank-2 toric code (R2TC), a prototypical archetype of the discrete rank-2 symmetric gauge theory, has properties that differ from those of the standard toric code. Specifically, it features a blending of UV and IR in its ground state, restricted mobility of its quasiparticles, and variations in the braiding statistics of its quasiparticles based on their position. In this paper, we investigate various aspects of $\Z_N$ rank-2 gauge theory in ${(2+1)}$-dimensional spacetime. Firstly, we demonstrate that $U(1)$ rank-2 gauge theory can arise from ${U(1)\times U(1)}$ rank-1 gauge theory after condensing the gauge charges in a specific way. This construction scheme of $U(1)$ rank-2 gauge theory carries over to the $\Z_N$ case simply by Higgsing $U(1)$ to $\Z_N$, after which the resulting rank-2 gauge theory can be tuned to the R2TC. The holonomy operators of R2TC are readily identified using this scheme and are given clear physical interpretation as the pair creation/annihilation of various monopoles and dipoles. Explicit tensor network construction of the ground states of R2TC are given as two copies of the ground states of Kitaev's toric code that are `sewn together' according to the condensation scheme. In addition, through a similar anyon condensation protocol, we present a double semion version of rank-2 toric code whose flux excitations exhibit restricted mobility and semionic statistics. Finally, we identify the generalized discrete symmetries of the R2TC, which are much more complex than typical 1-form symmetries. They include conventional and unconventional 1-form symmetries, such as framed 1-form symmetries and what we call sublattice 1-form symmetries. Using these, we interpret the R2TC's unique properties (UV/IR mixing, position-dependent braiding, etc.) from the modern perspective of generalized spontaneous symmetry breaking and 't Hooft anomalies.
\end{abstract}
\date{\today}

\maketitle

\tableofcontents 

\section{Introduction}

Long-range entangled phases of quantum matter are commonly described by fractionalized quasiparticles and emergent gauge fields which provide an effective description capturing the phase's universal properties~\cite{Wen04}. Indeed, canonical examples include fractional quantum Hall liquids~\cite{W9505} and quantum spin liquids~\cite{SB160103742}. Unsurprisingly, long-range entangled quantum matter with increasingly exotic properties is described by increasingly rich generalizations of conventional gauge theory. A particular example is abelian gauge theories whose gauge fields are symmetric tensor fields instead of vector fields. These higher-rank gauge theories\footnote{We will generally denote abelian \txti{symmetric} tensor gauge theory simply as higher-rank gauge theory. This is not to be confused with higher-form gauge theories whose gauge fields are differential forms---\txti{antisymmetric} tensors.} have attracted substantial interest recently in the study of fracton phases~\cite{P160405329, SPP180700827, PY200101722, barkeshli,MHC180210108} and topological order~\cite{barkeshli,MHC180210108,oh22a,pace-wen,oh22b, seiberg22a}.

One of the simplest archetype of discrete gauge theories can be obtained from Higgsing the $U(1)$ theory into $\Z_2$ by condensing charge-2 gauge charges. Following this protocol, we can obtain the rank-2 $\Z_2$ gauge theory starting from a rank-2 $U(1)$ gauge theory and Higgsing $U(1)$ down to $\Z_2$~\cite{barkeshli,MHC180210108}. In the zero-correlation length limit, the resultant gauge theory can be interpreted as an exactly solvable Hamiltonian, so-called rank-2 toric code (R2TC)~\cite{kitaev03,oh22a,pace-wen,oh22b}.

The R2TC features several interesting properties. One of them is the sensitive dependence of the ground state degeneracy (GSD) on the system size $L_x \times L_y$ against $N$, the Hilbert space dimension of the local spin state $|s\rangle$ ($s=0, \cdots, N-1$). The GSD varying from $N^3$ to $N^6$ was first discovered in Ref. \cite{oh22a} and was soon clarified as a rigorous formula~\cite{pace-wen}
\begin{align} {\rm GSD} = N^3 {\rm gcd} (L_x , N) {\rm gcd} (L_y , N) {\rm gcd} (L_x , L_y , N), \label{eq:GSD} \end{align} 
where gcd stands for the greatest common divisor among the two or three integers. The fact that the GSD, a macroscopic property, depends sensitively on the number of unit cells, a microscopic property, is a manifestation of what's known as UV/IR mixing~\cite{SS200310466, GLS210800020, pace-wen,seiberg22a}. The braiding process between a pair of quasiparticles showed interesting position dependence that is not seen in Kitaev's toric code, and requires a new form of field theory called the dipolar BF theory (dBF)~\cite{oh22b} for comprehensive understanding. A different interpretation of the dipolar braiding in terms of multi-component mutual Chern-Simons theory was given in Ref. \cite{pace-wen}. Lastly, quasiparticle excitations in the model showed restricted mobility such as the ability to hop only in multiples of $N$ lattice sites in certain directions~\cite{barkeshli,MHC180210108}. We note that many of these features were already apparent in the plaquette model of Wen~\cite{wen03}, and more recently several models sharing similar features were proposed~\cite{seiberg22a,delfino22,watanabe}.

The restricted mobility exhibited by quasiparticle excitations in R2TC is clearly shared in a more rigorous way in fracton models such as the X-cube model~\cite{x-cube}. A standard way of constructing these fracton models is to use the network construction scheme first proposed in Refs. \cite{hermele17,vijay}. In it, one starts with layers of 2D toric codes with fully mobile quasiparticle excitations, and produces a fractonic model with immobile excitations by imposing constraints among the layers. It is a natural question then if a similar scheme does exist to construct R2TC -a model based on rank-2 gauge theory - from the R1TC which is rooted in conventional rank-1 gauge theory. This paper answers this question in the affirmative. 

In accomplish this, we get to exploit the idea of coupling two gauge theories together through constraint, in a process often called the anyon condensation. Several past works have exploited the condensation idea to produce the X-cube model from layers of 2D toric codes~\cite{hermele17,vijay} (hereafter referred to as rank-1 toric code, or R1TC for short), or to produce rank-2 gauge theories from rank-1 theories~\cite{williamson19,radzihovsky}. Our condensation scheme share the similar spirit as these works, but differs greatly in details of how we implement the constraint. In particular, we make a clear comparison between the condensation scheme of Ref.~\cite{radzihovsky} and our own in Sec.~\ref{subsec:pre-proj} in an effort to emphasize the consequences of various condensation schemes. In the past, R2TC was obtained by Higgsing the symmetric rank-2 gauge field~\cite{barkeshli,oh22a} but the origin of this higher-rank gauge field was left obscure. We show here that it emerges naturally in the course of constraining the two copies of rank-1 gauge fields in a certain way. 

Furthermore, the GSD of Eq. (\ref{eq:GSD}) is closely related to the existence of six independent Wilson line operators in the model, which have been identified previously in the spin operator~\cite{oh22a} and the field theory language~\cite{pace-wen}, respectively. Still lacking was a clear physical picture accompanying these Wilson operators, such as the creation/annihilation of electric and magnetic quasiparticle pairs in the case of R1TC. It turns out that the condensation scheme provides a helpful guide in constructing the full set of holonomies~\footnote{In this paper, we use the words holonomies interchangeably with Wilson line operators.} needed to fully account for the GSD, which are also amenable to physically appealing interpretations. In addition to the winding of charges, winding of dipoles play an important role in accounting for the degeneracy of the R2TC ground states. 

As an added benefit of the condensation picture, we find some useful applications in explicitly constructing the first-quantized ground state wave function of R2TC in tensor network (TN) form, as two copies of R1TC ground states sewn together through some constraining tensor that directly reflects the constraint. As another example we construct the rank-2 generalization of a model with semionic flux statistics~\cite{levin05} by coupling two copies of the pristine double semion model through anyon condensation. 

As a final topic of the paper, we explore the generalized symmetries of the R2TC. Modern generalizations of symmetry~\cite{NOc0605316,GW14125148,TW191202817,KZ200514178,M220403045, CD220509545,FT220907471} have opened up an exciting frontier for the discovery of new phases of quantum matter~\cite{TK151102929,Y150803468, W181202517, TW190802613, JW200900023, MF220607725, PW220703544} and in the conceptual organization of both known and new quantum phases~\cite{M220403045}. For instance, these generalizations have allowed topological order to be understood in a symmetry framework~\cite{NOc0702377,W181202517,HS181204716,KKO210713091, M220403045}. It is therefore natural to wonder if the interesting properties of the R2TC can be understood in this unifying, modern point of view of topological quantum matter. Here, we construct all of the symmetry operators for the R2TC for general $N$. In the ground state sub-Hilbert space, the symmetries we identify are all 1-form symmetries. However, they are not all conventional 1-form symmetries: some rely on a framing structure of the lattice (framed 1-form symmetries~ \cite{QRH201002254}) and others on a sublattice structure of the lattice (sublattice 1-form symmetries). Furthermore, these symmetries have a rich mixed 't Hooft anomaly structure. We show that the R2TC ground state spontaneously breaks all of these 1-form symmetries. This allows us to interpret the unconventional properties of the R2TC (position-dependent braiding, UV/IR mixing, etc.) all in terms of these symmetries. 

Organization of the paper is as follows. In Sec. \ref{sec:cond-gauge-field} we outline the condensation scheme that leads one from two copies of rank-1 lattice gauge theory to the rank-2 lattice gauge theory and ultimately to the R2TC. In Sec.~\ref{sec:application} we discuss two applications of the condensation idea in the construction of the ground state of rank-2 toric code out of those of the rank-1 toric code, and the construction of `twisted' rank-2 gauge theory resulting in a new model with semionic flux statistics. In Sec.~\ref{sec:holonomies} we carefully go through the procedure by which all the holonomies in the R2TC can be derived. Physical interpretation of the holonomies thus constructed is given. In Sec.~\ref{sec:generalized-symmetry}, after first reviewing the generalized symmetries of the R1TC, we discuss the R2TC from the point of view of generalized symmetries. Additional themes such as instanton effects in rank-1 gauge theories in 2+1D (Sec.~\ref{subsec:instanton}), field-theoretic understanding of the holonomy and the position-dependent braiding (Secs.~\ref{subsec:field-theoretic-holonomy} and \ref{posDepBraidSubSec}) are discussed. The summary and outlook follows in Sec.~\ref{sec:discussion}. 

\section{Condensation Scheme}
\label{sec:cond-gauge-field}

We show how rank-2 gauge theoires can emerge from two copies of rank-1 gauge theory through the condensation of certain components of the gauge fields. We outline this procedure first from the perspective of U(1) gauge theory, followed by that of $\mathbb{Z}_N$ gauge fields. Discussion of instanton suppression in the rank-2 gauge theory is given as well.

\begin{figure}[tb]
\includegraphics[width=0.48\textwidth]{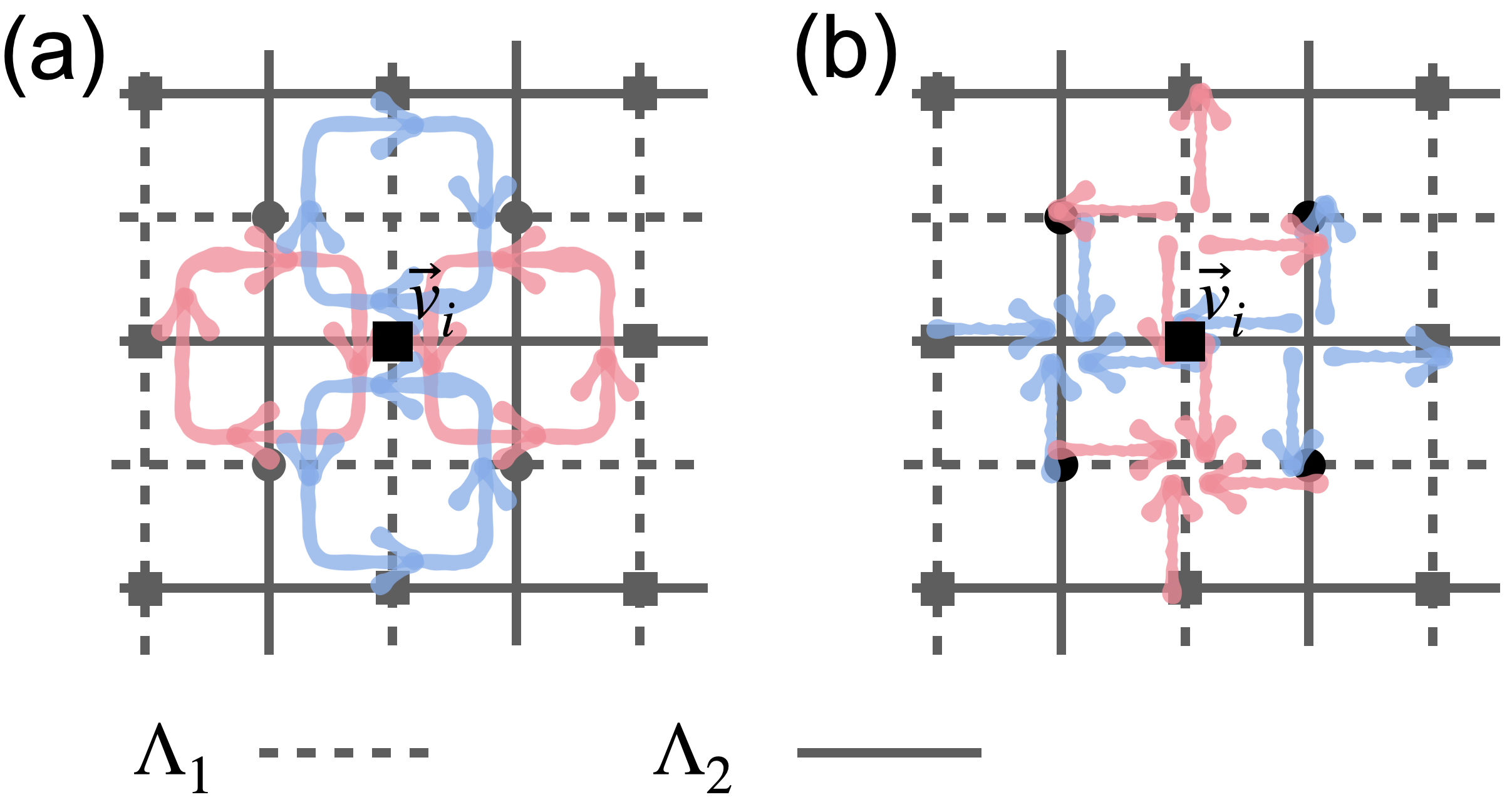}
\caption{
Illustration of two possible condensation processes leading to the appearance of new (a) magnetic and (b) electric flux operators given by Eq. (\ref{eq:cond-b}) and Eq. (\ref{eq:cond-g}), respectively. There are two inter-penetrating sublattices $\Lambda_1 , \Lambda_2$ shown by dashed and solid lines, respectively. The dark squares (circles) represent $\mathcal{V}_{\rm vh}$ ($\mathcal{V}_{\rm hv}$) sites. The coordinate $\vec{v}_i$ refers to the $\mathcal{V}_{\rm vh}$ sites. The pink and blue arrows represent rank-1 gauge field patterns from the sublattice $\Lambda_1$ and $\Lambda_2$, respectively, that combine to give new flux patterns in the rank-2 gauge theory. 
}
\label{fig:quadrupole}
\end{figure}

\subsection{Condensation of rank-2 U(1) lattice gauge fields}
Consider two interpenetrating square lattices denoted $\Lambda_1$ (dashed lines) and $\Lambda_2$ (solid lines) as in Fig. \ref{fig:quadrupole}. Each square lattice has gauge degrees of freedom $(A^\mu_{a} , E^\mu_{a})$ residing at the $\mu=x,y$-oriented links of the respective square sublattice labeled by $a=1,2$, satisfying the canonical commutation $[A_a^\mu , E_{a'}^{\mu'} ] = i \delta_{aa'} \delta_{\mu \mu'}$. The two square lattices are superposed in such a way that horizontal bonds in $\Lambda_1$  and vertical bonds in $\Lambda_2$ intersect at one set of sites belonging to $\mathcal{V}_{\rm hv}$, while vertical bonds of $\Lambda_1$ and horizontal bonds of $\Lambda_2$ cross at sites belonging to $\mathcal{V}_{\rm vh}$. In Fig. \ref{fig:quadrupole}, sites belonging to ${\cal V}_{\rm vh}$ and ${\cal V}_{\rm hv}$ are designated by dark squares and circles, respectively. The coordinate $\vec{v}_i = x_i \hat{x}  + y_i\hat{y}$, or sometimes just $i$, is used to label the ${\cal V}_{\rm vh}$ sites. (Note that what we call {\it sites} are the {\it links} in the individual sublattice.) To reduce the notational clutter, we will also use $i$ to label the vertex position $\vec{r}_{1,i}=\vec{v}_i-\hat{y}/2$ of the $\Lambda_1$ lattice and the vertex position $\vec{r}_{2,i} = \vec{v}_i-\hat{x}/2$ of the $\Lambda_2$ lattice as well.
According to this notation scheme, the $y$-oriented fields $(A_{1,i}^y, E_{1,i}^y)$ and the $x$-oriented fields $(A_{2,i}^x, E_{2,i}^x)$ both reside on the same site $\vec{v}_i = x_i \hat{x}  + y_i\hat{y}$. The sites in ${\cal V}_{\rm hv}$ and the fields defined on them are then assigned appropriate coordinates in reference to those given to ${\cal V}_{\rm vh}$ sites. 

Each square lattice $\Lambda_1 , \Lambda_2$ hosts its own gauge-invariant quantities ($a =1,2$),
\begin{align} 
G_{a}(\vec{r}_{a,i}) = ({\bm \nabla} \cdot {\bm E}_a )_i , 
~~~ B_{a}(\vec{r}_{a,i}) = ({\bm \nabla} \times {\bm A}_a )_i
, \label{eq:2.1}\end{align} 
which are the lattice divergence and lattice curl. 
Suppose that we impose the constraint $E^x_{1,i+\hat{y}} = E^y_{2, i+\hat{x}}$ at half the links, {\it e.g.} on the $\mathcal{V}_{\rm hv}$ sites (dark circles in Fig. \ref{fig:quadrupole}). In other words, we write the combined Hilbert spaces of the two lattice gauge theory's as $|\Psi \rangle = |\psi_1 \rangle \otimes |\psi_2 \rangle$ where $|\psi_a\rangle$ belongs to the Hilbert space of $\Lambda_a$, and insist that only the subset of Hilbert spaces obeying the following constraint survives:
\begin{align}
(  E^x_{1,i+\hat{y}} -  E^y_{2,i+\hat{x}} ) |\Psi \rangle = 0 .   \label{eq:constraint} 
\end{align}
To be clear, $i$ refers to all the ${\cal V}_{\rm hv}$ sites. Such constraint necessarily precludes operators that do not commute with it, such as $( {\bm \nabla}\times {\bm A}_a )_i$, while $({\bm \nabla} \cdot {\bm E}_a )_i$ is still allowed. In their place, a new operator that commutes with the constraint can be constructed by noting that $[E^x_{1,i+\hat{y}} - E^y_{2,i+\hat{x}} , A^x_{1,i+\hat{y}} + A^y_{2,i+\hat{x}} ] = 0$. We find 
\begin{align}
 B_i = \Delta_x ({\bm \nabla} \times {\bm A}_{1})_i - \Delta_y ({\bm \nabla} \times {\bm A}_{2} )_i
 \label{eq:cond-b}
\end{align}
indeed involves only the combination $A^x_{1} + A^y_{2}$ at all the $\mathcal{V}_{\rm hv}$ sites where the constraint Eq. (\ref{eq:cond-b}) is imposed, and therefore commutes with it~\footnote{It is easy to convince that no simpler operator exists that commutes with the constraint.}. The meaning of discrete derivatives $\Delta_x$ and $\Delta_y$ is clear from Fig. \ref{fig:quadrupole}. 

Due to the constraint, one must identify $E^x_{1,i+\hat{y}}  =  E^y_{2,i+\hat{x}}$ as one gauge field and accordingly introduce a new label
\begin{align} 
(E_{2,i}^x, E_{1,i}^y, E_{1,i+\hat{y}}^x = E_{2,i+\hat{x}}^y ) \rightarrow (E_i^{xx}, E_i^{yy}, E_i^{xy} ) . 
\label{eq:E-field-condensate}
\end{align} 
A similar re-labeling 
\begin{align}
(A_{2,i}^x , A_{1,i}^y , A_{1,i+\hat{y}}^x + A_{2,i+\hat{x}}^y ) \rightarrow (A_i^{xx}, A_i^{yy}, A_i^{xy})
\end{align} 
yields symmetric rank-2 gauge fields $(A_i^a, E_i^a)$ ($a=xx,xy,yy$) obeying the canonical relation $[A_i^a, E_j^b] = i \delta_{ij} \delta_{ab}$\footnote{This follows from the original gauge fields obeying the canonical relation $[A^\mu_{a,i} , E^\nu_{b,j} ] = i \delta_{ab} \delta_{ij} \delta_{\mu\nu}$.}. 

There are two electric charges $(e^x_i, e^y_i)$ and one vector charge $m_i$ in the projected Hilbert space obeying Eq. (\ref{eq:constraint}) given by
\begin{align}
e_i^x = & E_{i+\hat{x}}^{xx} - E_{i}^{xx} + E_{i}^{xy}- E_{i-\hat{y}}^{xy} 
, \nn
e_i^y = & E_{i}^{xy} - E_{i-\hat{x}}^{xy} + E_{i+\hat{y}}^{yy}- E_i^{yy} 
, \nn 
m_i = & A^{xx}_{i+\hat{y}}+ A^{xx}_{i-\hat{y}} -2 A^{xx}_{i} 
+ A^{yy}_{i+\hat{x}}+ A^{yy}_{i-\hat{x}} -2 A^{yy}_{i} \nn &   
-A^{xy}_i + A^{xy}_{i-\hat{x}} + A^{xy}_{i-\hat{y}} - A^{xy}_{i-\hat{x}-\hat{y}} . 
\label{eq:generators} 
\end{align} 
Here $m_i$ is simply the re-writing of $B_i$ in Eq. (\ref{eq:cond-b}). The symmetric rank-2 gauge fields as well as the new mutually commuting generators formed by them emerge naturally from the condensation process just outlined. Upon Higgsing, the three charge operators in Eq. (\ref{eq:generators}) become the three commuting spin operators of R2TC~\cite{oh22a}. Among the tensor gauge fields, $xx$ and $yy$ components reside at the $\mathcal{V}_{\rm vh}$ sites where no condensation has taken place, and the $xy$ component resides at the $\mathcal{V}_{\rm hv}$ sites where condensation reduces the degrees of freedom from two to one. 

The constraint expressed in Eq. (\ref{eq:constraint}) is by no means the unique one. Instead of condensing $E$, one can condense the $A$ fields through the constraint
\begin{align}
(A_{1,i+\hat{y}}^x - A_{2,i+\hat{x}}^y)|\Psi\rangle = 0 ,
\label{eq:a1_a2-proj}
\end{align}
at the $\mathcal{V}_{\rm hv}$ sites. In this case, $({\bm \nabla}\cdot {\bm E}_a )_i$ is no longer an allowed operator but a new quantity
\begin{align}
G_i  = \Delta_x ({\bm \nabla}\cdot {\bm E}_2)_i  + \Delta_y ({\bm \nabla} \cdot {\bm E}_1)_i 
\label{eq:cond-g}
\end{align}
emerges as a viable operator in the constrained Hilbert space - see Fig. \ref{fig:quadrupole}(b). After re-labeling 
\begin{align} 
(A_{2,i}^x, A_{1,i}^y , A_{1,i+\hat{y}}^x = A_{2,i+\hat{x}}^y ) & \rightarrow (A_i^{xx}, A_i^{yy}, A_i^{xy} )  \nn 
(E_{2,i}^x , E_{1,i}^y , E_{1,i+\hat{y}}^x + E_{2,i+\hat{x}}^y ) & \rightarrow (E_i^{xx}, E_i^{yy}, E_i^{xy}),  \label{eq:re-labeling} 
\end{align} 
one arrives at two magnetic charges $(m^x_i , m^y_i)$ and one electric charge $e_i$ defined by 
\begin{align}
m^{x}_i = & A_{i+\hat{x}}^{yy} - A_{i}^{yy} - A_{i}^{xy} + A_{i-\hat{y}}^{xy} 
, \nn
m^{y}_i = & A_{i}^{xy} - A_{i-\hat{x}}^{xy} - A_{i+\hat{y}}^{xx} + A_i^{xx} 
, \nn
e_i = & E^{yy}_{i+\hat{y}}+ E^{yy}_{i-\hat{y}} -2 E^{yy}_{i} 
+ E^{xx}_{i+\hat{x}}+ E^{xx}_{i-\hat{x}} -2 E^{xx}_{i} \nn  
& ~~ +E^{xy}_i - E^{xy}_{i-\hat{x}} - E^{xy}_{i-\hat{y}} + E^{xy}_{i-\hat{x}-\hat{y}} ,
\label{eq:quadrupole-operator}
\end{align} 
where $e_i$ is a mere re-writing of $G_i$ in Eq. (\ref{eq:cond-g}). Higging them leads to R2TC with a scalar electric charge and vector magnetic charges.

It is worth noting that the choice of how to condense two copies of rank-1 gauge fields, either through Eq. (\ref{eq:E-field-condensate}) or (\ref{eq:re-labeling}), results in rank-2 gauge fields with distinct gauge symmetries. One leads to the vector charge theory, while the other results in the scalar charge theory~\cite{P160405329}. These two theories are dual to each other~\cite{oh22a,pace-wen}.
In the ensuing discussion, we will adopt the version of R2TC that has scalar-electric and vector-magnetic charges unless otherwise specified.

\subsection{Condensation of stabilizers and holonomies}
\label{subsec:pre-proj}

The previous subsection showed which operators survive under the projection (condensation) of two rank-1 lattice gauge theory's  to the constrained Hilbert space. The operators that become the stabilizers in the R2TC emerged naturally. In this subsection we elaborate how the condensation idea plays out for the various spin operators and stabilizers. To be specific, we first construct the stabilizers and holonomies in the pre-projected Hilbert space consisting of two copies of R1TCs. Then we examine which of these operators survive, or become modified, under the projection. Stabilizers of the R2TC are recovered once again in this way. Although at first sight this discussion seems redundant in light of the aforementioned projection scheme outlined in the context of U(1) gauge fields, there is a nice benefit to the present discussion in that it paves the way for the efficient identification and construction of holonomy operators of R2TC in Sec.~\ref{sec:holonomies}. The insight gained in this subsection will also be pivotal in the construction of TN wave functions in Sec.~\ref{subsec:TN}.

As before we consider two interpenetrating square lattices $\Lambda_1$ and $\Lambda_2$, and place $\mathbb{Z}_N$ spins on the links. There are generalized Pauli operators satisfying $ZX = \omega XZ$ ($\omega= e^{2\pi i/N}$) at the links of each sublattice, which follow from the Higgsing formula~\cite{oh22a}: 
\begin{align} X= e^{2\pi i A}, ~~ Z= e^{2\pi i E/N}.
\end{align} 
We place R1TC on each of the sublattices $\Lambda_1$ and $\Lambda_2$, with the star ($a_{a}(\vec{r}_{a,i})$) and the plaquette ($b_{a}(\vec{r}_{a,i})$) operators defined, respectively, by ($a=1,2$)
\begin{align}
a_{a,i} = & Z_{a,x}(\vec{r}_{a,i}) Z_{a,x}(\vec{r}_{a,i}\!-\!\hat{x})^{\!-\!1}\!Z_{a,y}(\vec{r}_{a,i}) Z_{a,y}(\vec{r}_{a,i}\!-\!\hat{y})^{\!-\!1}, \nn 
b_{a,i} = & X_{a,x}(\vec{r}_{a,i}) X_{a,x}(\vec{r}_{a,i}\!+\!\hat{y})^{\!-\!1}\! X_{a,y}(\vec{r}_{a,i})^{\!-\!1}\! X_{a,y}(\vec{r}_{a,i}\!+\!\hat{x}) . \label{eq:star-and-plaquette}
\end{align}
Here, the subscript $i$ indicate the vertex of square lattice $\vec{r}_{a,i}$, and  the extra subscripts $x,y$ in the $X, Z$ operators indicates the direction of the bond on which the operators are defined. 

States in each R1TC are denoted as $|\psi\rangle_1$ and $|\psi \rangle_2$, respectively. 
The eigenstates of $X$ operator are $X|n\rangle = \omega^{n} |n\rangle$. The constraint, Eq. (\ref{eq:a1_a2-proj}), implies $A^x_{1,i+\hat{y}} |\psi \rangle = A^y_{2,i+\hat{x}} |\psi\rangle$ or, after Higgsing, 
\begin{align}\label{R1TCscondScheme}
X_{1,x}(\vec{r}_{1,i}+\hat{y}) |\psi \rangle = X_{2,y} (\vec{r}_{2,i} +\hat{x}) |\psi\rangle.
\end{align} 
In other words, only the following product of states in the pre-projection Hilbert space survives the projection, 
\begin{align}
|n\rangle_1 \otimes |n\rangle_2 \xrightarrow{\cal P} |n\rangle . 
\label{eq:spin-X-proj}
\end{align} 
Besides, Eq. (\ref{eq:re-labeling}) states that $E^x_{1,i+\hat{y}} + E^y_{2,i+\hat{x}}$ must be identified with $E^{xy}_i$ as well, which in the $\mathbb{Z}_N$ language means 
\begin{align} Z_{1,x}(\vec{r}_{1,i}+\hat{y}) Z_{2,y}(\vec{r}_{2,i}+\hat{x}) |\psi\rangle = Z (\vec{v}_i) |\psi\rangle.
\nonumber 
\end{align} 
This constraint can be expressed in the $Z$-basis $Z|m \rangle = \omega^m |m\rangle$ as the projection
\begin{align}
|m_1 \rangle_1 \otimes |m_2 \rangle_2 \xrightarrow{\cal P} |m_1 + m_2 \rangle . 
\label{eq:spin-Z-proj}\end{align} 
In both Eqs. (\ref{eq:spin-X-proj}) and (\ref{eq:spin-Z-proj}) the mapping acts only at the $\mathcal{V}_{\rm hv}$ sites where the gauge field constraint has been imposed. 

Regarding the pre-projection state $|\Psi\rangle$ that remains after the projection, denoted as ${\cal P} |\Psi \rangle = |\psi \rangle$, 
one can think of the operator projection as follows:
\begin{align}
{\cal P} \left[ {\cal O}  |\Psi\rangle  \right] = {\cal O}' \left[ {\cal P} |\Psi \rangle \right]   = 
{\cal O}' |\psi \rangle . 
\end{align}
Here $|\Psi\rangle$ and ${\cal O}$ refer to the pre-projected state and the operator, respectively, while $|\psi\rangle$ and ${\cal O}'$ are their post-projection counterparts. Based on the above consideration, one can identify the operator mapping
\begin{align}
X_{1,x}(\vec{r}_{1,i}+\hat{y})  & \xrightarrow{\cal P}
 X_{0}(\vec{v}_i), \nn 
X_{2,y}(\vec{r}_{2,i}+\hat{x}) & \xrightarrow{\cal P} X_{0}(\vec{v}_i),  \nn 
Z_{1,x}(\vec{r}_{1,i}+\hat{y}) Z_{2,y}(\vec{r}_{2,i}+\hat{x})  & \xrightarrow{\cal P} Z_{0}(\vec{v}_i), \nn 
Z_{1,x}(\vec{r}_{1,i}+\hat{y}) &  \xrightarrow{\cal P} 0 , \nn 
Z_{2,y} (\vec{r}_{2,i}+\hat{x}) & \xrightarrow{\cal P} 0 ,
\label{eq:x0-proj}
\end{align}
where the new subscript 0 indicates the condensed sites $\mathcal{V}_{\rm hv}$. 
Note that these arguments regarding the projection of operators are applicable solely to the Hilbert space that survives the projection.

%
Operators at the $\mathcal{V}_{\rm vh}$ sites are not affected by the projection and are simply re-labeled as
\begin{align}
X_{1,y}(\vec{r}_{1,i}) & \xrightarrow{\cal P} X_{2}(\vec{v}_i), & 
X_{2,x}(\vec{r}_{2,i}) & \xrightarrow{\cal P} X_{1}(\vec{v}_i), \nn 
Z_{1,y}(\vec{r}_{1,i}) & \xrightarrow{\cal P} Z_{2}(\vec{v}_i), &
Z_{2,x}(\vec{r}_{2,i}) & \xrightarrow{\cal P} Z_{1}(\vec{v}_i). 
\label{eq:xz12-proj} 
\end{align}
The post-projected $X, Z$ operators are defined with respect to the site $\vec{r}_i$, and carry three internal  indices 0,1,2. The pre-projection plaquette operators $b_{1}(\vec{r}_{1,i})$ and $b_{2}(\vec{r}_{2,i})$, with supports on $\Lambda_1$ and $\Lambda_2$ respectively, survive the projection ${\cal P}$ and become, after some re-labeling,
\begin{align}
b_{1,i} \xrightarrow{\cal{P}} \mathfrak{b}_i^x =&  X_{2}(\vec{v}_i)^{-1} X_{2}(\vec{v}_i + \hat{x}) X_{0}(\vec{v}_i)^{-1} X_{0}(\vec{v}_i-\hat{y}) ,\nn
b_{2,i} \xrightarrow{\cal{P}} \mathfrak{b}_i^y =&  X_{1}(\vec{v}_i) X_{1}(\vec{v}_i+\hat{y})^{-1} X_{0}(\vec{v}_i) X_{0}(\vec{v}_i-\hat{x})^{-1} . 
\label{eq:b1b2-proj}
\end{align}
Despite the re-labeling, they are the same stabilizers from the two underlying R1TCs. 

On the other hand, the pre-projection star operators $a_{1,i}$ and $a_{2,i}$ from $\Lambda_1$ and $\Lambda_2$ become zero under the projection as they contain only $Z_{1,x}$ or $Z_{2,y}$, but not both. To survive the projection,  $Z_{1,x}(\vec{r}_{1,i}+\hat{y})$ and $Z_{2,y}(\vec{r}_{2,i}+\hat{x})$ must appear simultaneously, as in the following operator
\begin{align}
a_i = & a_{1,i} a_{1,i-\hat{x}}^{-1} a_{2,i} a_{2,i-\hat{y}}^{-1}, 
\label{eq:pp-a}
\end{align}
which becomes, under the projection $a_i \xrightarrow{\cal{P}} \mathfrak{a}_i$,
\begin{align}\label{eq:r2tc-a-proj}
\mathfrak{a}_i  =& Z_{0}(\vec{v}_i) Z_{0}(\vec{v}_i\!-\!\hat{x})^{-1} 
Z_{0}(\vec{v}_i\!-\!\hat{y})^{-1} Z_{0}(\vec{v}_i\!-\!\hat{x}\!-\!\hat{y}) \nn 
& \otimes Z_{2}(\vec{v}_i\!-\!\hat{y}) Z_{2}(\vec{v}_i)^{-2} Z_{2}(\vec{v}_i\!+\!\hat{y})  \nn
& \otimes Z_{1}(\vec{v}_i\!-\!\hat{x}) Z_{1}(\vec{v}_i)^{-2} Z_{1}(\vec{v}_i\!+\!\hat{x}) .
\end{align}
The three post-projection stabilizers $\mathfrak{a}_i$, $\mathfrak{b}_i^x$, and $\mathfrak{b}_i^y$ 
are mutually commuting, and are none other than the stabilizers of the R2TC Hamiltonian. 

\begin{figure}[tb]
\includegraphics[width=0.45\textwidth]{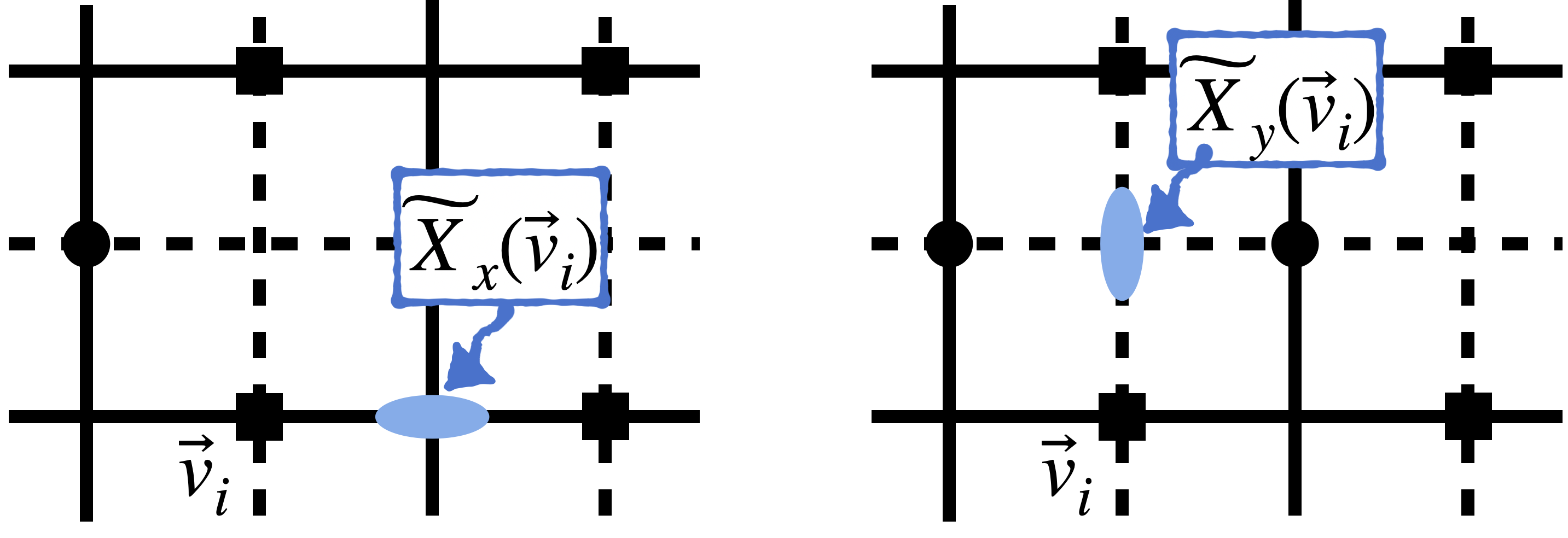}
\caption{
The emergent operator $\widetilde{X}_{x}(\vec{r}_i)$\,(left panel) is defined at $x$-bond (blue oval) with respect to ${\vec r}_i$, and $\widetilde{X}_{y}(\vec{r}_i)$\,(right panel) at the $y$-bond.  
}
\label{fig:b3-action}
\end{figure}

So far the discussion seems limited to the recovery of stabilizers that make up the R2TC. Importantly, though, there is an additional stabilizer one can identify in the pre-projected Hilbert space that is not given as a mere product of $a_i, b_{1,i}, b_{2,i}$. It is given by
\begin{align}
b_{3,i}  
 = \widetilde{X}_{x}(\vec{v}_i) \widetilde{X}_{y}(\vec{v}_i+\hat{x})  \widetilde{X}_{x}(\vec{v}_i+\hat{y})^{-1} \widetilde{X}_{y}(\vec{v}_i)^{-1} . 
\label{eq:pp-b3}\end{align}
The {\it emergent operators} $\tilde{X}_x , \tilde{X}_y$ are defined by
\begin{align}
\widetilde{X}_{x} (\vec{v}_i )
= &  
\bigl( X_{1,x}(\vec{r}_{1,i}+\hat{y}) \bigr)^{y_i-y_0-1} \bigl(X_{2,x} (\vec{r}_{2,i} +\hat{x})\bigr) ^{x_i-x_0} 
 \nn 
\widetilde{X}_{y}(\vec{v}_i) 
= & 
\bigl( X_{1,y} (\vec{r}_{1,i} +\hat{y})\bigr)^{y_i - y_0}
\bigl( X_{2,y}(\vec{r}_{2,i}+\hat{x}) \bigr)^{x_i - \!x_0 - 1} ,
\label{eq:super-x}\end{align}
and illustrated in Fig.~\ref{fig:b3-action}. The arbitrary constants $x_0$ and $y_0$ are kept here to simplify certain algebraic relations among the holonomies, and do not serve other purpose. Other stabilizers $b_{1,i}, b_{2,i}, a_i$ have the matching lattice gauge theory expressions given in Eq. (\ref{eq:quadrupole-operator}). As for $b_{3,i}$, the corresponding gauge field expression is the lattice curl 
\begin{align}
m'_{i} = (A' )_{i}^x - ( A' )_{i+\hat{y}}^x - ( A' )_{i}^y + ( A')_{i+\hat{x}}^y ,
\label{eq:new-combination}
\end{align}
where 
\begin{align}
( A' )_{i}^x = & (x_i-x_0 ) A_{i+\hat{x}}^{xx} + (y_i - y_0 - 1) A_{i}^{xy} , \nn 
( A' )_{i}^y = & (y_i-y_0 ) A_{i+\hat{y}}^{yy} + (x_i - x_0 - 1) A_{i}^{xy} .
\end{align}

Despite the apparent complexity of the definition of $b_{3,i}$, the virtue of this choice is that it allows us to express the product of $b_{3,i}$ as a product of boundary operators and thereby leads naturally to the new holonomies, as discussed thoroughly in Sec. \ref{sec:holonomies}. In fact, there is another choice, namely $b_{3,i} = X_{1,x}(\vec{r}_{1,i}+\hat{y}) X_{2,y}(\vec{r}_{2,i}+\hat{x})^{-1}$, which is composed of operators from both sublattices and commutes with $a_i, b_{1,i}, b_{2,i}$. Such a choice amounts to the condensation scheme adopted in Ref.~\cite{radzihovsky}. This choice, however, does not allow the transformation of the bulk product to the boundary product, hence no new holonomy operators can be generated.

The new field $m'_{i}$ commutes with $m^x_{i}, m^y_{i}, e_i$ and may seem to constitute the fourth charge in the rank-2 theory, but one can show that, after projection, $b_{3,i}$ becomes  $b_{3,i} \xrightarrow{\cal{P}} \left(\mathfrak{b}_{i+\hat{y}}^x\right)^{y_i -y_0 } \left(\mathfrak{b}_{i+\hat{x}}^y \right)^{x_i - x_0 }$ - a composite of existing stabilizers. The main use of identifying the stabilizer $b_{3,i}$ is that, through it, we come to identify the two emergent-$\widetilde{X}$ operators as given in Eq. (\ref{eq:super-x}). Naively, two copies of R1TC will generate only four holonomies, made of products of $X_{1,i}$ or $X_{2,i}$ along horizontal and vertical directions of the torus. The existence of the emergent operators allows the construction of two additional holonomies, as products of $\widetilde{X}_x$ along the $x$- and of $\widetilde{X}_y$ along the $y$-direction of the torus, and in total account for the six holonomies generating the GSD of R2TC, Eq. (\ref{eq:GSD}).

\subsection{Higher-order instanton and confinement \\ in the rank-2 U(1) gauge theory}
\label{subsec:instanton} 

While this paper focuses mainly on the $\mathbb{Z}_N$ gauge theory on a lattice, it is instructive to touch upon the physics of U(1) rank-2 compact gauge theory in the continuum for comparison. The Maxwell theory for the gauge fields of Eq.~(\ref{eq:generators}) is given by the effective Lagrangian,
\begin{align}
    \mathcal{L} = \left[  (E^{xx})^2+(E^{yy})^2+ 2 (E^{xy})^2 \right] -\frac{1}{2g} B^2 , 
\end{align}
with a quadratic dispersion $\omega \sim k^2$ due to the fact that $B$ is given by second spatial derivatives, $B = \partial_y^2 A^{xx} + \partial_x^2 A^{xx} - \partial_x \partial_y A^{xy}$. For a compact gauge theory with gapless fluctuations, the key question is whether the theory becomes confined due to the proliferation of instantons. To delineate the instanton event, we consider the pure gauge theory in the charge-neutral sector $e^x = e^y = 0$ in Eq.~(\ref{eq:generators}) that allows the solution 
\begin{align}
&E^{xx}= \partial_y^2 h, ~E^{yy}= \partial_x^2 h , ~ E^{xy}=-\partial_x \partial_y h . 
\end{align}

The $h$ field can be viewed as the height operator that is canonically conjugate with the flux $[B(\vec{r}),h({\vec r}')]=i\delta({\vec r}-{\vec r}')$ so that the instanton operator $e^{i2\pi h}$ creates a $2\pi$ flux~\cite{rasmussen}. Such an instanton event, once proliferated, can potentially lead to a confined phase. The low energy effective theory of the height field can be obtained by integrating out the gaussian fluctuation of $B$,
\begin{align} 
&\mathcal{L}_{h}= -g(\partial_t h)^2+ ({\bm \nabla}^2 h)^2 . 
\end{align}
The quantum theory of $h$ is defined in $2+1$D space-time with a quadratic dispersion reminiscent of the Rokhsar-Kivelson point in 2D compact gauge theory, suggesting that the instanton operator has a power-law decay correlation whose operator dimension depends on $g$. The relevance of $2\pi$ flux tunneling event and the proliferation of topological defects depends on the parameters of the theory. 

On the other hand, there exists another kind of higher-order instanton events that are more relevant. For instance the instanton operator $e^{i \partial_x h}$ creating a flux-dipole - a pair of $2\pi$ and $-2\pi$ fluxes spatially separated along the $x$-link - has the correlator~\footnote{The importance of flux-dipole tunneling events in governing the phase of matter was pointed out in the context of one-dimensional dipolar boson Hubbard model recently~\cite{lake,feldmeier}.} 
\begin{align} 
&e^{-(\partial_x h(0)\partial_x h({\vec r}))} \xrightarrow{r\rightarrow \infty}  ~\text{Const}  . \label{corr}
\end{align}
These higher-order instanton terms creating flux-dipole tunneling events display long-range order and thus can proliferate. As a result, the theory would be confined due to the proliferation of instanton-dipoles. This unique feature is due to the fact that the dipole flux is conserved in our higher-rank gauge theory and thus the $2\pi$ flux tunneling event must appear in a {\it quadrupolar process}, i.e., creating a pair of opposite flux-dipoles from the vacuum and separating them apart. The correlation function in Eq.~(\ref{corr}) implies the interaction between flux-dipoles are short-ranged so they will proliferate and gap out the low-energy modes. 

\subsection{Conservation laws}
\label{subsec:conservation}

Before the explicit construction of holonomies, it is useful to identity the full content of conserved charges in the theory. Physically, it is the winding of one of these conserved charges around the non-contractible loop of the torus that defines the holonomy. The discussion is most conveniently carried out in the continuum language. 

The three expressions in Eq. (\ref{eq:quadrupole-operator}) can be cast in the continuum as
\begin{align}
m^x = & \, \partial_x A^{yy} - \partial_y A^{xy},\nn 
m^y = & \, \partial_x A^{xy} - \partial_y A^{xx}, \nn 
e = & \, \partial_x^2 E^{xx} + \partial_y^2 E^{yy} + \partial_x \partial_y E^{xy} .
\label{eq:gauss-law}
\end{align}
The three charge densities obey the continuity equations as derived recently~\cite{oh22b},
\begin{align}
\partial_t {m}^{x}+ \partial_x J_m^{xx} + \partial_y J_m^{xy} & = 0, 
\nn 
\partial_t {m}^{y}+ \partial_x J_m^{xy} - \partial_y J_m^{yy} & = 0, 
\nn 
\partial_t e + \partial_x^2 J_e^{xx} +\partial_x \partial_y J_e^{xy} + \partial_{y}^2 J_e^{yy} &  = 0 , 
\label{eq:continuity}
\end{align}
where $J_m^{ab}$ and $J_e^{ab}$ ($a,b = x,y$) are symmetric rank-2 current densities for the magnetic and electric charges, respectively\,\footnote{
The continuity equations derived in \cite{oh22b} were for the vector electric charge. Here we are dealing with the dual theory with vector magnetic charge. The two theories are dual to each other~\cite{oh22a}, and the continuity equations for the quasiparticles have the same structure.}. By assuming vanishing currents at the boundary, one can show that all three monopole charges are conserved:
\begin{align}
\partial_t \int e dV  = \partial_t \int m^x dV  = \partial_t \int m^y dV = 0.
\end{align}

In addition, we have three dipole conservation laws 
\begin{align}
& \partial_t \int x e dV \! = \! -\!  \int d^2 {\bm r} ~x \left[\partial_x^2 J_e^{xx} +\partial_x \partial_y J_e^{xy} + \partial_{y}^2 J_e^{yy} \right]\! = 0, \nn
& \partial_t \int y e dV \! = \! -\!  \int d^2 {\bm r} ~y \left[\partial_x^2 J_e^{xx} +\partial_x \partial_y J_e^{xy} + \partial_{y}^2 J_e^{yy} \right]\! 
= 0, \nn
& \partial_t \int (x m^y + y m^x ) dV   \! 
 = - \int d^2 {\bm r} \left[ x \partial_x J^{xy}_m - y \partial_y J^{xy}_m \right]  = 0.
\end{align}

Altogether we have the conservation of three monopoles and three dipoles. We will now construct the six magnetic and six electric holonomies associated with the $x$- and $y$-winding around the torus of the six conserved quantities.

\section{Applications of the Condensation Scheme}
\label{sec:application}
Two useful applications of the condensation idea are considered. One is the explicit construction of the tensor network wave function for the ground state of R2TC. The second is the construction of the rank-2 version of the double-semion model. 

\subsection{Tensor network representation of \\ $\mathbb{Z}_N$ R2TC wavefunctions}
\label{subsec:TN}

In this section, we show that $\mathbb{Z}_N$ R2TC wave function can be obtained by stacking two copies of $\mathbb{Z}_N$ R1TC wave function followed by a certain isometric operation that reflects the gauge-field constraint of the previous section. 
To this end, we begin with the tensor network\,(TN) representation of the R1TC ground state wave function~\cite{verstraete06,Aguado2008,Vidal2009} that is composed of two types of tensors $g$ and $T$ as below:
%
\begin{align}
    \includegraphics[width=0.3\textwidth]{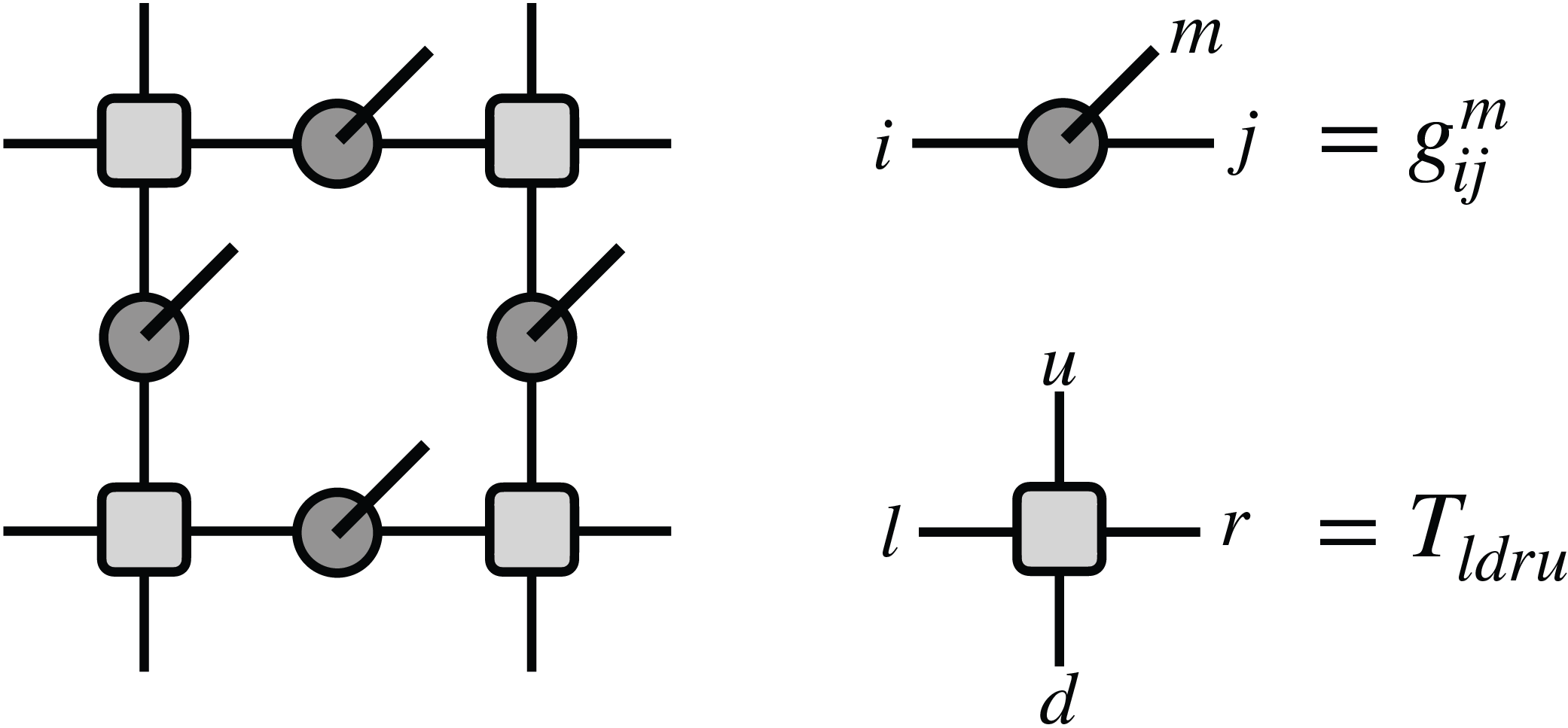},   
\end{align}
where
\begin{align} 
g_{ij}^m & = \delta_{i,j} \delta_{j,m} , \nn
T_{lurd} & = \delta_{r+u, l+d} .  \label{eq:g-and-T} 
\end{align} 
The delta function in the second line is implemented mod $N$. The physical index $m$ represents the qudit state in the $Z$-basis, i.e., $Z|m\rangle = \omega^m |m\rangle$, and all subscripts denote the virtual indices of dimension $N$. One can easily show that $g$ and $T$ tensors satisfy the following relations:
\begin{align}
    &[Z^n]_{mm'} g_{ij}^{m'} = [Z^{n'}]_{ii'}[Z^{n-n'}]_{jj'} g_{i'j'}^m,\\
    &[X^n]_{mm'} g_{ij}^{m'} = [X^{-n}]_{ii'}[X^{-n}]_{jj'} g_{i'j'}^m,\\
    &[Z^n]_{ll'}[Z^{-n}]_{uu'}[Z^{-n}]_{rr'}[Z^n]_{dd'}  T_{l'u'r'd'} = T_{lurd},
\end{align}
and 
\begin{align}
    &[X^{n_l}]_{ll'}[X^{n_u}]_{uu'}[X^{n_r}]_{rr'}[X^{n_d}]_{dd'}  T_{l'u'r'd'} = T_{lurd},
\end{align}
if $(n_r-n_l +n_u-n_d)\,\,{\rm mod} \,\, N = 0$. Graphical representations of the above equations are the following:
\begin{align}
    \includegraphics[width=0.4\textwidth]{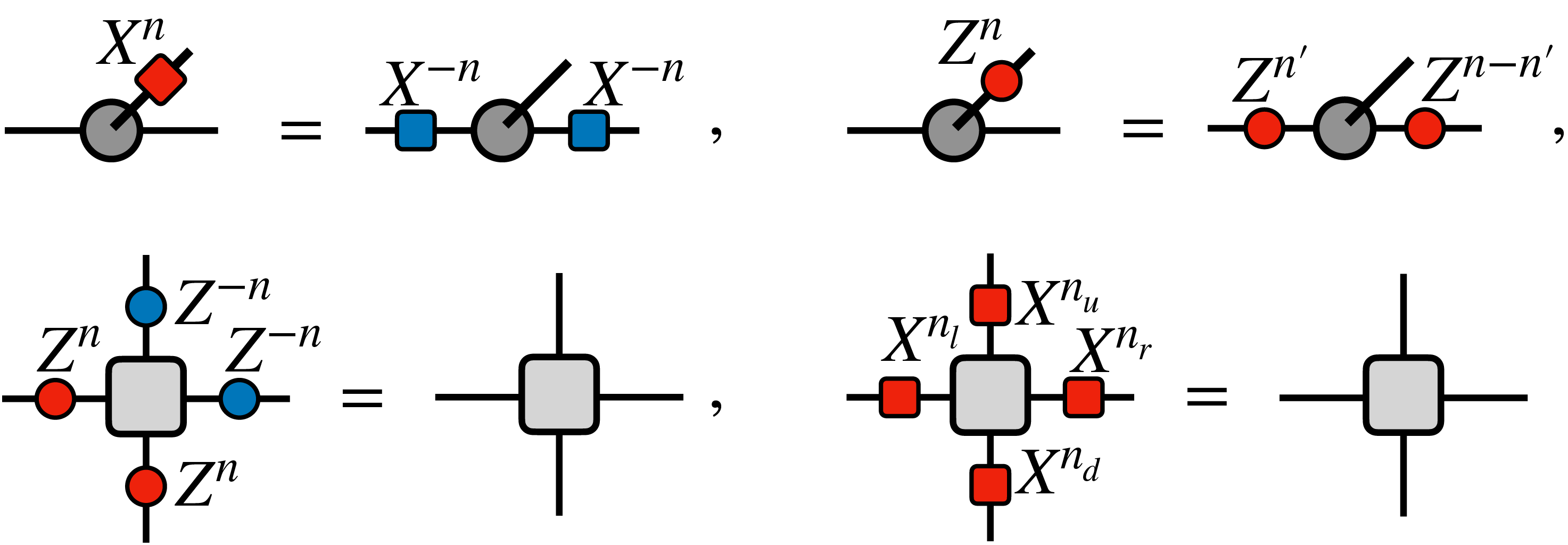}
    \label{eq:tensor_relations}
\end{align}

Note that the $T$-tensor generates the string-net configurations corresponding to the domain wall configurations of the $N$-state Potts model on the square lattice. For example, $\mathbb{Z}_2$ R1TC wave function is depicted as a superposition of closed-loop configurations, i.e., the domain wall of the Ising model. Using the above relations, one can easily verify that the TN wave function $|\psi\rangle$, obtained by contracting all the virtual indices, is the ground state of the $\mathbb{Z}_N$ R1TC Hamiltonian, i.e., $a_i|\psi\rangle = |\psi\rangle$ and $b_i|\psi\rangle = |\psi\rangle$, or graphically as below,
\begin{align}
    \includegraphics[width=0.4\textwidth]{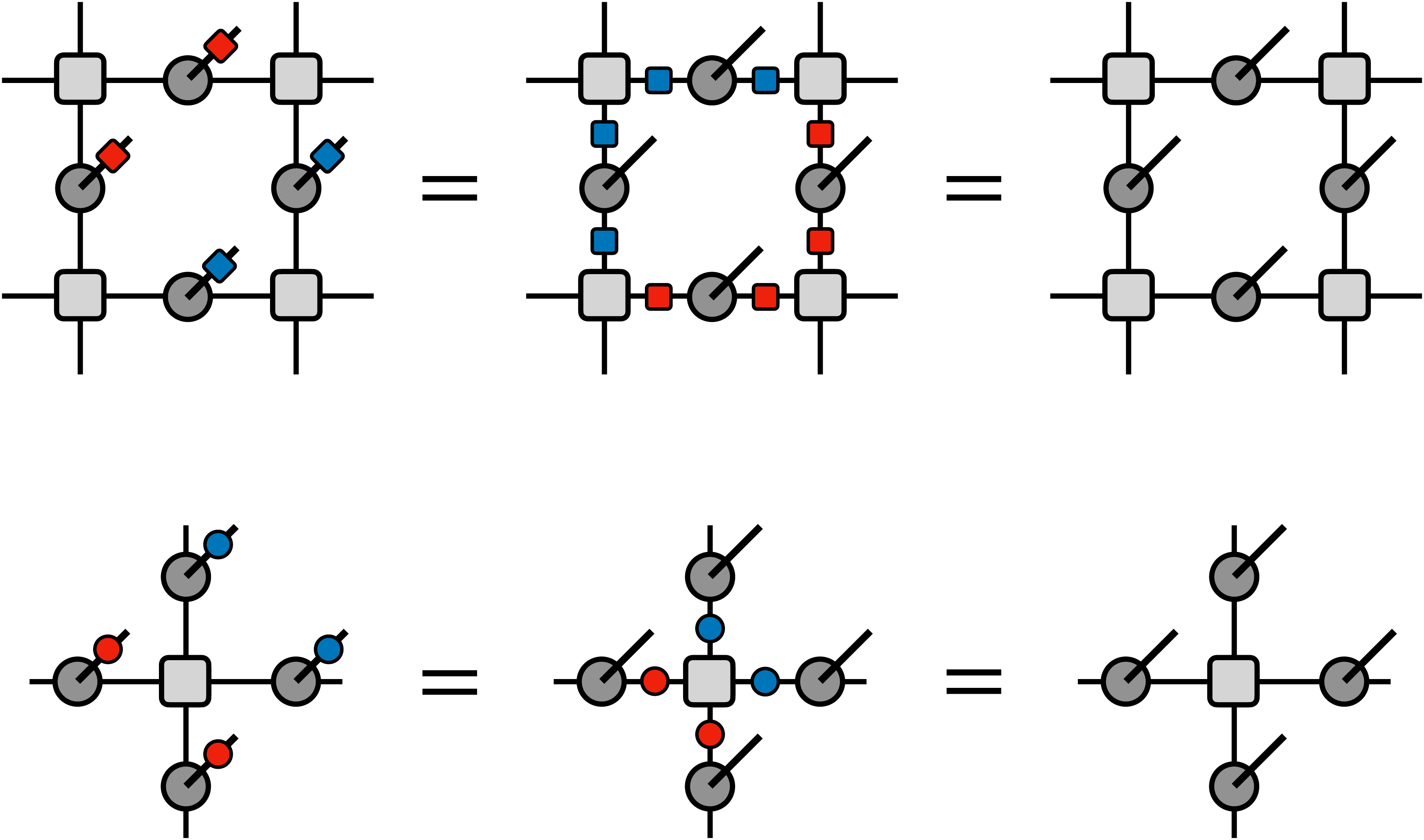}.
\end{align}

Now, we consider the square lattice\,($\Lambda_1$) and its dual\,($\Lambda_2$) together, and accommodate the $\mathbb{Z}_N$ R1TC wave function on each lattice, i.e., $|{\rm R1TC}\rangle_{\Lambda_1} \otimes |{\rm R1TC}\rangle_{\Lambda_2}$. Then, there are two types of vertices in the system: $\mathcal{V}_{\rm hv\,(vh)}$ at which horizontal\,(vertical) bonds in $\Lambda_1$ and vertical\,(horizontal) bonds in $\Lambda_2$ cross each other. Generally, two unentangled qudits live on the vertex $\mathcal{V}_{\rm hv} \oplus \mathcal{V}_{\rm vh}$. Now we impose the following isometry on the two qudits labeled by quantum numbers $(m_1 , m_2)$ residing on the $\mathcal{V}_{\rm vh}$ vertices: 
\begin{align}
    P_{m_1 m_2}^m =  \delta_{m, m_1 + m_2},
    \label{eq:isometry-tensor}
\end{align}
where the delta function is implemented mod $N$, and $ P_{m_1 m_2}^m P_{m_1 m_2}^{m'} = \delta_{m m'} $. The two-qudit state is mapped to a single-qudit state through isometry and, furthermore, the resulting TN exactly represents the ground state of $\mathbb{Z}_N$ R2TC. The TN wave function thus constructed is written in the $Z$-basis, $Z|m\rangle = \omega^m |m\rangle$, and the constraint Eq. (\ref{eq:spin-Z-proj}) is faithfully reflected through the isometry tensor $P^{m}_{m_1 m_2} = \delta_{m, m_1 + m_2}$. 

The TN representation for the R2TC ground state is illustrated below:
\begin{align}
    \centering
    \includegraphics[width=0.4\textwidth]{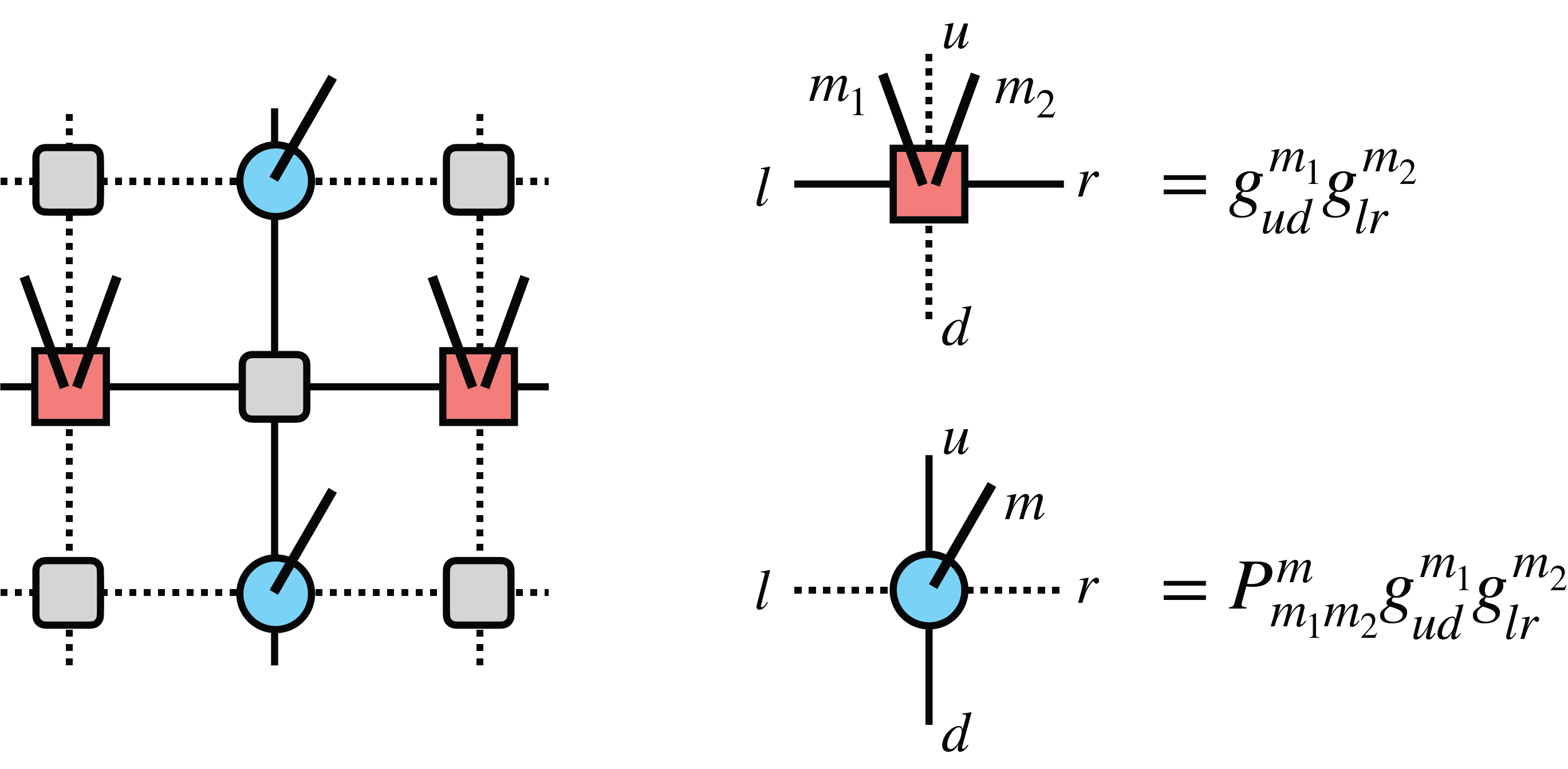},
\end{align}
where the square lattice\,(dual lattice) in solid\,(dotted) line denotes $\Lambda_{1\,(2)}$, and the gray square stands for the $T$-tensor given in Eq. (\ref{eq:g-and-T}). The isometry $P$ satisfies the relations,
\begin{align}
    & [Z^n]_{mm'} P_{m_1 m_2}^{m'} 
    = [Z^n]_{m_1 m_1'} [Z^n]_{m_2 m_2'} P_{m_1' m_2'}^{m},\nonumber\\ 
    & [X^n]_{mm'} P_{m_1 m_2}^{m'} 
    = [X^{-n'}]_{m_1 m_1'} [X^{n'-n}]_{m_2 m_2'} P_{m_1' m_2'}^{m},
\end{align}
or graphically 
\begin{align}
    \includegraphics[width=0.38\textwidth]{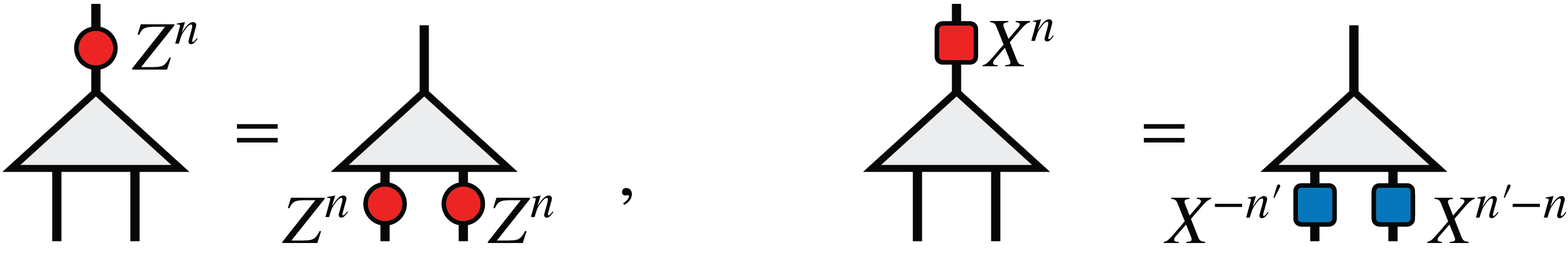}.
    \label{eq:p_relations}
\end{align}
Using Eqs.\,\eqref{eq:tensor_relations} and \eqref{eq:p_relations}, it is straightforward to derive the following relation:
\begin{align}
    &[X^n]_{mm'} P_{m_1 m_2}^{m'} g_{ud}^{m_1} g_{lr}^{m_2} \nonumber\\
    & = P_{m_1 m_2'}^{m} g_{ud}^{m_1} [X^{-n}]_{ll'} [X^{-n}]_{rr'} g_{l'r'}^{m_2},\nonumber\\
    & [Z^n]_{mm'} P_{m_1 m_2}^{m'} g_{ud}^{m_1} g_{lr}^{m_2} \nonumber\\
    & = P_{m_1 m_2}^{m} g_{u'd}^{m_1} g_{l'r}^{m_2} [Z^n]_{uu'}[Z^n]_{ll'} \nonumber\\
    & = P_{m_1 m_2}^{m} g_{u'd}^{m_1} g_{lr'}^{m_2} [Z^n]_{uu'}[Z^n]_{rr'} \nonumber\\
    & = P_{m_1 m_2}^{m} g_{ud'}^{m_1} g_{lr'}^{m_2}[Z^n]_{dd'}  [Z^n]_{rr'}\nonumber\\
    & = P_{m_1 m_2}^{m} g_{ud'}^{m_1} g_{l'r}^{m_2} [Z^n]_{dd'}[Z^n]_{ll'},
\end{align}
or graphically
\begin{align}
    \includegraphics[width=0.4\textwidth]{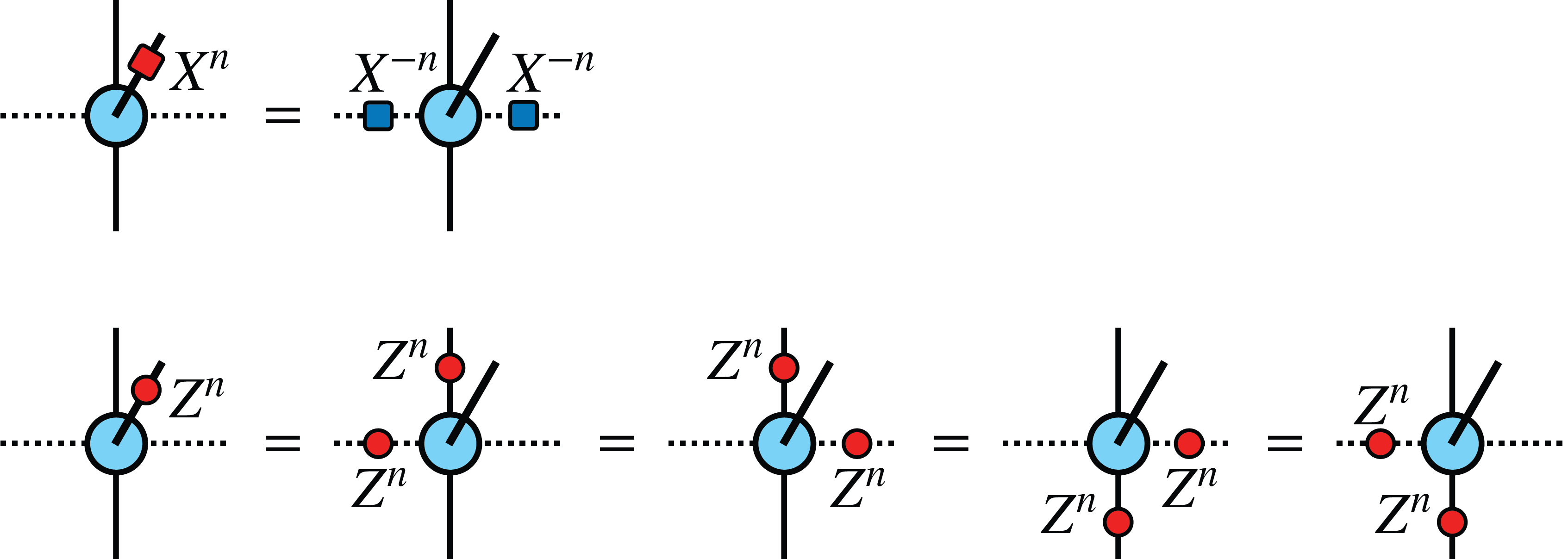}.
    \label{eq:b_relations}
\end{align}

Now, using Eqs.\,\eqref{eq:tensor_relations} and \eqref{eq:b_relations}, we show that the above TN wave function, $|\psi\rangle$, is the ground state of the $\mathbb{Z}_N$ R2TC Hamiltonian, i.e., $\mathfrak{b}_i^x |\psi\rangle = |\psi\rangle$, $\mathfrak{b}_i^y |\psi\rangle = |\psi\rangle $ as below
\begin{align}
\includegraphics[width=0.4\textwidth]{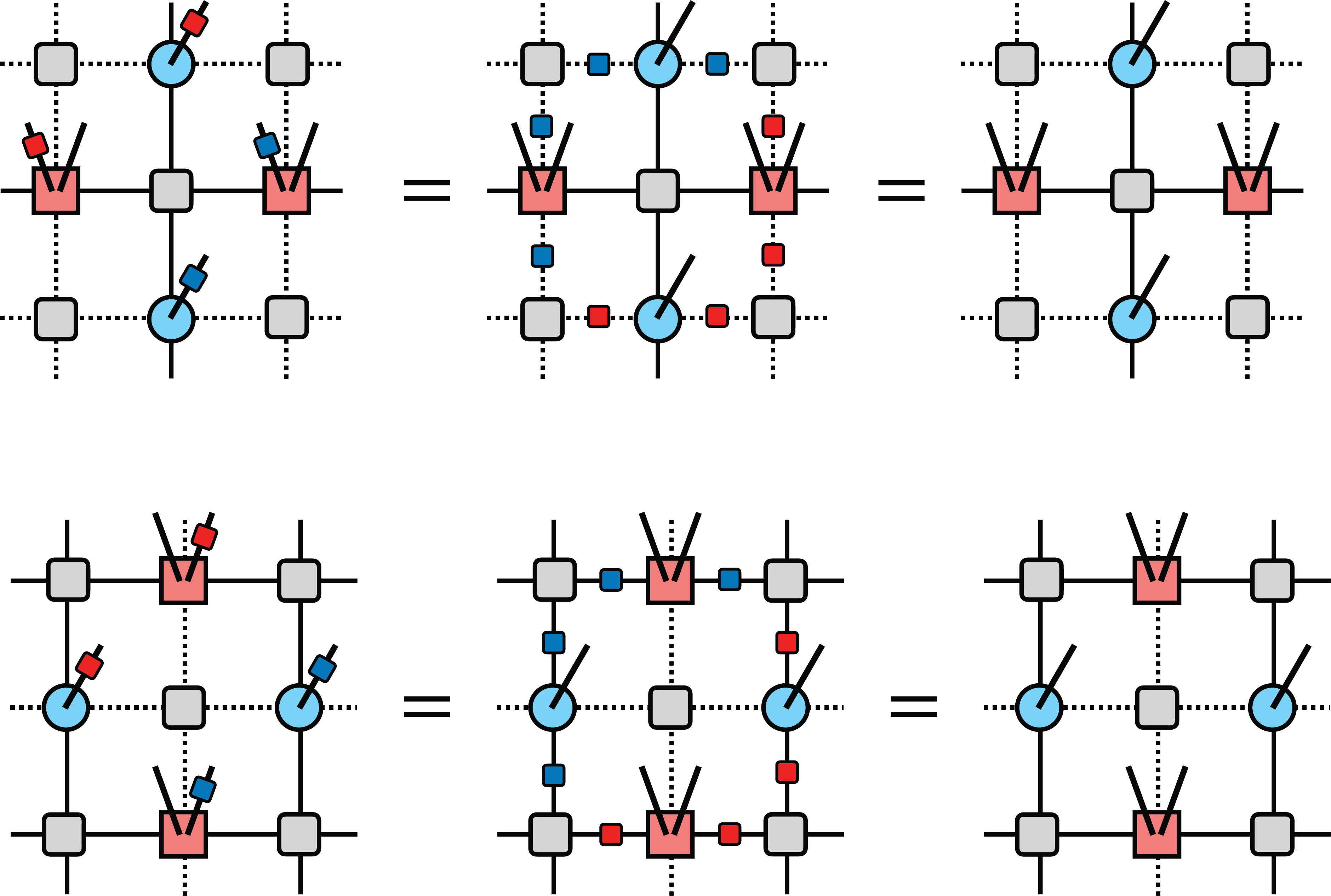},
\label{eq:peps-b-eig}
\end{align}
and $\mathfrak{a}_i |\psi\rangle = |\psi\rangle $ in the following way,
\begin{align}
\includegraphics[width=0.5\textwidth]{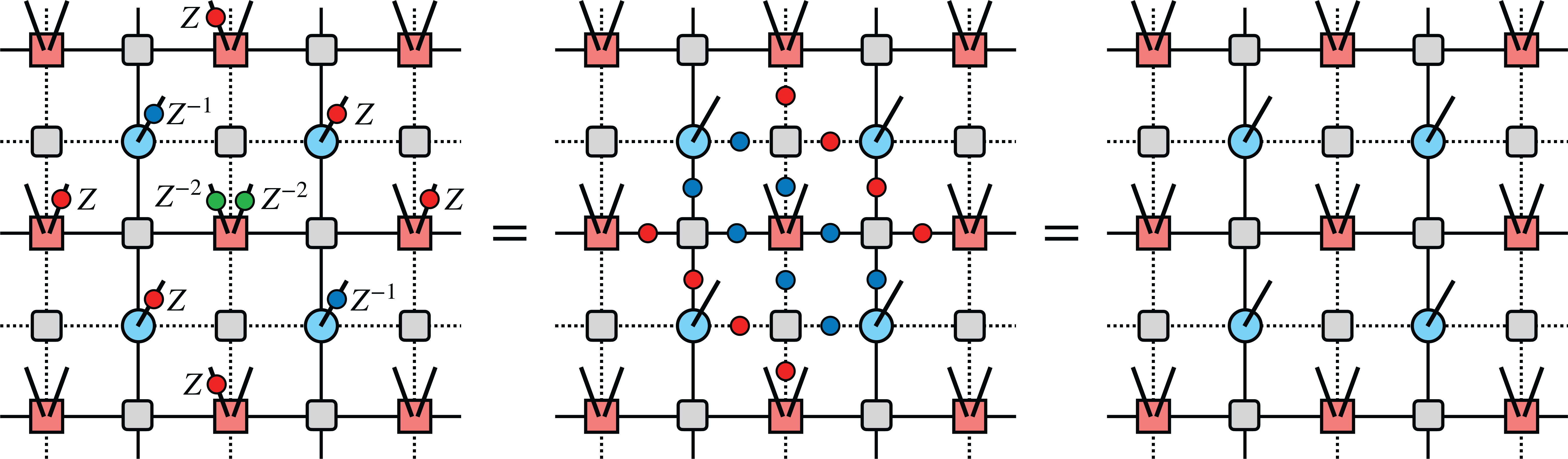}.
\label{eq:peps-a-eig}
\end{align}

This completes the proof that the TN ground state wave function of R2TC is given as two copies of those of R1TC with an additional isometry. To summarize, the ground state wave function of the R1TC is constructed using the well-known tensors given in Eq. (\ref{eq:g-and-T}). Two copies of such TN wave functions are introduced one for each of the two interpenetrating square lattices. Then the isometry operation $P^m_{m_1 m_2}$ given in Eq. (\ref{eq:isometry-tensor}) acts on half of the overlapping sites (the ${\cal V}_{\rm vh}$ sites) to reduce the two qudits $(m_1 , m_2)$ to a single qudit $m = m_1 + m_2$. 

There are a large number of ground states given by the GSD formula, Eq. (\ref{eq:GSD}), and our TN construction captures only one of them. 
By employing a similar approach to that used in Eq. (\ref{eq:peps-b-eig}) and (\ref{eq:peps-a-eig}), one can verify that the TN ground state is an eigenstate of holonomies $W_1$, $W_2$, $W_3$, $W_4$, $\widetilde{W}_5$, and $\widetilde{W}_6$, which will be derived in the next section, and that all eigenvalues are equal to $1$.
The rest of the states can be generated by applying holonomy operators to the existing TN wave function. 

\subsection{Twisted rank-2 gauge theory from anyon condensation}
\label{subsec:twisted-R2TC}

In this section, we utilize the coupled layer construction protocol to build a twisted rank-2 gauge theory in 2D with dipole conservation, whose gauge flux turns out to have semionic statistics.  The strategy is to combine two intersecting $\mathbb{Z}_2$ twisted gauge theories from string-net models~\cite{levin05} and implement anyon condensation to impose restricted mobility for quasiparticle excitations. To avoid technical complexities, the discussion in this section is limited to $N=2$.

\begin{figure}[t!]
\centering
\includegraphics[width=.48\textwidth]{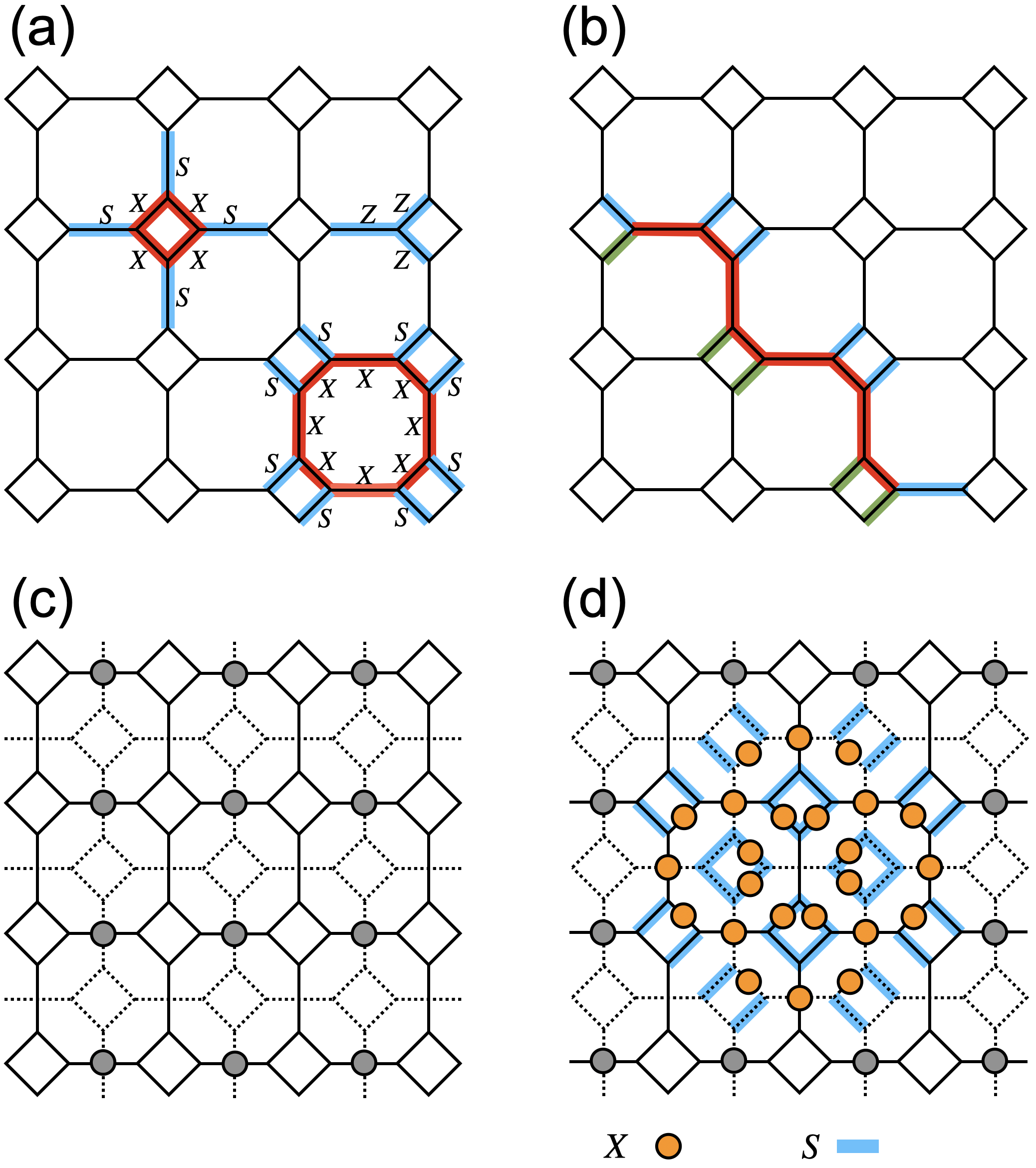}
\caption{(a) The double semion model on the Fisher lattice. The flux operator $\prod_{i \in Y} Z_i$ is defined on the vertex with three $Z$ operators on the adjoining links. The charge operators $\prod X$ are defined on the diamond and the octagon. (b) The string operator in the double semion model (see text for details). (c) Intersecting bilayers of the Fisher lattice, illustrated as one solid and one dashed lines. The circles are the intersection between $x$-link from the first layer and the $y$-link from the second layer where we put a strong coupling term $-J_z Z^{1} Z^{2}$. (d) The charge operator after perturbation contains the product of four octagon operators.
}
\label{fig:ds}
\end{figure}

To construct the ``semionic'' version of $\mathbb{Z}_2$ gauge theory from commuting projectors, we will need to start from trivalent 2D lattices such as the Fisher lattice shown in Fig.~\ref{fig:ds}. A small diamond shape is added at each vertex of the square lattice so that every vertex is connected to three links with $\mathbb{Z}_2$ qubits living on them.

Now we begin with the conventional double semion model that manifests a twisted $\mathbb{Z}_2$ gauge theory~\cite{levin05} on the Fisher lattice with the Hamiltonian,
\begin{align} 
&H=-\sum_Y \prod_{i \in Y} Z_i - \sum_{O} F_0 \prod_{i \in O } X_i  - \sum_{D}  F_1 \prod_{i \in  D} X_i   \nn 
&F_0= \prod_{i \in V_{O}} S_i , ~~ F_{1}= \prod_{i \in V_{D}} S_i, ~~~
S_i=\begin{pmatrix}
1 & 0 \\
0 & i 
\end{pmatrix}
\label{gauge}
\end{align}
Here $Y$ is any set of three co-planar links entering a vertex, $O$ is the octagon, and $D$ is the diamond on the Fisher lattice. $V_{O}$ ($V_D$) refers to the eight (four) links pointing outward from an octagon (diamond), and are indicated by $S$ in Fig. \ref{fig:ds}(a). In the $Z$ basis, the first term in $H$ imposes a condition that the parity of the gauge flux entering any vertex be even. The remaining two terms in $H$ provide dynamics to the gauge field while preserving this parity at each vertex. Specifically, they effectively bind the charge, given by $\prod X$, to the gauge flux as measured by $F_0$ and $F_1$. 

This charge-flux binding has important consequences for the braiding statistics, as creating a flux excitation would simultaneously generate half $Z_2$ charge. The string operator $L$ that creates a pair of flux is
\begin{align} 
&L_{m}=\prod_{a \in {\rm  red}} X_a \prod_{a \in {\rm blue}}  S_a
\prod_{a \in {\rm green}} (-1)^{G_a},\nonumber\\
&G_a=\frac{1}{4}(1-Z_a) , 
\end{align}
where the $X$ operator along the red links in Fig. \ref{fig:ds}(b) flips the spins along the string, analogous to the flux operator in the toric code. The red string has a product of $X$ operators while the operators on blue (green) lines living on either side of the string embellish it with an additional sign structure that endows the semionic statistics between the flux, as well as ensuring that $L_m$ commutes with the Hamiltonian except near the endpoints of the line. Two $L_{m}$ operators anti-commute with each other when they intersect. The extra sign structure embellished a half-charge with the flux, and, as a result, the flux excitation carries a half gauge charge and thus displays semion statistics.

Now we take two intersecting layers of the Fisher lattice with the $x$-link from the first layer intersecting the $y$-link from the second layer and vice versa, as shown in Fig.~\ref{fig:ds}(c). Links that form the diamonds do not overlap between the layers.
We then strongly couple the qubits from distinct layers on the circled intersections in Fg.~\ref{fig:ds} through the interaction,
\begin{align} 
&-J_z Z^{1}_i Z^{2}_i
\end{align}
with $J_z \gg 1$. In this strong coupling limit, the vertex operators on $Y$-junctions and the charge operators $\prod X$ on diamonds are unaffected. However, charge operators around octagons do not commute with the $J_z$ term, and instead, a product of four such terms appearing at the third order in perturbation theory does. In Fig.~\ref{fig:ds}(d), we illustrated the new charge operator after perturbation, composed of the product of four octagon operators from $J_z=0$. The resulting Hamiltonian takes a form very similar to the R2TC, except that the diamond plaquette terms and the quadrupolar octagon terms are supplemented with a product of $S$ operators over outward-pointing edges. It can be checked that all the terms in the new Hamiltonian commute with one another.

What is the excitation structure of this semionic version of R2TC? The charge excitations share a similar character as the scalar charge theory in Eq.~(\ref{eq:quadrupole-operator}). The novelty comes from the vector flux excitation. For example, consider a 1D particle moving in the $x$-direction which creates a flux $m^x$. Suppose that the line on which this 1D excitation move intersects that of another flux $m^x$ moving in the $y$- direction, which can only hop on the even-numbered sites. The string operators associated with the two 1D particles anti-commute. This anti-commutation of string operators is related to the fact that two such flux excitations can undergo a full braiding, so their mutual statistics is well defined. In this case, the two types of flux have mutual statistics $\theta = \pi$, which contrasts with trivial mutual statistics 
$\theta = 0$ in the original R2TC model.

One can also find the TN representation for the rank-2 double semion\,(DS) wave function in a similar manner. The ground state of the rank-1 DS model in Eq.\,\eqref{gauge} is a loop gas state with a particular sign structure:
\begin{align}
    |\Psi_{\rm DS}\rangle = \sum_{\Gamma} (-)^{N_{\Gamma}} |\Gamma\rangle,
\end{align}
where $N_\Gamma$ denotes the number of loops of which the length is $4n+2$ in the configuration $\Gamma$. 
To obtain the TN representation of $|\Psi_{\rm DS}\rangle$,\cite{Gu2008} one needs to place the $g$ and $\bar{T}$ tensors on the edges and vertices of the Fisher lattice, respectively, where 
\begin{align}
    \bar{T}_{ijk} = 
    \begin{cases}
    i \quad {\rm if}\quad (i+j+k)\,\, {\rm mod}\,\,2 = 0 \\ 
    0 \quad {\rm otherwise}.
    \end{cases}
\end{align}
Next, two TN states should be placed on the original and its dual lattices, as depicted in Fig.\,\ref{fig:ds}\,(c), and two qubits on the edges indicated by the gray circle should be projected in a similar way as in the R2TC case. This procedure results in the TN representation of the rank-2 DS wavefunction.

\section{Holonomy Construction}
\label{sec:holonomies}

\subsection{Pre-projection holonomies}\label{sec:preholonomies}
There are six pre-projection holonomies consisting of the product of $X$ operators. The first four holonomies are taken directly from those of two independent R1TC's,  
\begin{align}
W_{1}^{\rm pp}(y_i) = & \prod_{x_i=1}^{L_x} X_{1,x}(\vec{r}_{1,i}), & 
W_{2}^{\rm pp}(x_i) = & \prod_{y_i=1}^{L_y} X_{1,y}(\vec{r}_{1,i}), \nn 
W_{3}^{\rm pp}(y_i) = & \prod_{x_i=1}^{L_x} X_{2,x}(\vec{r}_{2,i}), & 
W_{4}^{\rm pp}(x_i) = & \prod_{y_i=1}^{L_y} X_{2,y}(\vec{r}_{2,i})
\label{eq:x-hol-14} \end{align}
for a torus of size $L_x \times L_y$. Here, $\vec{r}_{1,i} = \vec{v}_i-\hat{y}/2$, and $\vec{r}_{2,i} = \vec{v}_i - \hat{x}/2$, where $\vec{v}_i = (x_i,y_i)$. All of them commute with the pre-projection stabilizers $a_i, b_{1,i}, b_{2,i}$ and $b_{3,i}$ introduced in Sec. \ref{subsec:pre-proj}. The two additional holonomies are constructed from the emergent operators $\widetilde{X}_x$ and $\widetilde{X}_y$:
\begin{align}
W_{5}^{\rm pp}(y_i) = & \prod_{x_i=1}^{{\rm lcm}(L_x,N) } \!\widetilde{X}_{x}(\vec{v}_i), \nn 
W_{6}^{\rm pp}(x_i) = & \prod_{y_i=1}^{{\rm lcm}(L_y,N) } \!\widetilde{X}_{y}(\vec{r}_i) . 
\label{eq:x-hol-56}
\end{align}
Note that in these two cases the product over $x_i$\,($y_i$) goes around the torus multiple times, i.e. by 
\begin{align} 
c_x = &  {\rm lcm}(L_x,N)/L_x = N/{\rm gcd}(L_x , N ) ,  \nn   
c_y =& {\rm lcm}(L_y,N)/L_y = N/{\rm gcd}(L_y, N ) , 
\label{eq:winding-numbers} 
\end{align} 
to ensure that the holonomy action on a ground state returns another ground state with no residual excitations~\cite{pace-wen}. As a consequence we have 
\begin{align}
\left[W_{5}^{{\rm pp}} \right]^{{\rm gcd}(L_x, N)} = \left[W_{6}^{{\rm pp}} \right]^{{\rm gcd}(L_y , N)} = 1 
\nonumber 
\end{align}
while for other holonomies it is $(W )^N = 1$. 
The six holonomies are seemingly coordinate-dependent, but this dependence goes away when their actions on the ground state are examined. We omit the proof, which is purely technical, since the result is well-anticipated. 

The holonomies we constructed can be motivated in a different way. One can show that the product of $b_{1,i}$ or $b_{2,i}$ over all sites inside a rectangle ${\cal S} = [x_1,x_2] \times [y_1,y_2]$ is equal to the product of $X$'s or $X^{-1}$'s along its four boundaries as all terms in the interior cancel out. These boundary operators precisely take the form of $W^{\rm pp}_1$ through $W^{\rm pp}_4$. In a similar fashion, product of $b_{3,i}$ over a closed area leads to the cancellation of all terms in the interior, leaving only the product of emergent-$X$ operators along the boundary. These boundary operators motivate the $W^{\rm pp}_5, W^{\rm pp}_6$ holonomies. 

Notably, the holonomies $W^{\rm pp}_5$ and $W^{\rm pp}_6$ are not obtainable through linear combinations of the remaining four holonomies, $W^{\rm pp}_1$, $W^{\rm pp}_2$, $W^{\rm pp}_3$, and $W^{\rm pp}_4$. Consequently, there exist six holonomies that are independent of one another. Collectively we refer to the six logical operators in Eqs. (\ref{eq:x-hol-14}) and (\ref{eq:x-hol-56}) as $X$-holonomies.

The six $Z$-holonomies are constructed by following a similar reasoning. One can show that the product of $a_i$ in Eq. (\ref{eq:pp-a}) inside a closed region reduces to the boundary product, which motivates the two $Z$-holonomies:
\begin{align}
\widetilde{W}_{5}^{\rm pp}(x_i) = & \prod_{y_i = 1}^{L_y} Z_{2,x}(\vec{r}_{2,i}+\hat{x}) Z_{2,x}(\vec{r}_{2,i})^{-1} , \nn 
\widetilde{W}_{6}^{\rm pp}(y_i) = &  \prod_{x_i = 1}^{L_x} Z_{1,y}(\vec{r}_{1,i}+\hat{y}) Z_{1,y}(\vec{r}_{1,i})^{-1} . 
\label{eq:pre-Z56}
\end{align}
Hereafter we drop the explicit coordinate dependence from the holonomy operators. Taking the product of $(a_i)^{y_i-y_0'}$ on a closed region and extracting the boundary terms gives two other $Z$-holonomies:
\begin{align}
\widetilde{W}_{1}^{\rm pp} &= \! \prod_{y_i = 1}^{{\rm lcm}(L_y,N)} \! \widetilde{Z}_{1,x}(\vec{v}_{i}),  &
\widetilde{W}_{2}^{\rm pp} &= \! \prod_{x_i = 1}^{L_x} \widetilde{Z}_{1,y}(\vec{v}_{i}) . 
\label{eq:pre-Z12}
\end{align}
Finally, the product of $(a_i)^{x_i-x_0'}$ gives
\begin{align} 
\widetilde{W}_{3}^{\rm pp} &=\! \prod_{y_i = 1}^{L_y} \widetilde{Z}_{2,x}(\vec{v}_{i}), 
&
\widetilde{W}_{4}^{\rm pp} &=\! \prod_{x_i = 1}^{{\rm lcm}(L_x,N)}\! \widetilde{Z}_{2,y}(\vec{v}_{i}),
\label{eq:pre-Z34}
\end{align}
Various emergent-$Z$ operators appearing in the holonomies are 
\begin{align}
\widetilde{Z}_{1,x}(\vec{v}_i) = & Z_{1,x}(\vec{r}_{1,i}) Z_{2,y}(\vec{r}_{2,i}+\!\hat{x} - \hat{y}) 
\nn & \otimes 
\left(Z_{2,x}(\vec{r}_{2,i}) Z_{2,x}(\vec{r}_{2,i} + \hat{x})^{ - 1} \right)^{y_i - y_0'-2} , \nn 
\widetilde{Z}_{1,y}(\vec{v}_i) = & Z_{1,y}(\vec{r}_{1,i}) \left(Z_{1,y}(\vec{r}_{1,i}) Z_{1,y}(\vec{r}_{1,i}+\hat{y})^{-1}\right)^{y_i-y_0'-1},  \nn 
\widetilde{Z}_{2,x}(\vec{v}_i) = & Z_{2,x}(\vec{r}_{2,i}) \left(Z_{2,x}(\vec{r}_{2,i}) Z_{2,x}(\vec{r}_{2,i}+\hat{x})^{-1} \right)^{x_i-x_0'-1}, \nn 
\widetilde{Z}_{2,y}(\vec{v}_i) = & Z_{2,y}(\vec{r}_{2,i}) Z_{1,x }(\vec{r}_{1,i} - \hat{x} + \hat{y})
\nn & \otimes 
\left( Z_{1,y}(\vec{r}_{1,i}) Z_{1,y}(\vec{r}_{1,i} + \hat{y})^{-1} \right)^{x_i - x_0'-2} . \end{align}
Note that $\widetilde{W}_{1}$ and $\widetilde{W}_{4}$ involve the product of the emergent operators over the circumference of the torus $c_y$ and $c_x$ times, respectively. Arbitrary constants $x'_0 , y'_0$ are introduced for generality and for simplifying certain aspect of the holonomy algebra.

Invoking $ZX = \omega XZ$, one can verify the following Heisenberg algebra among the $X$- and $Z$-holonomies:
\begin{align}
\left[ \widetilde{W}_1^{\rm pp} , W_1^{\rm pp} \right]  & = \omega^{c_y} ,   
&
\left[ \widetilde{W}_2^{\rm pp}, W_2^{\rm pp}  \right] & = \omega  ,
\nn
\left[ \widetilde{W}_3^{\rm pp} , W_3^{\rm pp} \right] & = \omega  ,
&
\left[ \widetilde{W}_4^{\rm pp} , W_4^{\rm pp} \right] & = \omega^{c_x}  , 
\nn
\left[ \widetilde{W}_5^{\rm pp},  W_5^{\rm pp} \right] & = \omega^{c_x} , 
&
\left[ \widetilde{W}_6^{\rm pp} , W_6^{\rm pp} \right] & = \omega^{c_y} .
\label{eq:pp-ha-1}
\end{align}
The commutator here means $[A, B] = AB A^{-1} B^{-1}$. In addition, the following set of holonomies show nontrivial commutation:
%
\begin{align}
\left[ \widetilde{W}_1^{\rm pp} , W_5^{\rm pp}  \right] = & \omega^{ c_x c_y  \left[ y_0' - y_0  -  \frac{1}{2}(N - {\rm gcd}(L_y,N) ) \right]} 
\nn 
\left[ \widetilde{W}_2^{\rm pp} , W_6^{\rm pp}   \right]  =& \omega^{c_y\left[y_0' - y_0  + \frac{1}{2} (N -  {\rm gcd}(L_y,N) ) \right]} 
\nn
\left[ \widetilde{W}_3^{\rm pp}  , W_5^{\rm pp}  \right]   =& \omega^{c_x\left[x_0' - x_0  +  \frac{1}{2}(N -  {\rm gcd}(L_x,N) ) \right]} 
\nn 
\left[ \widetilde{W}_4^{\rm pp} , W_6^{\rm pp}  \right]  =& \omega^{c_xc_y \left[x_0'-x_0 - \frac{1}{2} (N-{\rm gcd}(L_x,N)) \right]} . 
\label{eq:pp-ha-2}
\end{align}
However, the following choice removes the non-trivial phase factors among them\,\footnote{This has been the sole purpose of keeping the arbitrary constants in the definition of the emergent operators.},
\begin{align}\label{eq:preProjCoordFix}
y_0' = & y_0 + \frac{1}{2}\bigl(N- {\rm gcd}(L_y,N) \bigr), \nn 
x_0' = & x_0 + \frac{1}{2}\bigl(N- {\rm gcd}(L_x,N) \bigr) , 
\end{align}
and the nontrivial Heisenberg algebra is spanned entirely by Eq. (\ref{eq:pp-ha-1}). The holonomies of the R2TC are then obtained by projection of the pre-projection holonomies constructed here. 

\subsection{Post-projection holonomies} 
\label{subsec:pph} 

As one can see from the projection schemes, Eqs. (\ref{eq:x0-proj})-(\ref{eq:xz12-proj}), the $X$-operators remain intact through the projection except some re-labeling. The pre-projection $X$-holonomies of Eqs. (\ref{eq:x-hol-14}) and (\ref{eq:x-hol-56}) become, after re-labeling, 
\begin{align}
W_1  & = \prod_{x_i=1}^{L_x} X_{0}(\vec{v}_i), ~~~~~~
W_2  = \prod_{y_i=1}^{L_y} X_{2}(\vec{v}_i)\nn
W_3 & = \prod_{x_i=1}^{L_x} X_{1}(\vec{v}_i), ~~~~~~
W_4 = \prod_{y_i=1}^{L_y} X_{0}(\vec{v}_i) \nn
W_5 & = \prod_{x_i=1}^{\lcm{L_x}{N}}\bigl( X_1(\vec{v}_i+\hat{x} ) \bigr)^{x_i-x_0}
\bigl( X_0(\vec{v}_i)\bigr)^{y_i-y_0-1},
\nn 
W_6 & = \prod_{y_i=1}^{{\rm lcm}(L_y ,N) }\bigl( X_2(\vec{v}_i+\hat{y}) \bigr) ^{y_i-y_0}
\bigl(X_0(\vec{v}_i)\bigr)^{x_i-x_0-1},
\label{eq:x-hol}
\end{align}
after the projection. At first it seems the number of distinct holonomy actions is $N^4 {\rm gcd}(L_x , N) {\rm gcd} (L_y , N)$ since $W_1$ through $W_4$ has $(W)^N =1$ but 
\begin{align} (W_5 )^{\gcd{L_x}{N}} = (W_6 )^{{\rm gcd}(L_y ,N)}= 1 , \nonumber \end{align} 
at odds with the GSD formula in Eq. (\ref{eq:GSD}). A delicate consideration is required to see that the number of independent actions among $W_1$ and $W_4$, which are both products of $X_0$'s, is not $N^2$ but $N {\rm gcd}(L_x , L_y , N)$. 

One begins with labeling the holonomy $(W_1 )^{n_1} (W_4 )^{n_4}$ by $(n_1, n_4)$ and invoking the identity~\footnote{The proof of the identity is simple. Since ${W_1 (y) = \prod_{x_i = 1}^{L_x}X_0(x_i,y)}$ and ${W_4 (x) = \prod_{y_i = 1}^{L_y}X_0(x,y_i)}$, it then follows that $\prod_{y=1}^{L_y}W_1 (y) = \prod_{x=1}^{L_x} W_4(x)$. Furthermore, since the actions of $W_1 (y)$ and $W_4 (x)$ do not actually depend on $y$ and $x$, respectively, when acting on the ground states, we obtained the claimed relation.}
\begin{align}
( W_1 )^{L_y} |{\rm GS} \rangle = ( W_4 )^{L_x} |{\rm GS} \rangle . 
\label{eq:w1w4-relation}
\end{align}
This implies the equivalence relation $(n_1 , n_4 ) \sim (n_1 + L_y, n_2 - L_x )$ among the holonomies. We need to carefully figure out how the points $(n_1, n_4)$ become equivalent by Eq. (\ref{eq:w1w4-relation}) when the $\mathbb{Z}_N$ nature is considered at the same time. 

Invoking the two winding numbers $c_x , c_y$ defined in Eq. (\ref{eq:winding-numbers}), 
\begin{align}
( W_1 )^{c_y L_y} |{\rm GS} \rangle  & = ( W_4 )^{c_y L_x} |{\rm GS} \rangle  = |{\rm GS} \rangle , \nn 
(W_1 )^{c_x L_y} |{\rm GS} \rangle  & = (W_4 )^{c_x L_x} |{\rm GS} \rangle = |{\rm GS} \rangle .  
\end{align}
Applying the Euclidean argument for identifying the gcd of two integers, we conclude
\begin{align}
(W_1 )^{N_y} |{\rm GS} \rangle = (W_4 )^{N_x } |{\rm GS} \rangle = |{\rm GS} \rangle  
\end{align}
where 
\begin{align}
N_x \equiv &  {\rm gcd}(c_y L_x , N), & N_y \equiv & {\rm gcd}(c_x L_y , N) . \label{eq:torus-shrinks}
\end{align}
From Eqs. (\ref{eq:w1w4-relation}) and (\ref{eq:torus-shrinks}) we deduce the equivalence relation
\begin{align}
n_1 \sim n_1 + L_y \sim n_1 + N_y  , \nn
n_2 \sim n_2 + L_x \sim n_2 + N_x  . 
\end{align} 
Invoking the Euclidean argument again, the number of inequivalent integers $n_1$ for fixed $n_4$ becomes ${\rm gcd}(L_y, N_y)$, and the number of inequivalent $(n_1 , n_4)$ equals $N_x {\rm gcd}(L_y, N_y)$. It can be simplified further to
\begin{align}
N_x {\rm gcd}(L_y, N_y) 
& = {\rm gcd} \left( L_x c_y , N \right) {\rm gcd} \left( L_y , N  \right) \nn 
    & = N {\rm gcd}\left( L_x, L_y, N\right),
\label{eq:wz3wz4-gsd-simp}
\end{align}
by employing several number-theoretic identities
\begin{align}
{\rm{gcd}}(a,{\rm{gcd}}(b,c)) & ={\rm{gcd}}({\rm{gcd}}(a,b),c) ={ \rm{gcd}}(a,b,c) , \nn 
    m{\rm{gcd}}(a,b) & = {\rm{gcd}} (ma , mb). 
\end{align}
The number of independent holonomy actions among $W_1$ and $W_4$ is $N{\rm gcd}(L_x , L_y , N)$. The GSD formula in Eq. (\ref{eq:GSD}) breaks down to
\begin{align}
N^2 \cdot (N {\rm gcd} (L_x , L_y , N) ) \cdot {\rm gcd} (L_x , N) \cdot {\rm gcd} (L_y , N)  .  
\label{eq:GSD-breakdown}
\end{align}
Here, the first $N^2$ are coming from $W_2$ and $W_3$, $N {\rm gcd} (L_x , L_y , N)$ from $W_1$ and $W_4$, and ${\rm gcd} (L_x , N) \cdot {\rm gcd} (L_y , N)$ from $W_5$ and  $W_6$, respectively.

\begin{figure*}[tb]
\includegraphics[width=0.95\textwidth]{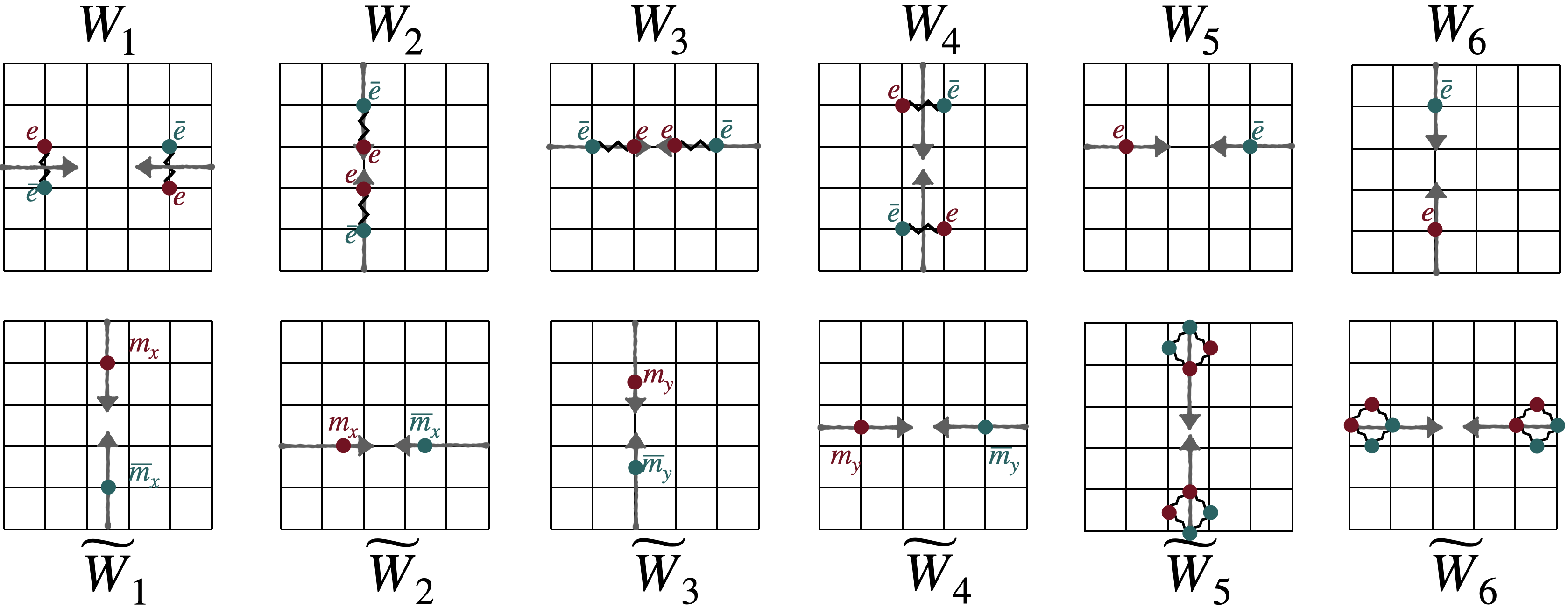}
\caption{
Pictorial representation of the holonomies. (first row) Six $X$-holonomies as creation/annihilation of $e$ dipole-antidipole pairs oriented either horizontally or vertically ($W_1$ through $W_4$), and of $e$ monopole-antimonopole pairs ($W_5$ and $W_6$). (second row) Six $Z$-holonomies as creation/annihilation of $m^x$ or $m^y$ monopole-antimonopole pairs ($\widetilde{W}_1$ through $\widetilde{W}_4$), and of $m$ dipole-antidipole pairs ($\widetilde{W}_5$ and $\widetilde{W}_6$ Monopole winding processes are accompanied by the motion of auxiliary dipoles to preserve the total dipole moment, but they are omitted from the figure for the sake of clarity.
}
\label{fig:holonomy-pics}
\end{figure*}

As for the $Z$-holonomies, projection of the pre-projection $Z$-holonomies of Eqs. (\ref{eq:pre-Z56})-(\ref{eq:pre-Z34}) leads to
\begin{align} 
\widetilde{W}_1 = & \prod_{y_i = 1}^{{\rm lcm}(L_y,N)} Z_0(\vec{v}_i - \hat{y} ) \bigl( Z_1(\vec{v}_i )  Z_1 (\vec{v}_i  +\hat{x})^{-1} \bigr) ^{y_i-y_0}, 
\nn
\widetilde{W}_2 = & \prod_{x_i = 1}^{L_x} Z_2(\vec{v}_i ) \bigl( Z_2(\vec{v}) Z_2(\vec{v}_i +\hat{y})^{-1}\bigr)^{y_i-y_0}, 
\nn 
\widetilde{W}_3 = & \prod_{y_i = 1}^{L_y} Z_1(\vec{v}_i )\bigl( Z_1(\vec{v}_i ) Z_1(\vec{v}_i +\hat{x})^{-1} \bigr)^{x_i-x_0},
\nn
\widetilde{W}_4 = & \prod_{x_i = 1}^{{\rm lcm}(L_x,N)} Z_0(\vec{v}_i -\hat{x})\bigl(Z_2(\vec{v}_i ) Z_2 (\vec{v}_i +\hat{y})^{-1}\bigr) ^{x_i-x_0}, 
\nn
\widetilde{W}_5 = & \prod_{y_i = 1}^{L_y}  Z_1(\vec{v}_i + \hat{x}) Z_1(\vec{v}_i )^{-1}, 
\nn
\widetilde{W}_6 = & \prod_{x_i = 1}^{L_x}  Z_2(\vec{v}_i + \hat{y}) Z_2(\vec{v}_i )^{-1}.
\label{eq:post-projection-Z-holonomy} 
\end{align}
One can check the following non-trivial commutators among the post-projection holonomies. 
\begin{align}
\left[\widetilde{W}_1, W_1\right]   & = \omega^{c_y}, 
&
\left[\widetilde{W}_4, W_4 \right] & = \omega^{c_x} , 
\nn
\left[ \widetilde{W}_2, W_2 \right] & = \omega,
&
\left[ \widetilde{W}_3, W_3 \right] & = \omega ,
\nn
\left[ \widetilde{W}_5, W_5 \right] & = \omega^{c_x} ,
&
\left[ \widetilde{W}_6 ,  W_6 \right] & = \omega^{c_y}.
\label{eq:hol-spin-nontriv}
\end{align}
The commutator here means $[A, B] = AB A^{-1} B^{-1}$. There exist four more nontrivial commutations relations: 
\begin{align}
\left[ \widetilde{W}_1 , W_5  \right] = & \omega^{ c_x c_y  \left[ y_0' - y_0  -  \frac{1}{2}(N - {\rm gcd}(L_y,N) ) \right]} 
\nn 
\left[ \widetilde{W}_2 , W_6   \right]  =& \omega^{c_y\left[y_0' - y_0  + \frac{1}{2} (N -  {\rm gcd}(L_y,N) ) \right]} 
\nn
\left[ \widetilde{W}_3 , W_5  \right]   =& \omega^{c_x\left[x_0' - x_0  +  \frac{1}{2}(N -  {\rm gcd}(L_x,N) ) \right]} 
\nn 
\left[ \widetilde{W}_4 , W_6  \right]  =& \omega^{c_xc_y \left[x_0'-x_0 - \frac{1}{2} (N-{\rm gcd}(L_x,N)) \right]} ,
\label{eq:hol-spin-nontriv-2}
\end{align}
which is nothing but the projection of Eq. (\ref{eq:pp-ha-2}). Hence, applying the condition in Eq. (\ref{eq:preProjCoordFix}) removes this nontrivial phase factors as well.

One can read off the GSD from the Heisenberg algebra of the holonomies. For instance $W_1$ acting on a ground state changes the eigenvalues of $\widetilde{W}_1$ by $\omega^{c_y}$, generating in total $N/c_y = {\rm gcd} (L_y , N)$ distinct ground states. Naively applying the reasoning to the first two pairs of commutators (1 and 4) gives the GSD equal to $(N/c_x ) (N/c_y) = \mathrm{gcd} (L_x, N)\mathrm{gcd} (L_y , N)$, the next two pairs (2 and 3) yields $N^2$, and the final two pairs (5 and 6) yields another $\mathrm{gcd}(L_x , N) \mathrm{gcd} (L_y , N)$. In total, this gives the GSD count $N^2 [ {\rm gcd}(L_x , N) {\rm gcd} (L_y , N) ]^2$ that is {\it less} than the correct GSD formula, Eq. (\ref{eq:GSD}), by ${\rm gcd}(L_x , N) {\rm gcd} (L_y , N)/N {\rm gcd} (L_x , L_y , N)$. In other words, the holonomies constructed above {\it underspans} the space of ground states. 

The deficiency comes from the fact that $\widetilde{W}_4$ given in Eq.~(\ref{eq:post-projection-Z-holonomy}) is not the most minimal choice of the holonomy. The correct holonomy expression can be found by referring to Ref.~\cite{pace-wen}:
\begin{align}
\widetilde{W}^{\rm min}_4 = & \prod_{x_i = 1}^{n_x L_x} Z_0(\vec{v}_i -\hat{x})\bigl(Z_2(\vec{v}_i ) Z_2 (\vec{v}_i +\hat{y})^{-1} \bigr) ^{x_i-x_0}, \nn 
&\otimes \prod_{y_i = 1}^{n_y L_y} Z_0(\vec{v}_i - \hat{y} ) \bigl( Z_1(\vec{v}_i )  Z_1 (\vec{v}_i  +\hat{x})^{-1} \bigr) ^{y_i-y_0},
\label{eq:w-t-4}
\end{align}
where the integers $n_x, n_y$ are given by
\begin{align}
n_x = &  \frac{{\rm gcd} (L_x,N)}{{\rm gcd}(L_x,L_y,N)}, & 
n_y = &  \frac{{\rm lcm}(L_x,{\rm gcd}(L_y,N)) + k N}{L_y}.
\end{align}
Here $k$ is a minimal integer that makes $n_y$ an integer~\cite{pace-wen}. With the new definition of $\widetilde{W}_4 \rightarrow \widetilde{W}^{\rm min}_4$ we obtain a new commutator 
\begin{align}
\left[\widetilde{W}^{\rm min}_4 , W_4 \right] & = \omega^{n_x} . 
\end{align} 
The GSD coming from this sector equals $N/n_x = N {\rm gcd}(L_x,L_y,N) / {\rm gcd} (L_x,N)$ and indeed, we recover the full GSD simply from the Heisenberg algebra, with the modified $\widetilde{W}_4$. Replacing $\widetilde{W}_4$ by $\widetilde{W}_4^{\rm min}$ gives us an orthogonal set of six $X$-holonomies $\{W_1, \cdots, W_6\}$ and six $Z$-holonomies $\{\widetilde{W}_1, \cdots, \widetilde{W}_6 \}$ that fully span the ground states of R2TC. 

In making physical interpretations of the $\widetilde{W}_4$ holonomy, though, we will continue to adopt the simpler (albeit slightly incorrect) representation as the horizontal winding of $m^y$ quasiparticle as shown in Fig. \ref{fig:holonomy-pics}. The interpretation of $\widetilde{W}_4^{\rm min}$ involves a mix of the horizontal winding of $m^y$ and the vertical winding of $m^x$, as can be seen from its definition in Eq.~(\ref{eq:w-t-4}). 

With the explicit construction of the holonomies, we can check the quantum numbers of the TN ground state wave function we have constructed in Sec.~\ref{subsec:TN}. Following the similar procedure as in Eqs. (\ref{eq:peps-b-eig}) and (\ref{eq:peps-a-eig}), one can verify that our TN wave function $|\psi\rangle$ on the torus is the simultaneous eigenstate of the four $X$-holonomies $W_1$ through $W_4$, as well as two $Z$-holonomies $\widetilde{W}_5$ and $\widetilde{W}_6$, with eigenvalue $+1$. The remaining six holonomies, $\widetilde{W}_1$ through $\widetilde{W}_4$ and $W_5, W_6$, then act to shift the ground state into orthogonal ground states. One can also construct a TN wave function for the eigenstate of all six of the $X$-holonomies $W_1$ through $W_6$, but it requires a `double layer structure' of TN that goes beyond the present construction and will be presented elsewhere~\cite{oh-forthcoming}.

\subsection{Physical interpretation of the holonomies}

It is well known that the $X$- and $Z$-holonomies in the R1TC has a concise physical picture as the creation and subsequent annihilation of a pair of $e\overline{e}$ or $m\overline{m}$ (bar denotes the anti-particle) anyons after one anyon is wound around one of the non-contractible paths of the torus. A total of four holonomies form two conjugate pairs and span the $N^2$ degenerate ground states. To account for the GSD of R2TC, which reaches the maximum value of $N^6$, one requires a total of twelve holonomies breaking up into two groups. Six of them bear obvious physical interpretations as the winding of $m^x , m^y, e$ particles around either of the two circumferences of the torus. We provide the physical interpretations of the remaining six holonomies. 

Each action of $X$-holonomies corresponds to the winding of the three electric quantities that are conserved. Figure~ \ref{fig:holonomy-pics} illustrates these processes. 
The physical action of $W_1$ ($W_2$) among the $X$-holonomies in Eq. (\ref{eq:x-hol}) is to create a $y$-oriented $e$-dipole and its anti-dipole, then to move one of the dipoles along the horizontal (vertical) non-contractible path of the torus. For $W_3$ ($W_4$), it is $x$-oriented $e$-dipole winded horizontally (vertically). The $W_5$ ($W_6$), on the other hand, moves the $e$ monopole horizontally (vertically).
Note that an auxiliary dipole is attached to the $e$ monopole during its adiabatic motion, to ensure the total dipole moment conservation in the process, and disappears at the end of completing the loop. We omit the auxiliary dipoles from Fig.~\ref{fig:holonomy-pics} for simplicity.

Each action of $X$-holonomies corresponds to the winding of the three magnetic quantities that are conserved.
The action of the first four $Z$-holonomies in Eq. (\ref{eq:post-projection-Z-holonomy}) is to create a monopole and anti-monopole pair of either $m^x$ or $m^y$ and wind them. Specifically, $\widetilde{W}_1 , \widetilde{W}_2$ ($\widetilde{W}_3 , \widetilde{W}_4$) wind $m^x$ ($m^y$) along the $y$- and $x$-oriented non-contractible loops. To enforce the total dipole moment $y m^x + x m^y = 0$, some auxiliary dipoles are attached during the vertical motion of $m^x$ as well as the horizontal motion of $m^y$~\cite{oh22b}. The last two $Z$-holonomies, $\widetilde{W}_5$ and $\widetilde{W}_6$, correspond to the winding of $m$-dipole along the $y$ and $x$ non-contractible loops of the torus. 
The list of holonomies and their physical interpretations are summarized in Table \ref{tab:holonomies}. 

\begin{table}[h]
    \centering
    \begin{tabular}{|c |c | c | }
    \hline 
$(W_1 ,  \widetilde{W}_1 )$ & ($e$-dipole, h) & ($m_x$-monopole, v) \\
\hline
$(W_2 ,  \widetilde{W}_2 )$ & ($e$-dipole, v) & ($m_y$-monopole, h) \\
\hline
$(W_3 ,  \widetilde{W}_3 )$ & ($e$-dipole, h) & ($m_y$-monopole, v) \\
\hline
$(W_4 ,  \widetilde{W}_4 )$ & ($e$-dipole, v) & ($m_x$-monopole, h) \\
\hline
$(W_5 ,  \widetilde{W}_5 )$ & ($e$-monopole, h) & ($m_y$-dipole, v) \\
\hline 
$(W_6 ,  \widetilde{W}_6 )$ & ($e$-monopole, v) & ($m_x$-dipole, h) \\ 
\hline
    \end{tabular}
    \caption{(left) Pair of holonomies (logical operators) with non-trivial commutation relations. (middle) nature of $e$-excitations and the direction of winding associated with a given holonomy $W$. (right) nature of $m$-excitations and the direction of winding associated with a given holonomy $\widetilde{W}$. (h=horizontal, v=vertical)}
    \label{tab:holonomies}
\end{table}

\subsection{Field-theoretic derivation of the holonomies}
\label{subsec:field-theoretic-holonomy}

The holonomy construction thus far proceeded from a known microscopic Hamiltonian, i.e., R2TC model whose quasiparticle excitations are well-explored. Historically, the holonomies engendered by the Wilson line operators manifest the global flux sectors to which the ground state on a torus belongs. Building on this line of thinking, we show how to obtain the Wilson operators pertinent to the R2TC from the underlying rank-2 gauge theory. 

For higher-rank gauge theories, the Wilson operators creating immobile quasiparticle excitations turn out to be richer and more diverse than in the conventional $\mathbb{Z}_N$ gauge theory for the following reasons. 1) Due to the restricted mobility of the quasiparticles, some of the Wilson lines need to be straight and geometrically oriented in a specific direction\cite{slagle2017quantum,you2020fractonic}. 2) There might exist other Wilson operators defined on a non-contractable manifold, such as membrane, cage, or fractal, that are responsible for the holonomies of higher-rank gauge theories~\cite{haah2011local,you2020fractonic,prem2019cage}. 3) Different Wilson operators that are parallel to each other may not render the same value, as opposed to the conventional $\mathbb{Z}_N$ gauge theory whose Wilson line operators are invariant under translation. For higher-rank gauge theory, the dipole and quadruple moments transform nontrivially under translation, and so does the global flux sector. Consequently, two parallel flux lines might return different values.

Recall that in the usual 2D $\mathbb{Z}_N$ gauge theory, the magnetic flux is given by $m = \partial_x A_y - \partial_y A_x$ and the total flux on the half cylinder ${\cal A}$ with boundaries at $x=x_0$ and $x=x_n$ is  characterized by parallel Wilson line operators
\begin{align} \int m dV= \oint A_y(x_n , y ) dy-\oint A_y(x_0 , y ) dy=0 ,  \nonumber \end{align} 
with the integral $\oint$ going around the full circumference of the cylinder. The net flux condition ($\int m dV = 0$) implies that the two parallel Wilson lines render the same value. Since the two Wilson lines are spatially separated while the Hamiltonian is local, each $\oint A_y(x , y) dy$ must commute with all local terms in the Hamiltonian and can be treated as a global flux operator that characterizes the holonomy. One obtains another Wilson line operator along the $y$-direction from the charge sector, i.e. $\oint E_y (x, y ) dy$. These two comprise all possible Wilson lines along the $y$-loop.

Now we apply this protocol to R2TC. Begin with the definition of three monopole charges given in Eq. (\ref{eq:quadrupole-operator}) in the continuum limit,   
\begin{align}
m^x & =  \partial_x A^{yy} - \partial_y A^{xy}, \nn
m^y & = \partial_x A^{xy} - \partial_y A^{xx} , \nn
e & = \partial_x^2 E^{xx} + \partial_y^2 E^{yy} + \partial_x \partial_y E^{xy} . 
\label{r2tc-2}
\end{align} 
As noted in Sec.~\ref{subsec:conservation}, the magnetic charges $m^x , m^y$  demonstrate a number of conservation laws
\begin{align} 
\int m^x dV = \int m^y dV = \int (x m^y + y m^x ) dV=0.  \label{eq:conservation-1} 
\end{align} 
The first two yields
\begin{align}
&\int m^x dV =  \oint A^{yy}(x_n , y )  dy-\oint A^{yy}(x_0 , y)  dy = 0 ,\nonumber\\
&\int  m^y dV = \oint A^{xy}(x_n , y)  dy-\oint A^{xy}(x_0 , y)  dy = 0  .  \label{eq:qxqy-conserve-2}
\end{align}
Following the aforementioned argument, one can define two Wilson line operators,
\begin{align}
W_2(x) & =  \oint A^{yy}(x , y )  dy, \nn  
W_4(x) & =  \oint A^{xy}(x , y )  dy . 
\end{align}
Due to the flux conservation law, Eq. (\ref{eq:qxqy-conserve-2}), they are both uniform along the $x$-coordinate: $\partial_x W_2 (x) =\partial_x W_4 (x) =0$. The subscripts 2, 4 are intended to match the definitions of post-projection Wilson operators in Eq. (\ref{eq:x-hol})~\footnote{Note that we use the same symbol $W_1 \cdots W_6$ for the holonomies in both continuum and the $\mathbb{Z}_N$ theories although, strictly speaking, the $\mathbb{Z}_N$ holonomies are obtained by raising the continuum holonomies to an exponential.}

In addition, we have 
\begin{align}
&\int_{\cal A}   (y m^x + x m^y) dV =  0 
=   \oint y A^{yy}(x_n , y)  dy \nn 
&  -\oint y A^{yy}(x_0 , y) dy 
+ \oint^{x_n}_{x_0} \left( \int  A^{xy}(x , y ) dy \right) dx \nn
&=  \oint y A^{yy}(x_n , y ) dy - \oint y A^{yy}(x_0 , y )dy + (x_n - x_0 ) W_4 .  \label{eq:W3-2} 
\end{align}
In arriving at the last equality we used the fact that the Wilson line operator $W_4= \oint A^{xy}(x,y)  dy$ is uniform in $x$. We arrive at another Wilson line operator,
\begin{align}
&W_6 (x) =   \oint y A^{yy}(x , y )  dy + x W_4, ~~ \partial_x W_6 (x) = 0 ,
\end{align}
which matches the definition of $W_6$ in Eq. (\ref{eq:x-hol}) after Higgsing. 

As the theory is rotationally symmetric, the other set of Wilson line operators follow as integrals along the $x$-loop: 
\begin{align}
W_1 & =  \oint A^{xy}(x, y)  dx, ~~~~
W_3   =  \oint A^{xx}(x, y) dx, \nn
W_5 & =  \oint x A^{xx}(x, y) dx + y W_1  ,
\end{align}
with matching definitions in Eq. (\ref{eq:x-hol}) after Higgsing. Their coordinate independence follows readily. 

The previous holonomies $W_1$ through $W_6$ were derived on the basis of conservation laws of the magnetic charges. Alternatively, the holonomies can be derived from the electric charge conservation, 
\begin{align}
&e = \partial^2_x E^{xx}+\partial^2_y E^{yy} + \partial_x \partial_y E^{xy} , \nn  
& \int e \, dV = \int x e  \, dV = \int y e \, dV = 0 ,  \label{eq:conservation-2} 
\end{align} 
and hence 
\begin{align}
\int_{\cal A} e \, dV =  \oint \partial_x E^{xx}(x_0 , y ) dy-\oint \partial_x E^{xx}(x_n , y) dy  = 0. 
\end{align}
This yields the first holonomy 
\begin{align}
&\widetilde{W}_5  (x) =  \int \partial_x E^{xx}(x , y) dy , ~~ \partial_x \widetilde{W}_5 (x) = 0 .  
\end{align}
From the other two conservation laws we find 
\begin{align}
\int y e \, dV =& \int  (y \partial_y E^{xy} +  y \partial_x E^{xx})(x_n,y) dy  \nn 
& -\int  (y \partial_y E^{xy} +  y \partial_x E^{xx})(x_0,y)dy , \nn 
\int_{\cal A} x e \,  dV  =&   \oint  E^{xx}(x_0 , y ) dy 
-\int  E^{xx}(x_n , y ) dy \nn
&+ (x_n-x_0 ) \widetilde{W}_5 . 
\end{align}
We arrive at two additional Wilson line operators 
\begin{align}
\widetilde{W}_1 & = - \oint (y \partial_y E^{xy} +  y \partial_x E^{xx}) dy  \nn 
& =  \oint (E^{xy} -  y \partial_x E^{xx}) dy  ,\nn
\widetilde{W}_3 & =   - \oint  E^{xx} dy + x \widetilde{W}_5 . 
\end{align}
The other three Wilson line operators are obtained from rotational symmetry:
\begin{align}
\widetilde{W}_2  & =   - \oint  E^{yy} dx + y \widetilde{W}_6 , \nn 
\widetilde{W}_4  & = \oint (E^{xy} -  x \partial_y E^{yy}) dy , \nn
\widetilde{W}_6  & = \oint \partial_y E^{yy} dx.
\end{align}
Coordinate independence of all Wilson operators can be verified easily. After Higgsing, $\widetilde{W}_1$ through $\widetilde{W}_6$ match the six $Z$-holonomies of Eq. (\ref{eq:post-projection-Z-holonomy}). Physical interpretation of the holonomy operators has been given in Table~\ref{tab:holonomies}. 

For completeness we briefly mention that in a theory with vector-electric and scalar-magnetic charges such that %
\begin{align}
&e^x =  \partial_x E^{xx} +\partial_y E^{xy}\nonumber\\
&e^y = \partial_x E^{xy} +\partial_y E^{yy}\nonumber\\
&m = \partial^2_x A^{yy}+\partial^2_y A^{xx} - \partial_x \partial_y A^{xy} ,
\label{r2tc}
\end{align} 
we can construct the relevant holonomies based on a different set of conservation laws
\begin{align} 
& \int e^x dV = \int e^y dV = \int (y e^x- x e^y ) dV=0 , \nn  
& \int m  dV = \int x m dV = \int y m dV = 0 . 
\end{align} 

As the derivation in this subsection clearly shows, the construction of holonomies are firmly rooted in the conservation laws such as Eqs. (\ref{eq:conservation-1}) and (\ref{eq:conservation-2}). The existence of dipole-like conservations in addition to the usual charge conservations for $m^x , m^y, e$ monopoles plays a crucial role in constructing the full set of holonomies for the rank-2 gauge theory as well as its Higgs descendant, which is the R2TC. We suspect that a similar scheme can be exploited for the holonomy construction in other rank-2 gauge theories.

\subsection{Understanding the position-dependent braiding}\label{posDepBraidSubSec}

The seemingly puzzling feature of R2TC was the position-dependent statistical phase obtained when one quasiparticle is braided around another~\cite{oh22a,pace-wen,oh22b}. While various elaborate arguments for why this should be so has been given already~\cite{oh22a,pace-wen,oh22b}, it turns out the field-theoretic holonomies just constructed can provide a simple picture for it. 

\begin{figure}[t!]
\centering
\includegraphics[width=.45\textwidth]{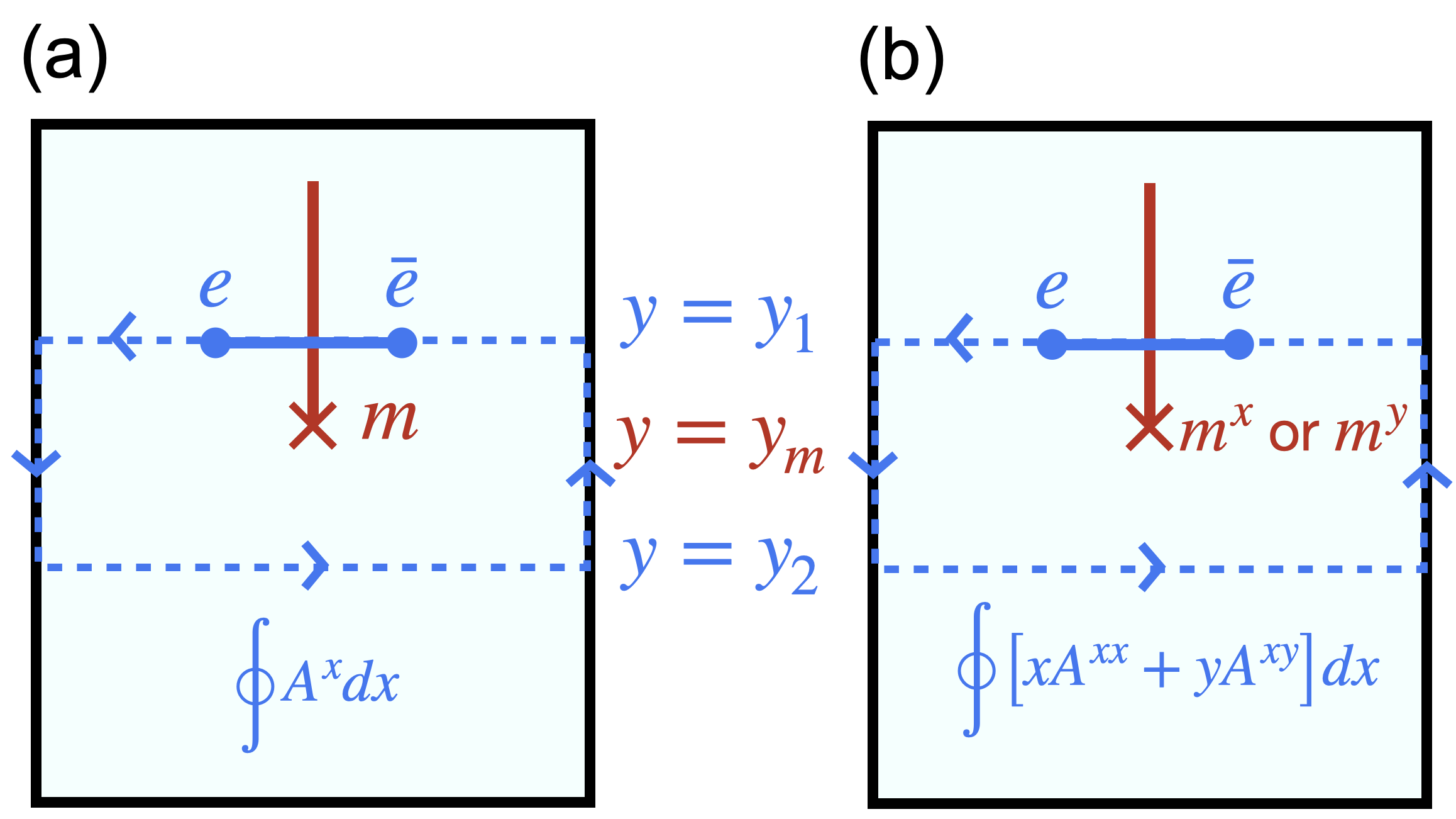}
\caption{(a) Braiding charge $e$ around the flux $m$ in the conventional $\mathbb{Z}_N$ gauge theory. The trajectory of the braiding loop corresponds to the total flux inside the enclosed area. (b) Braiding charge $e$ around the flux $m_x$ or $m_y$ in R2TC. The trajectory of the braiding loop corresponds to the total dipolar flux inside the enclosed area. 
}
\label{fig:braid}
\end{figure}

To do so, we first review how the adiabatic braiding process relates to the statistical phase. We begin with the prominent $\mathbb{Z}_N$ gauge theory example where the charge $e$ and flux $m$ have nontrivial statistics. Creating a pair of $m$ flux excitations is implemented at the two endpoints of an open string $e^{i\int_{0}^{x}  E_x dx}$. To braid the charge around the flux, we create a pair of charge ($e$) and anti-charge ($\overline{e}$) connected by an open string, wind the $e$ particle around $m$ and annihilate it with the anti-charge as shown in  Fig.~\ref{fig:braid}. The trajectory of the $e$ particle is associated with the Aharonov-Bohm (AB) phase $\exp\left( i \oint \vec{A} \cdot d\vec{r} \right)$ which corresponds to the total flux $\int m dV$ ($m = \nabla \times A$) inside the area enclosed by the loop. As a result, the braiding of charge excitation creates a flux loop that measures the total flux inside so their braiding phase is just the AB phase.

Now let us go back to the R2TC theory with vector-magnetic and scalar-electric charges as in Eq.~(\ref{eq:generators}). The flux~\footnote{Here we use the word magnetic `flux' interchangeably with the magnetic `charge'.} $m^x$ or $m^y$ excitations are created by open-string operators such as 
\begin{align} 
\widetilde{W}_1^{\rm open} & \sim  \exp\left[ i \int_{0}^{y_m} \left(E^{xy} - y \partial_x E^{xx}\right) dy \right] ~ {\rm  or} \nn 
\widetilde{W}_3^{\rm open} & \sim \exp \left[i \int_{0}^{y_m} \left( x \partial_x E^{xx}- E^{xx} \right)dy \right] , \nonumber 
\end{align} 
respectively. They are none other than open-ended versions of the holonomies constructed in Sec.~ \ref{subsec:field-theoretic-holonomy} and have physical interpretations of creating a $m^x \overline{m}^x$ or a $m^y \overline{m}^y$ pair separated along the vertical direction as shown in Fig.~\ref{fig:braid}. 

To braid the $m$ flux, we create a pair of charge $e$ and anti-charge $\bar{e}$ connected by an open string shown as the horizontal blue segment in Fig. \ref{fig:braid}(b), wind the $e$ particle around $m^x$ or $m^y$  as shown by two horizontal dashed lines in Fig.~\ref{fig:braid}(b), and annihilate it with the anti-charge. The trajectory of the $e$ particle is associated with the phase factor 
\begin{align} W_5 \sim e^{i \int \left[ (x A^{xx} + yA^{xy})(x, y_1)  - (x A^{xx} + yA^{xy})(x, y_2)\right] dx } . \nonumber \end{align}
As one can see from Table~\ref{tab:holonomies}, $W_5$ is associated with the horizontal winding of $e$ particle. 

For simplicity, we choose the braiding trajectory consisting of two parallel lines along the $x$-direction above and below the $m$ flux, {\it i.e.} at $y= y_1$ and $y=y_2$, $y_1 <  y_m < y_2$, and $y_m$ indicating the $y$-coordinate of the $m$ flux. We can further simplify the braiding operator as
%
\begin{align}
& e^{ i \int \bigl[(x A^{xx} + yA^{xy})(x, y_1) - (x A^{xx} + yA^{xy})(x, y_2)\bigr] dx } \nn 
& = e^{ 
i \int \bigl[ (x A^{xx} \!+\! yA^{xy})(x, y_1) - (x A^{xx}\!+\!yA^{xy})(x, y_2)\bigr] dx } \nn 
& \times e^{i \int_{y_1}^{y_2} \bigl[ (y A^{yy} + xA^{xy})(x_1, y) \!-\! (y A^{yy}\! +\!x A^{xy})(x_2, y )\bigr] dy }  
\nn 
& =e^{i \int (y m^x + x m^y ) dV } .    
\label{eq:braid-2}
\end{align}
%
In the second line we inserted some $y$-oriented integrals that cancel each other due to the periodic boundary condition and $x_2=x_1+L_x$~\footnote{We assume periodicity of the fields: $A^{xy} (x_1 + L , y) = A^{xy} (x_1 ,y)$, etc. Further, $\int_{y_1}^{y_2} A^{xy} (x, y) dy = 2\pi \mathbb{Z}/L_x$ is assumed.  }. Now one can understand the braiding operation as the line integral of the vector field
\begin{align}
( x A^{xx} + y A^{xy} , x A^{xy} + y A^{yy} ) .
\end{align}
The third line in Eq. (\ref{eq:braid-2}) follows from Stokes' theorem and the definition of $m^x , m^y$ in Eq.~(\ref{r2tc-2}). It shows that the braiding operation measure not the flux, but the `dipolar flux' that depends on the $x$-position of $m^y$ and the $y$-position of $m^x$ that the $e$ particle braids around. The statistical phase becomes accordingly position-dependent. Dipolar braiding among other quasiparticles can be understood in similar ways. The derivation of dipolar braiding statistics in terms of field-theoretic Wilson lines given here has some overlap with earlier consideration~\cite{oh22a,oh22b} of the dipolar braiding, but here we give a more clarified picture of how this seeminingly peculiar braiding statistics arises rather naturally in rank-2 gauge theories. It also suggests that the dipolar braiding phase is not unique to R2TC, but may be a general feature of rank-2 gauge theories and its Higgsed descendants. 

To put it in broader perspective, we comment that a position-dependent braiding process is also present elsewhere. Indeed, while typically not emphasized, even in Wen's $\Z_2$ Plaquette 2D model~\cite{wen03} where there is a single type of stabilizer and, according to the terminology used here, one quasiparticle species whose self-statistics depends on its initial position. Another 2D topologically ordered example, one more complicated than Wen's plaquette model yet simpler than the R2TC, is the model considered by Delfino \txti{et al.} in \Rf{delfino22}. In these 2D topologically ordered examples, a general reason for position-dependent braiding is that lattice translations induce nontrivial automorphisms on the anyon lattice~\cite{BZ14104540}. Consequentially the anyon types are labeled by their position which causes their braiding to become position-dependent~\cite{pace-wen}. In 3D, position-dependent braiding has been discovered in fracton models~\cite{you2020fractonic,shirley2020twisted}. In particular, for 3D twisted fracton theory, the flux excitations denoted as lineons, with restricted mobility along 1D lines, only exhibit nontrivial braiding statistics between the lineons on adjacent planes. That says that if we shift the braiding trajectory of the lineon between the layers, the resultant Berry phase from statistical angles can change.

\section{Generalized Symmetries}
\label{sec:generalized-symmetry}

The holonomies constructed in Sec.~\ref{sec:holonomies} are a piece of a more general structure present in the R2TC: its generalized symmetries. The generalized symmetries of some rank-2 gauge theories have been discussed previously in the literature~\cite{SS200310466, QRH201002254, seiberg22a,HYA220700854}. Given the rich properties of the R2TC, the exactly solvable point of scalar charge rank-2 $\Z_N$ gauge theory, it is interesting to wonder what its generalized symmetries are. In this section, we will identify its symmetries and discuss them in the context of spontaneous symmetry breaking and 't Hooft anomalies. We will consider the general $N$ case. This requires defining a branching and framing structure of the lattice, which we review in appendix Sec.~\ref{sec:diffgeoLat}.

\begin{figure}[t!]
\centering
    \includegraphics[width=.48\textwidth]{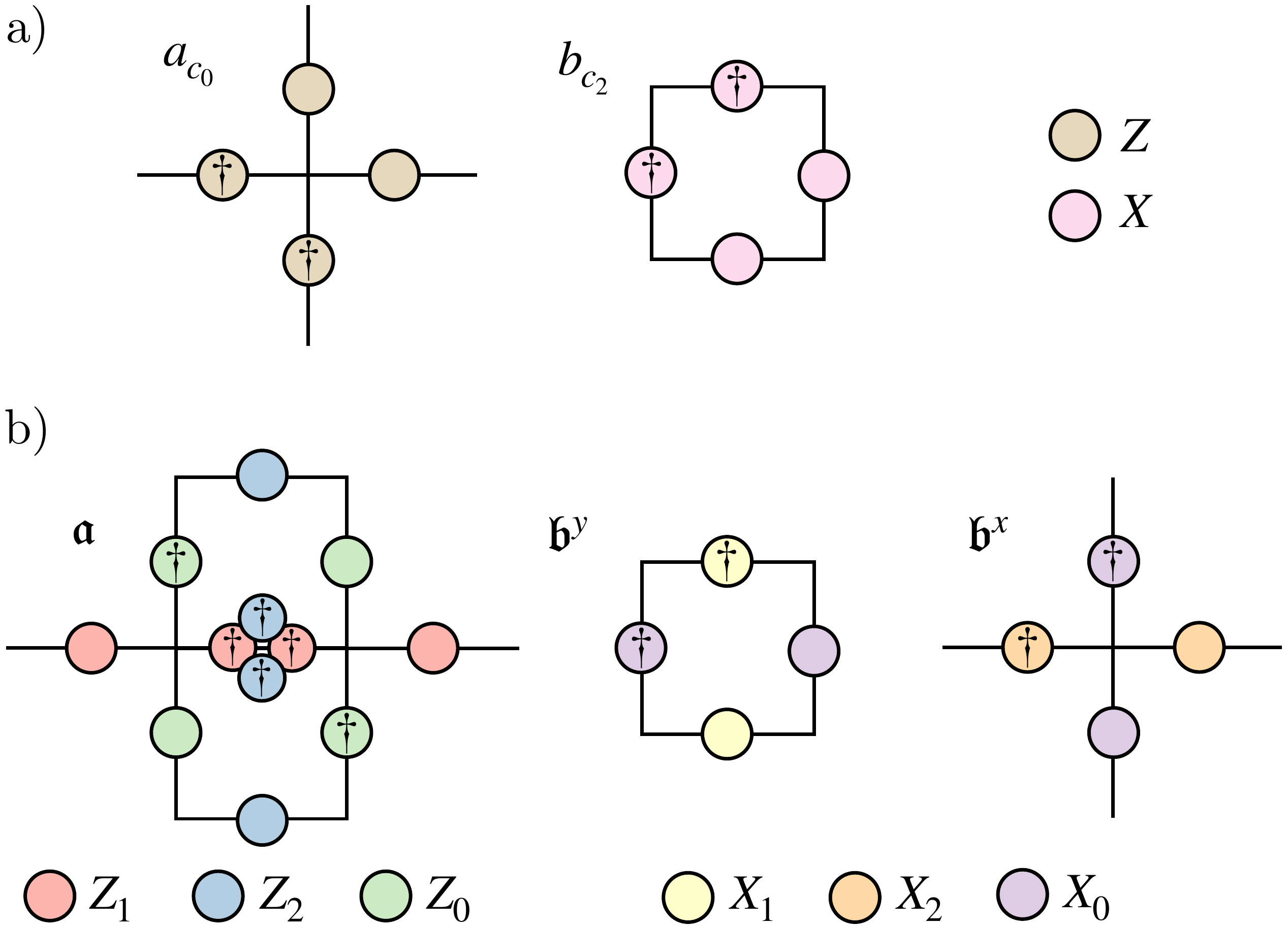}
    \caption{Graphical representations of (a) the $a_{c_0}$ and $b_{c_2}$ operators in the R1TC Hamiltonian Eq.~\eqref{Eq:R1TCHamiltonian} and (b) the $\mathfrak{a}$, $\mathfrak{b}^{y}$, and $\mathfrak{b}^{x}$ operators in the R2TC Hamiltonian Eq.~\eqref{R2TCHam}. The disks are color-coded to represent $X_i$ and $Z_i$ operators, according to the legend. Furthermore, disks with a $\da$ represent the Hermitian conjugate of the corresponding operator.}
    \label{fig:ops}
\end{figure}

\subsection{Reviewing the 1-form symmetries of the R1TC}\label{R1TCsymReview}

Let us first review the generalized symmetries in the ${\mathbb{Z}_N}$ R1TC on a spatial square lattice. The ${(2+1)}$D $\Z_N$ R1TC Hamiltonian can be written as
\begin{equation}
\begin{aligned}
H &= -\sum_{c_0}A_{c_0} - \sum_{c_2} B_{c_2},\label{Eq:R1TCHamiltonian}\\
& A_{c_0} = \frac{1}{N} \sum_{j=1}^N (a_{c_0} )^j , \quad\quad\quad B_{c_2} = \frac{1}{N} \sum_{j=1}^N (b_{c_2} )^j ,
\end{aligned}
\end{equation}
where $a_{c_0}$ and $b_{c_2}$ are the star and plaquette operators
\begin{equation}
    a_{c_0} = \prod_{c_1\in \delta c_0}Z_{c_1},\quad\quad\quad b_{c_2} =\prod_{c_1\in\partial c_2}X_{c_1}.
\end{equation}
We denote the square lattice's sites as ${c_0}$, its edges as ${c_1}$, and its plaquettes as ${c_2}$. In the definitions of ${a_{c_0}}$ and ${b_{c_2}}$, ${\delta c_0}$ denotes the coboundary of ${c_0}$---an oriented sum of edges whose boundary includes ${c_0}$---and ${\partial p}$ denotes the oriented boundary of ${c_2}$. The precise definitions of $\del$ and $\pp$ are given by Eqs.~\eqref{coboundaryDef} and~\eqref{boundaryDef}, respectively. Graphical representations of $a_{c_0}$ and $b_{c_2}$ are shown in Fig.~\ref{fig:ops}a, from which it is clear that they commute for all $c_0$ and $c_2$. We note that these expressions for $a_{c_0}$ and $b_{c_2}$ are equivalent to Eq.~\eqref{eq:star-and-plaquette}.

\begin{figure}[t!]
\centering
    \includegraphics[width=.48\textwidth]{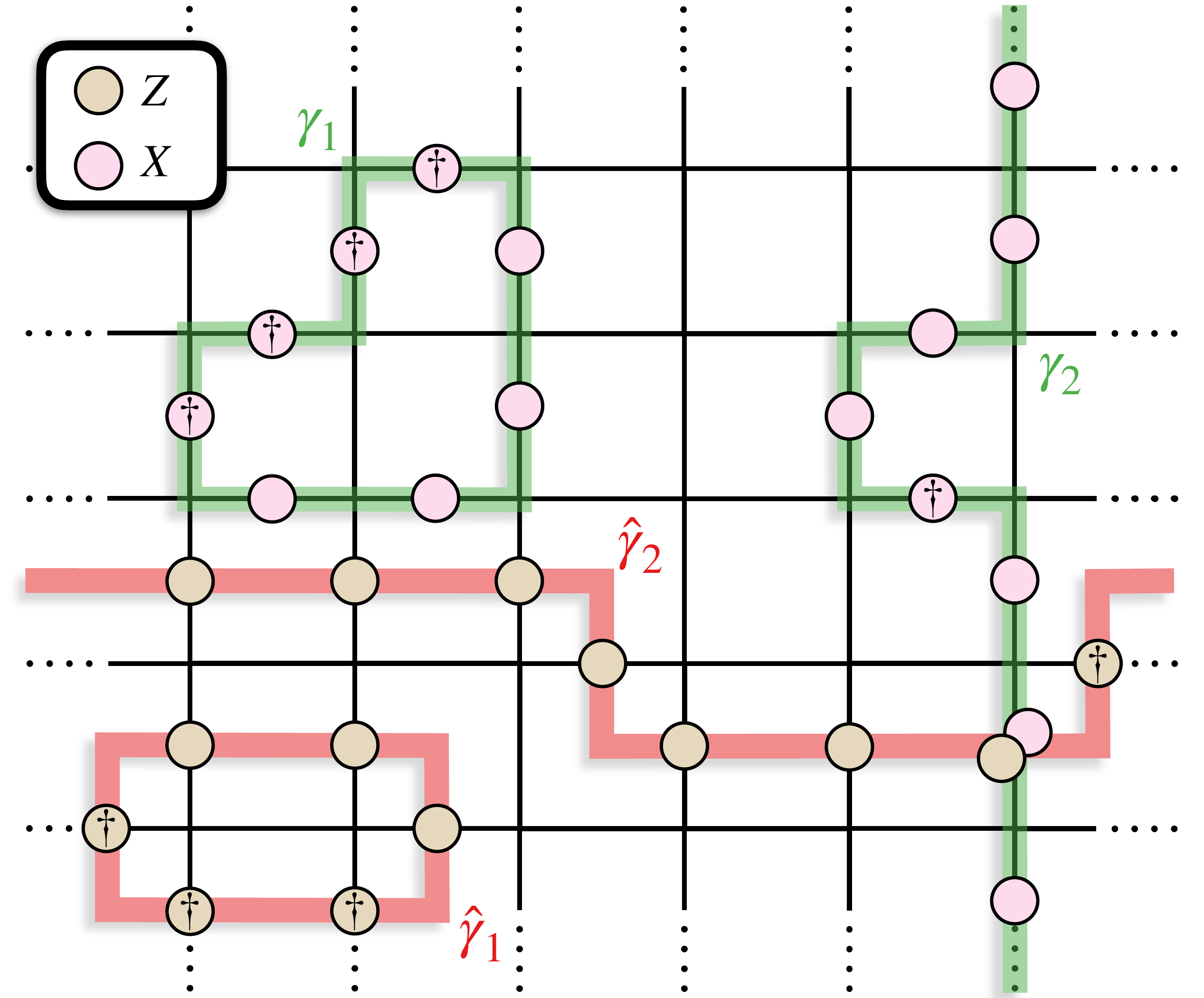}
    \caption{The symmetry operators $U(\ga)$ and $\t U(\h\ga)$ of the R1TC act on closed loops of the direct and dual lattice. Examples of $U(\ga)$ (see Eq.~\eqref{r1TCSYM1}) acting on loops of the direct lattice ${\gamma_1}$ and ${\gamma_2}$ are shown in green while examples of $\t U(\h\ga)$ (see Eq.~\eqref{r1TCSYM2}) acting on loops of the dual lattice ${\hat{\gamma}_1}$ and ${\hat{\gamma}_2}$ are in red. ${\gamma_1}$ and ${\hat{\gamma}_1}$ are contractible loops while, assuming periodic boundary conditions, ${\gamma_2}$ and ${\hat{\gamma}_2}$ are non-contractible.}
    \label{fig:loops}
\end{figure}

There are two independent operators that commute with the R1TC Hamiltonian Eq.~\eqref{Eq:R1TCHamiltonian}, each corresponding to a symmetry. First consider the unitary
\begin{equation}\label{r1TCSYM1}
U(\gamma) = \prod_{c_1\in\gamma}X_{c_1},
\end{equation}
where ${\gamma}$ is an oriented closed loop made of the lattice's edges (e.g., $\gamma_1$ and $\gamma_2$ in Fig.~\ref{fig:loops}) and ${X_{c_1}}$ satisfies ${X_{-c_1} = X_{c_1}^\da}$. ${U(\gamma)}$ trivially commutes with ${b_{c_2}}$ for all ${\gamma}$ and ${c_2}$. Furthermore, ${U(\gamma)}$ commutes with ${a_{c_0}}$ since for each site ${c_0}$, ${\gamma}$ is made up of an even number of elements of ${\delta c_0}$ with relative orientations such that all phases $\ee^{\ii 2\pi/N}$ cancel. Therefore, ${[U(\gamma),H] = 0}$ for all loops ${\gamma}$. Next, consider the unitary
\begin{equation}\label{r1TCSYM2}
\widetilde{U}(\hat{\gamma}) = \prod_{\h c_1 \in \hat{\gamma}}Z_{\hstar \h c_1},
\end{equation}
where ${\hat{\gamma}}$ is now an oriented closed loop made of the dual lattice's edges (e.g., $\hat{\gamma}_1$ and $\hat{\gamma}_2$ in Fig.~\ref{fig:loops}), $\h c_1$ is a dual lattice edge, and $\hstar \h c_1$ is the edge of the direct lattice that crosses $\h c_1$ (up to a differing sign, see Eq.~\eqref{stardualcell}). ${\widetilde{U}(\hat{\gamma})}$ trivially commutes with ${a_{c_0}}$ for all ${\hat{\gamma}}$ and ${c_0}$. Furthermore, ${\widetilde{U}(\hat{\gamma})}$ commutes with ${b_{c_2}}$ since for each plaquette ${c_2}$, ${\hat{\gamma}}$ is made up of an even number of elements of ${\partial c_2}$ with relative orientations such that all phases $\ee^{\ii 2\pi/N}$ cancel. Therefore, ${[\widetilde{U}(\hat{\gamma}),H] = 0}$ for all loops ${\hat{\gamma}}$.

Since ${U}$ and ${\widetilde{U}}$ commute with ${H}$, and since they transform the qubits nontrivially, they correspond to symmetries. Indeed, they generate the transformations
\begin{align}
U(\gamma) ~Z_{c_1} ~ U^{\dagger}(\gamma) &= \om^{\#(c_1,\gamma)}~Z_{c_1},\label{eq:Z2SymTr1}\\
\widetilde{U}(\hat{\gamma})~X_{c_1} ~ \widetilde{U}^{\dagger}(\hat{\gamma}) &= \om^{-\#(c_1,\hat{\gamma})}~X_{c_1}\label{eq:Z2SymTr2},
\end{align}
where ${\om = \ee^{\ii 2\pi/N}}$ and, for instance, ${\#(c_1,\gamma)}$ is the signed intersection number of ${c_1}$ and ${\gamma}$. Because ${[U(\gamma)]^N = [\widetilde{U}(\hat{\gamma})]^N = 1}$, they are the generators of a ${\mathbb{Z}_N \times \mathbb{Z}_N}$ symmetry. However, this is not quite an ordinary global symmetry since ${U}$ and ${\widetilde{U}}$ act on closed loops instead of the entire lattice. Instead, they correspond to \txti{non-topological} $\Z_N$ 1-form symmetries. Physically, this symmetry reflects the lack of dynamics of $e$ and $m$ anyons in the R1TC is absent throughout the rest of the deconfined phase of $\mathbb{Z}_N$ gauge theory.

In the ground state sub-Hilbert space, the operators ${a_{c_0}}$ and ${b_{c_2}}$ obey the constraints ${\langle a_{c_0}\rangle_{\text{gs}} = 1}$ and ${\langle b_{c_2}\rangle_{\text{gs}} = 1}$, where ${\langle~~\rangle_{\text{gs}}}$ denotes the expectation value with respect to the ground states. Consequentially, when ${\gamma}$ and ${\hat{\gamma}}$ are contractible loops, ${\langle U(\gamma) \rangle_{\text{gs}}=1}$ and ${\langle \widetilde{U}(\hat{\gamma})\rangle_{\text{gs}} = 1}$, which follows from
\begin{align}
    U(\ga = \pp M) &= \prod_{c_2\in M}b_{c_2},\\
    \t U(\h\ga = \pp\h M) &= \prod_{\h{c}_2\in\h M}a^\da_{\hstar\h c_2}.
\end{align}
In fact, ${U(\gamma)}$ and ${\widetilde{U}(\hat{\gamma})}$ are so-called topological operators in the ground state sub-Hilbert space, since their vacuum expectation values depend only on the topology---the homology class---of ${\gamma}$ and ${\hat{\gamma}}$, respectively. In other words, in the ground state sub-Hilbert space, the symmetry operators are nontrivial only when $\ga$ and $\h\ga$ are noncontractible loops. Furthermore, one can choose $\ga$ and $\h\ga$ such that $U$ and $\t U$ are the R1TC ``holonomies'' discussed in Sec.~\ref{sec:preholonomies}. 

In the ground state sub-Hilbert space, ${X_{c_1}}$ and ${Z_{c_1}}$ are not allowed operators since they excite ${e}$ and ${m}$ anyons, respectively, violating the ${\langle a_{c_1}\rangle = 1}$ and ${\langle b_{c_2}\rangle = 1}$ constraints. The allowed operators are, instead, ${U(\gamma)}$ and ${\widetilde{U}(\hat{\gamma})}$. The aforementioned generalized  ${\mathbb{Z}_N \times \mathbb{Z}_N}$ symmetry transformations, Eqs.~\eqref{eq:Z2SymTr1} and~\eqref{eq:Z2SymTr2}, in the ground state sub-Hilbert space are replaced with
\begin{align}
U(\gamma) ~\widetilde{U}(\hat{\gamma}) ~ U^{\dagger}(\gamma) &= \om^{\#(\hat{\gamma},\gamma)}~\widetilde{U}(\hat{\gamma}),\label{eq:Z21formSymTr1}\\
\widetilde{U}(\hat{\gamma})~U(\gamma)  ~ \widetilde{U}^{\dagger}(\hat{\gamma}) &= \om^{-\#(\gamma,\hat{\gamma})}~U(\gamma) \label{eq:Z21formSymTr2}.
\end{align}
These now correspond to ${\mathbb{Z}_N}$ 1-form---${\mathbb{Z}_N^{(1)}}$---symmetries since their symmetry operators are topological operators supported on codimension 1 closed subspaces and their charged operators are supported on 1-dimensional closed subspaces. In fact, this ${\mathbb{Z}_N^{(1)}\times \mathbb{Z}_N^{(1)}}$ symmetry is also a symmetry~\cite{KS14010740} of the topological quantum field theory description of the R1TC ground states~\cite{KLW0834}. Unlike the non-topological ${\mathbb{Z}_N\times \mathbb{Z}_N}$ 1-form symmetry of Eqs.~\eqref{eq:Z2SymTr1} and~\eqref{eq:Z2SymTr2}, the ${\mathbb{Z}_N^{(1)}\times \mathbb{Z}_N^{(1)}}$ symmetry exists as an emergent symmetry in the ground state sub-Hilbert space throughout the entire deconfined phase of $\mathbb{Z}_N$ gauge theory~\cite{HW0541, PW230105261}.

Just like ordinary global symmetries, 1-form symmetries can spontaneously break~\cite{GW14125148, L180207747}. The order parameter of a 1-form symmetry spontaneous breaking is the vacuum expectation value of its charged operator supported on a contractible loop. Recall that ${\langle U(\gamma) \rangle_{\text{gs}}=1}$ and ${\langle \widetilde{U}(\hat{\gamma})\rangle_{\text{gs}} = 1}$ when ${\gamma}$ and ${\hat{\gamma}}$ are contractible loops. Since $\widetilde{U}$ is charged under the $\mathbb{Z}_N^{(1)}$ symmetry generated by $U$ (see Eq.~\eqref{eq:Z21formSymTr1}) and vice versa, the R1TC ground states spontaneously break the ${\mathbb{Z}_N^{(1)}\times \mathbb{Z}_N^{(1)}}$ symmetry. This reproduces the well known property that there is a ground state degeneracy depending on the topology---the 1st cohomology---of the spatial lattice. In fact, the ${\mathbb{Z}_N^{(1)}\times \mathbb{Z}_N^{(1)}}$ symmetry is anomalous, meaning both $\mathbb{Z}_N^{(1)}$ symmetries cannot be simultaneously gauged. The ground state degeneracy (GSD) arising when this anomalous ${\mathbb{Z}_N^{(1)}\times \mathbb{Z}_N^{(1)}}$ symmetry is spontaneously broken is ${\txt{GSD} = N^2}$ for the square lattice with periodic boundary conditions.

A manifestation of this mixed 't Hooft anomaly is that the symmetry operators obey the Heisenberg algebra\footnote{\label{anomalyFootnoteExplain}Gauging a symmetry $U$ is the procedure of adding additional degrees of freedom such that the theory becomes invariant under the gauged symmetry operator ${U_{\txt{gauged}}}$. ${U_{\txt{gauged}}}$ acts on both open and closed subspaces and physical states must satisfy ${U_{\txt{gauged}}\ket{\psi}=\ket{\psi}}$. A contradiction arises when different ${U_{\txt{gauged}}}$ no longer commute, reflecting an obstruction to gauging the symmetry (a 't Hooft anomaly). For example, consider ${U^{(1)}_{\txt{gauge}}U^{(2)}_{\txt{gauge}} = -U^{(2)}_{\txt{gauge}}U^{(1)}_{\txt{gauge}}}$. Since ${U^{(1,2)}_{\txt{gauge}}\ket{\psi}=\ket{\psi}}$, this leads to the contradiction ${\ket{\psi} = -\ket{\psi}}$.}
\begin{equation}\label{eq:R1TCanomaly}
\widetilde{U}(\hat{\gamma})  U^{\dagger}(\gamma) = (\ee^{\ii 2\pi/N})^{\#(\hat{\gamma},\gamma)}~U^\dagger(\gamma) \widetilde{U}(\hat{\gamma}),
\end{equation}
and therefore $U$ and $\widetilde{U}$ form a projective representation of ${\mathbb{Z}_N^{(1)}\times \mathbb{Z}_N^{(1)}}$. The mixed 't Hooft anomaly ensures that the ground state cannot be a trivial product state 
(see, e.g., Refs.~\onlinecite{KR180505367, JW200900023, GKK170300501}), 
and instead the R1TC must be in either a gapless or an SSB phase. Therefore, the mixed 't Hooft anomaly protects the spontaneous symmetry breaking pattern and, therefore, the $\Z_N$ topological order. Furthermore, the mixed 't Hooft anomaly is also present at higher energies, affecting the non-topological $\Z_N$ 1-form symmetries. Its manifestation Eq.~\eqref{eq:R1TCanomaly} gives rise to nontrivial mutual statistics between $e$ and $m$ anyons~\cite{HS181204716}.

\subsection{Symmetries of the R2TC}

Having summarized the symmetries in the R1TC, let us now consider the R2TC. It is convenient to choose a slightly different, but physically equivalent, square lattice where the $(X_1,Z_1)$ and $(X_2,Z_2)$ $\mathbb{Z}_N$ spins reside on horizontal links while the $(X_0,Z_0)$ $\mathbb{Z}_N$ spins reside on vertical links. In fact, this is the lattice $\La_2$ in Fig.~\ref{fig:quadrupole}. The $\Z_N$ R2TC Hamiltonian is then given by
\begin{align}
    H &= -\sum_{c_1^{(h)}} \mathfrak{A}_{c_1^{(h)}} -\sum_{c_0} \mathfrak{B}^{x}_{c_0} -\sum_{c_2} \mathfrak{B}^{y}_{c_2},\label{R2TCHam}\\
    & \mathfrak{B}^x_{c_0} = \frac{1}{N} \sum_{j=1}^N (\mathfrak{b}^x_{c_0} )^j , \quad\quad\quad \mathfrak{B}^y_{c_2} = \frac{1}{N} \sum_{j=1}^N (\mathfrak{b}^y_{c_2} )^j ,\nonumber\\
    & \mathfrak{A}_{c_1^{(h)}} = \frac{1}{N} \sum_{j=1}^N (\mathfrak{a}_{c_1^{(h)}} )^j,\nonumber
\end{align}
where $c_1^{(h)}$ denotes a horizontal link, $c_0$ a lattice site, and $c_2$ a plaquette. Fig.~\ref{fig:ops}b shows graphical representations of the operators $\mathfrak{a}$, $\mathfrak{b}^{x}$, and $\mathfrak{b}^{y}$, which are also defined in Eqs.~\eqref{eq:r2tc-a-proj} and~\eqref{eq:b1b2-proj}, respectively. From Fig.~\ref{fig:ops}b, it is clear these operators are mutually commuting, and therefore the ground state satisfies ${\mathfrak{a}=1}$, ${\mathfrak{b}^{x}=1}$, and ${\mathfrak{b}^{y}=1}$.

The R2TC Hamiltonian operators $\mathfrak{a}$, $\mathfrak{b}^{x}$, and $\mathfrak{b}^{y}$ have a rich and complicated structure. Consequently, the theory can have many interesting generalized symmetries. We will construct its symmetries in~\ref{constructR2TCsymOp}, which will include mostly technical details. Afterwards, in~\ref{r2TCsymDisc}, we will discuss these symmetry operators, analyzing how the R2TC's interesting properties can be interpreted from a symmetry point of view and comparing the symmetry operators to conventional 1-form symmetries.

\subsubsection{Construction of symmetry operators}\label{constructR2TCsymOp}

\begin{figure}[t!]
\centering
    \includegraphics[width=.48\textwidth]{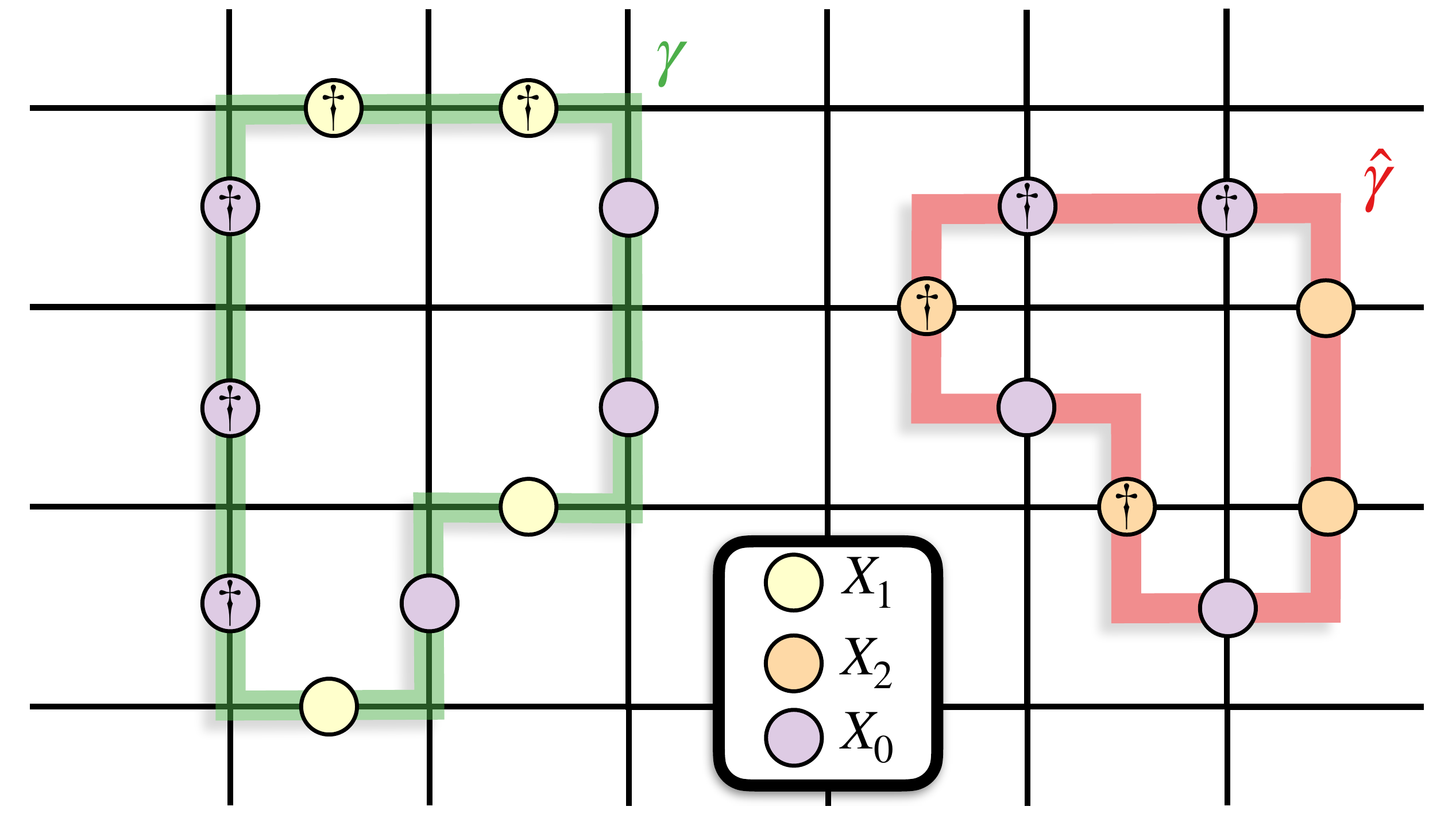}
    \caption{The R2TC symmetry operators $U_1(\ga)$ and $U_2(\h\ga)$, defined by Eqs.~\eqref{eq:r2tcSymOp1} and~\eqref{eq:r2tcSymOp2} respectively, act on closed loops of the direct and dual lattice. Here we show graphical representation of an example of $U_1(\ga)$ acting on a loop of the direct lattice ${\gamma}$ (drawn in green) and of $U_2(\h\ga)$ acting on a loop of the dual lattice ${\hat{\gamma}}$ (drawn in red).}
    \label{fig:r2tcsym2}
\end{figure}

Let us first identify the symmetries which are generated by operators built out of only $X_0$, $X_1$, and $X_2$. To do so, we define the lattice vector fields $\eX_1$ and $\eX_2$ which are related to $X_0$, $X_1$, and $X_2$ by
\begin{align}
    \eX_{1,c_0}^{i} &= (X_{1,c_0}^x,X_{0,c_0}^y)\\
    \eX_{2,\h{c}_0}^{i} &= (X_{0,|\hstar\h c_0|+\h x}^y,X_{2,|\hstar \h c_0|+\h y}^x).
\end{align}
Notice that while $\eX_1$ is specified by the links of the direct lattice, $\eX_2$ is instead specified by the links of the dual lattice. As elaborated on in appendix section~\ref{sec:diffgeoLat}, the position of ${|\hstar \h c_0|}$ is related to a direct lattice site $c_0$ by ${c_0 = |\hstar \h c_0|-\h x /2 - \h y/2}$, where $|\hstar \h c_0|$ is just the absolute value of $\hstar \h c_0$. Using $\eX_1$ and $\eX_2$, we construct the unitary operators
\begin{align}
        U_1(\ga) &= \prod_{c_{1}\in\ga}\eX_{1,c_{1}},\label{eq:r2tcSymOp1}\\
    U_2(\h\ga) &= \prod_{\h c_{1}\in\h\ga}\eX_{2,\h{c}_{1}},\label{eq:r2tcSymOp2}
\end{align}
where $\ga$ and $\h\ga$ are oriented loops on the direct and dual lattice, respectively (see Fig.~\ref{fig:r2tcsym2}). $U_1$ and $U_2$ trivially commute with $\mathfrak{B}_{c_0}^x$ and $\mathfrak{B}_{c_2}^y$ in the R2TC Hamiltonian Eq.~\eqref{R2TCHam}. $U_1$ and $U_2$ also commute with $\mathfrak{A}_{c_1^{(h)}}$, which can be confirmed directly or simply by comparing the graphical representations shown in Figs.~\ref{fig:ops}b and~\ref{fig:r2tcsym2}. Therefore, for all $\ga$ and $\h\ga$, ${[U_1(\ga),H] = [U_2(\h\ga),H] = 0}$, and $U_1$ and $U_2$ correspond to symmetry operators.

When $\ga$ and $\h\ga$ are contractible, the symmetry operators $U_1$ and $U_2$ can be written as
\begin{align}
        U_1(\ga = \pp M) &= \prod_{c_{2}\in M}\mathfrak{b}^y_{c_{2}},\\
    U_2(\h\ga = \pp\h M) &= \prod_{\h c_{2}\in\h M}\mathfrak{b}^x_{|\hstar\h{c}_{2}|}.
\end{align}
Consequentially, in the ground state subspace where ${\mathfrak{b}^{x,y} = 1}$, $U_1$ and $U_2$ are topological operators, depending only on the homology class of $\ga$ and $\h\ga$. Since ${U_1^N = U_2^N = 1}$, we therefore find that these symmetry operators generate an emergent ${\ZN{1}\times \ZN{1}}$ symmetry in the IR. We note that when $\ga$ is a loop of links in the $x$-direction ($y$-direction), $U_1$ becomes the ``holonomy'' $W_3$ ($W_4$) from Eq.~\eqref{eq:x-hol}. Similarly, when $\h\ga$ is a loop of dual links in the $x$-direction ($y$-direction), $U_2$ becomes the holonomy $W_1$ ($W_2$) from Eq.~\eqref{eq:x-hol}.

There is one more symmetry operator which can be constructed from the $X$ operators. Let us define the operator $\eX_3$ which acts only on the horizontal links $c_1^{(h)}$. $\eX_3$ is interpreted as a lattice vector field whose $x$ component acts on the horizontal links of $c_1^{(h)}$ but whose $y$ component acts on the vertical links of the dual lattice ${\hstar\hat{c}_1^{(v)} =  c_1^{(h)}}$. However, the horizontal links form their own square lattice ${\cV_{\txt{vh}}}$ whose sites
$v \equiv (v_x,v_y)$ are squares in Fig.~\ref{fig:quadrupole}. We will formulate this symmetry on the ${\cV_{\txt{vh}}}$ lattice where it turns out to be most naturally defined. However, this can also be formulated on the direct lattice if the framing structure is utilized, which makes the following symmetry a so-called framed-symmetry~\cite{QRH201002254}. $\eX_3$ is related to $X_1$ and $X_2$ by
\begin{equation}
    \eX^i_{3,v}  = (X_{1,c_1^{(h)}}, X_{2,c_1^{(h)}}),
\end{equation}
where the ${\cV_{\txt{vh}}}$ site $v$ on the left hand side is the center of the edge $c_1^{(h)}$ on the right hand side.

\begin{figure}[t!]
\centering
    \includegraphics[width=.48\textwidth]{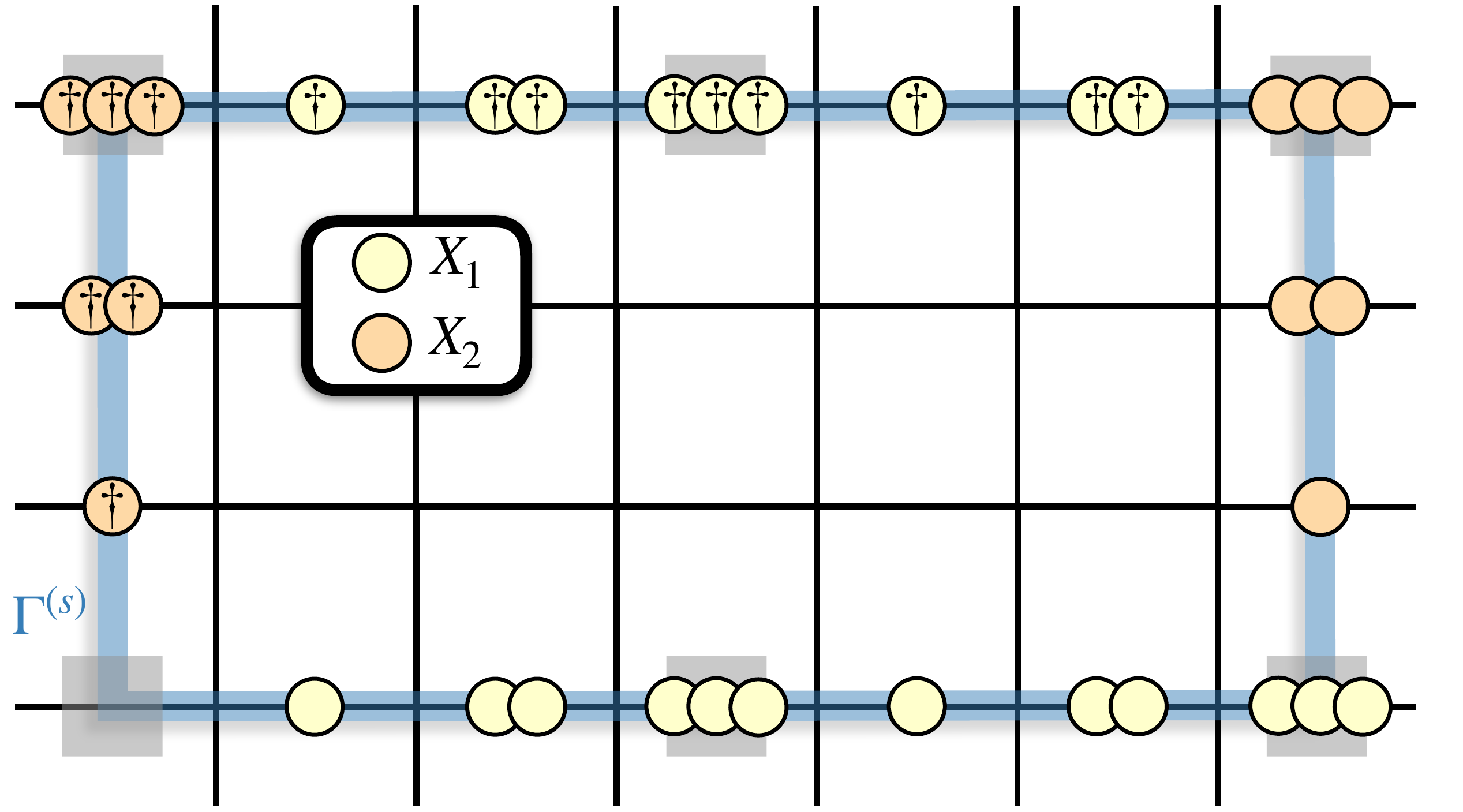}
    \caption{The R2TC symmetry operator $U_3(\Ga^{(s)})$ defined by Eq.~\eqref{eq:r2tcSymOp3} acts on closed loops $\Ga^{(s)}$ of the $\cV_{\txt{vh}}$ lattice. Here we show a graphical representation of $U_3(\Ga^{(s)})$ acting on a particular loop $\Ga^{(s)}$ drawn in blue with ${N=3}$. The $\cV_{\txt{vh}}$ lattice sites belonging to the $s$ sublattice are denoted by gray squares, and we sometimes include the operators $(X_1)^3$ and $(X_2)^3$ despite them being the identity.}
    \label{fig:r2tcsym3}
\end{figure}

Using $\eX_3$, we can construct a unitary which commutes with the R2TC Hamiltonian. To do so, we first reconsider the ${\cV_{\txt{vh}}}$ square lattice as a Bravais lattice with a basis. The conventional unit cell is an $N\times N$ square surrounding $N^2$ lattice sites, each of which belong to their own sublattice. We introduce the index ${s \in \{1,2,\cdots,N^2 -1,N^2\}}$ which labels each sublattice. Let us denote a generic oriented closed loop of the ${\cV_{\txt{vh}}}$ lattice as $\Ga$, and specify loops made of only length $N$ segments connecting sites of the sublattice $s$ as $\Ga^{(s)}$. With this set up, we now consider the unitary operator
\begin{equation}\label{eq:r2tcSymOp3}
    U_3(\Ga^{(s)}) = \prod_{v \in \Ga^{(s)}} (\eX_{3,r})^{(v_x-r^{(s)}_x)+(v_y-r^{(s)}_y)},
\end{equation}
where $v$ is a $\cV_{\txt{vh}}$ lattice site\footnote{In Eq.~\eqref{eq:r2tcSymOp3}, the notation $v \in \Ga^{(s)}$ simply means all $\cV_{\txt{vh}}$ lattice sites $v$ which the loop $\Ga^{(s)}$ crosses. Here, $\Ga^{(s)}$ should not be considered as a 1-cycle---an integer sum of 1-chains in the kernel of the boundary operator $\pp_1$. We will commit similar abuses of notation throughout this section.} and $r^{(s)}$ is the basis vector (in the crystallography sense) of sublattice $s$. Fig.~\ref{fig:r2tcsym3} shows an example $U_3(\Ga^{(s)})$ for ${N=3}$.

The operator $U_3$ trivially commutes with $\mathfrak{B}_{c_0}^x$ and $\mathfrak{B}_{c_2}^y$ in the R2TC Hamiltonian Eq.~\eqref{R2TCHam}. Furthermore, $U_3(\Ga^{(s)})$ also commutes with $\mathfrak{A}_{c_1^{(h)}}$ for all $\Ga^{(s)}$, which can be confirmed by direct computation or simply from inspecting the graphical representations shown in Figs.~\ref{fig:ops}b and~\ref{fig:r2tcsym3}. Therefore, ${[U_3(\Ga^{(s)}), H] = 0}$ for all $\Ga^{(s)}$, and $U_3$ indeed corresponds to a symmetry operator. When $\Ga^{(s)}$ is contractible, the symmetry operator $U_3$ can be written as
\begin{equation}
    U_3(\Ga^{(s)} = \pp M) \hspace{-2pt}= \hspace{-2pt}\prod_{c_0\in M}\hspace{-2pt}(\mathfrak{b}^x_{c_0})^{(c_0)_y}\hspace{-2pt}\prod_{c_2\in M}\hspace{-2pt}(\mathfrak{b}^y_{c_2})^{(c_2)_x}.
\end{equation}
Here, $(c_0)_y$ is the distance of $c_0$ from $\Ga^{(s)}$ in the $-y$-direction. Similarly, $(c_2)_x$ is the distance of $c_2$ from $\Ga^{(s)}$ in the $-x$-direction. Since ${\mathfrak{b}^x_{c_0} = \mathfrak{b}^y_{c_2} = 1}$ in the ground state subspace, $U_3(\Ga^{(s)})$ is a topological operator and corresponds to a 1-form symmetry in the IR. However, this is not an ordinary 1-form symmetry since $\Ga^{(s)}$ is not allowed to be any loop on the $\cV_{\txt{vh}}$ lattice. As a result $U_3(\Ga^{(s)})$ is not a fully topological operator on the $\cV_{\txt{vh}}$ lattice, but is on the $s$ sublattice. Therefore, we refer to $U_3(\Ga^{(s)})$ as a sublattice 1-form symmetry.

The precise nature of this sublattice 1-form symmetry depends on both the topology and geometry of the lattice in a sensitive way. Without periodic boundary conditions, this is a sublattice $\Z_N$ 1-form symmetry. With periodic boundary conditions, the previous ${N\times N}$ conventional unit cell shrinks to a ${\operatorname{gcd}(L_x,N)\times \operatorname{gcd}(L_y,N)}$ unit cell (but $\Ga^{(s)}$ is stll made of only length $N$ segments). Consequently, $\Ga^{(s)}$ must wrap around system ${N/\operatorname{gcd}(L_i,N)}$ times in the $i$-direction in order to close. Therefore, on a torus, $U_3$ is a $\Z_{\operatorname{gcd}(L_x,N)}^{(1)}\times \Z_{\operatorname{gcd}(L_y,N)}^{(1)}$ sublattice 1-form symmetry, where the noncontractible $\Ga^{(s)}$ of the $\Z_{\operatorname{gcd}(L_i,N)}^{(1)}$ sublattice symmetry is understood winding only in the $i$-direction. We note that the $\Z_{\operatorname{gcd}(L_x,N)}^{(1)}$ and $\Z_{\operatorname{gcd}(L_y,N)}^{(1)}$ symmetry operators are related to the $W_5$ and $W_6$ ``holonomies,'' respectively, from Eq.~\eqref{eq:x-hol}.

\begin{figure}[t!]
\centering
    \includegraphics[width=.48\textwidth]{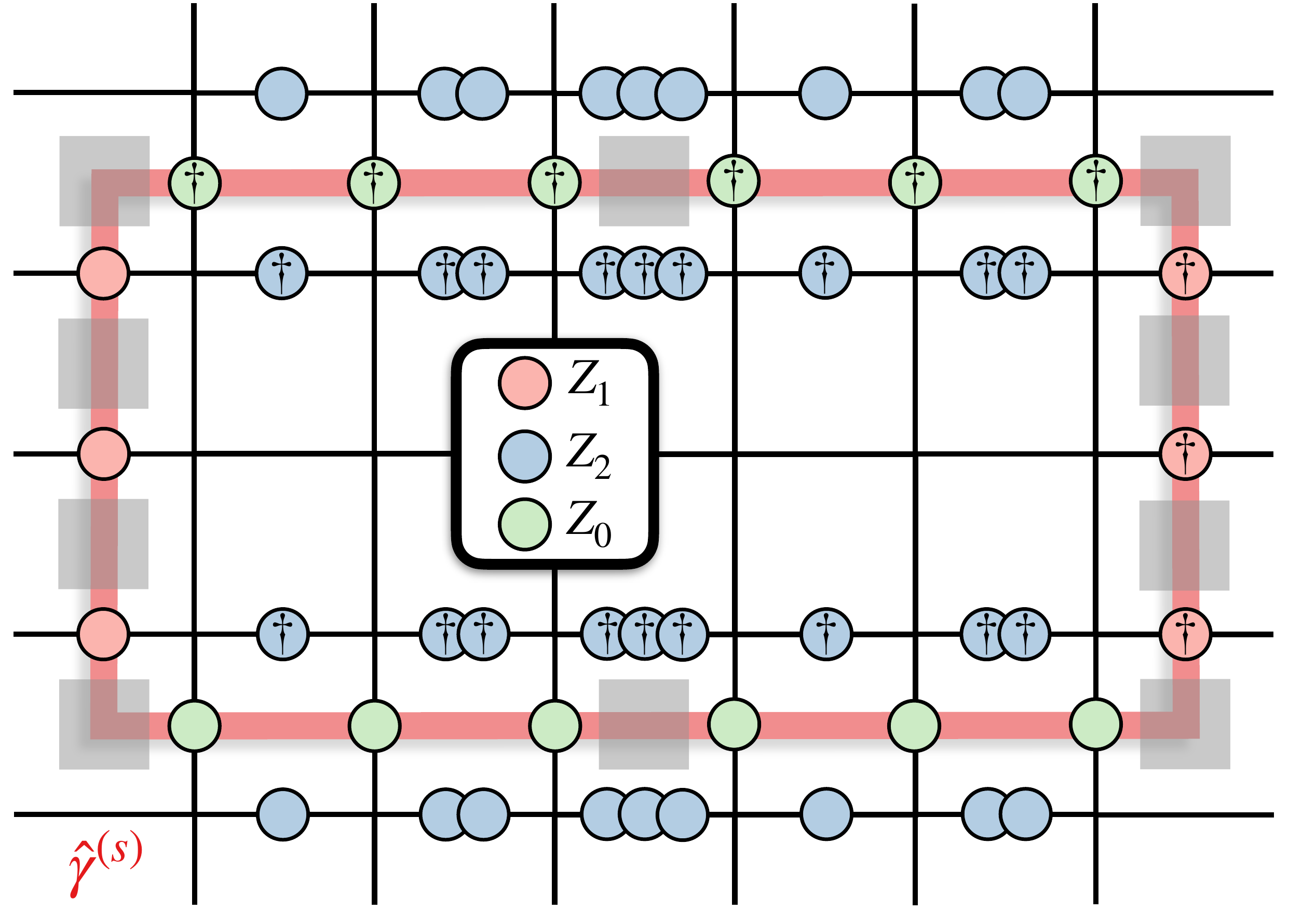}
    \caption{The R2TC symmetry operator $\t U_1(\h\ga^{(s)})$ defined by Eq.~\eqref{eq:r2tcSymOp4} acts on closed loops $\h\ga^{(s)}$ of the dual lattice. Here we show a graphical representation of $\t U_1(\h\ga^{(s)})$ acting on a particular loop $\h\ga^{(s)}$ drawn in red with ${N=3}$. The dual lattice sites belonging to the $s$ sublattice are denoted by gray squares, and we sometimes include the operator $(Z_2)^3$ despite it being the identity.}
    \label{fig:r2tcsym5}
\end{figure}

We now move on to discuss the symmetry operators constructed from only $Z_0$, $Z_1$, and $Z_2$. We will find that there are three symmetry operators, two of which correspond two sublattice 1-form symmetries and one is a conventional 1-form symmetry.

To construct the first symmetry operator, we must reconsider unit cell of the lattice as a $N\times 1$ unit cell with a basis labeled by $s\in\{1,\cdots,N\}$. A loop of the dual lattice made of only length $N$ segments in the horizontal direction connecting the sites of sublattice $s$ is denoted as $\h\ga^{(s)}$. We then introduce the lattice vector field $\eZ_1$ acting on the links of the dual lattice, $\h c_1$. It is related to the $Z_0$, $Z_1$, and $Z_2$ operators by
\begin{equation}
\begin{aligned}
        \hspace{-10pt}\eZ_{1,\h c_1}  &= (Z_{0, \hstar \h c_1}(Z^\da_{2,\hstar \h c_1+\h x/2 + \h y/2}Z_{2,\hstar \h c_1+\h x/2 - \h y/2})^{x-x^{(s)}}\hspace{-5pt},\\
        &\hspace{150pt}Z_{1,\hstar \h c_1})
\end{aligned}
\end{equation}
where $x$ is the $x$-coordinate of the dual lattice site in the $x$-direction of $\h c_1$ and $x^{(s)}$ is the $x$-coordinate of the basis vector for sublattice $s$. With this defined, we can then consider the unitary
\begin{equation}\label{eq:r2tcSymOp4}
    \t U_1(\h\ga^{(s)}) = \prod_{\h c_1\in\h\ga^{(s)}}\eZ_{1,\h c_1},
\end{equation}
an example of which is shown in Fig.~\ref{fig:r2tcsym5} for ${N=3}$. It is straightforward to check that for all $\h\ga^{(s)}$, $\t U_1$ commutes with $\mathfrak{a}$, $\mathfrak{b}^x$, and $\mathfrak{b}^y$, and therefore $[\t U_1,H] = 0$.

\begin{figure}[t!]
\centering
    \includegraphics[width=.48\textwidth]{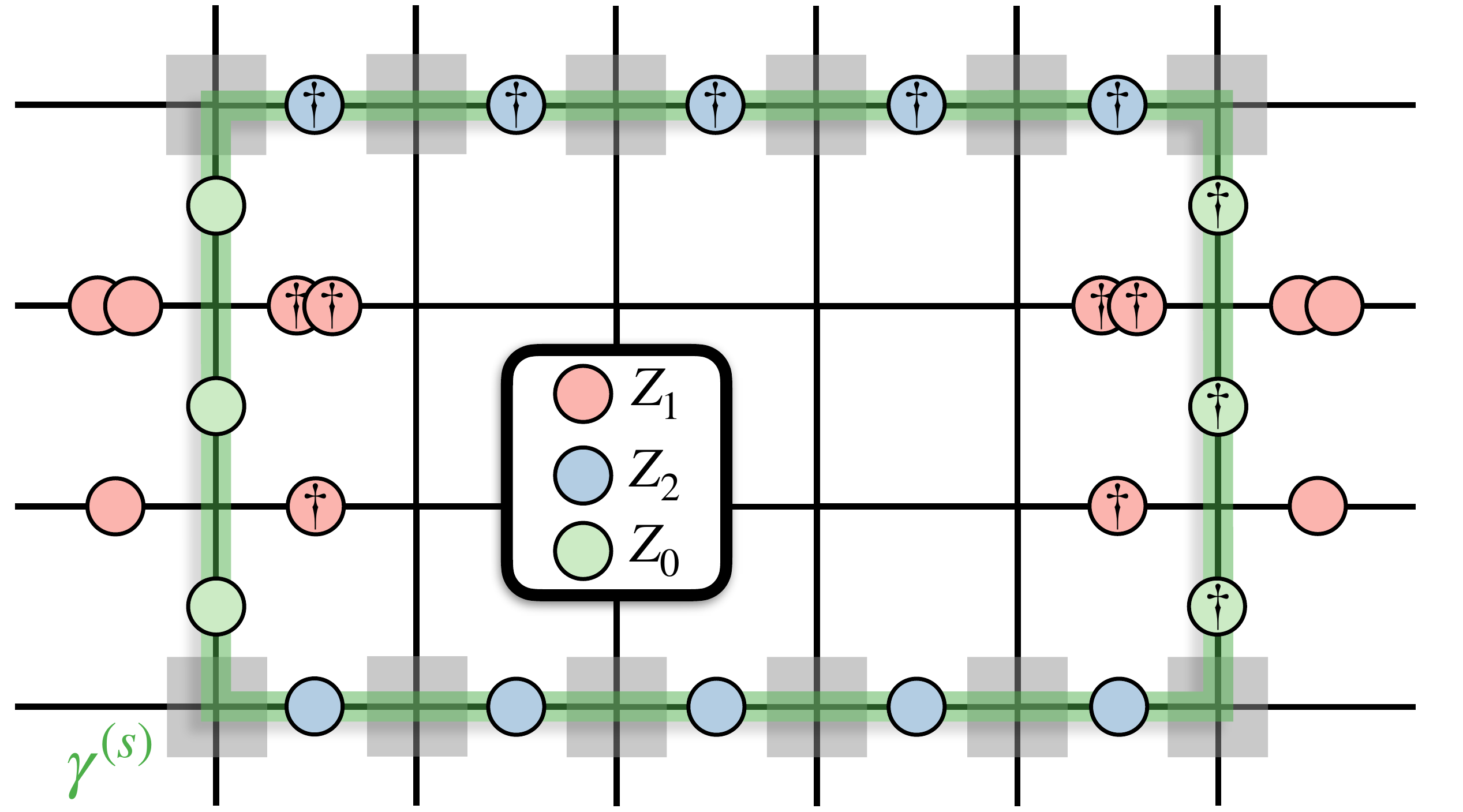}
    \caption{The R2TC symmetry operator $\t U_2(\ga^{(s)})$ defined by Eq.~\eqref{eq:r2tcSymOp5} acts on closed loops $\ga^{(s)}$ of the direct lattice. Here we show a graphical representation of $\t U_2(\ga^{(s)})$ acting on a particular loop $\ga^{(s)}$ drawn in green with ${N=3}$. The direct lattice sites belonging to the $s$ sublattice are denoted by gray squares.}
    \label{fig:r2tcsym4}
\end{figure}

The second symmetry operator is rather similar to the first one. Now we instead consider a ${1\times N}$ units cell with a basis again labeled by $s\in\{1,\cdots, N\}$. A Loop of the direct lattice made of only length $N$ segments in the vertical direction connecting the sites of sublattice $s$ is denoted as $\ga^{(s)}$. We then introduce the lattice vector field $\eZ_2$ acting on the links of the direct lattice, $c_1$. It is related to the $Z_0$, $Z_1$, and $Z_2$ operators by
\begin{equation}
\begin{aligned}
        \eZ_{2,c_1}  &= (Z_{1, c_1},\\
        &\hspace{10pt}Z^\da_{0, c_1}(Z^\da_{2, c_1-\h x/2 + \h y/2}Z_{2, c_1+\h x/2 + \h y/2})^{y-y^{(s)}})
\end{aligned}
\end{equation}
where $y$ is the $y$-coordinate of the lattice site in the $y$-direction of $c_1$ and $y^{(s)}$ is the $y$-coordinate of the basis vector for sublattice $s$. With this defined, we can then consider the unitary
\begin{equation}\label{eq:r2tcSymOp5}
    \t U_2(\ga^{(s)}) = \prod_{c_1\in \ga^{(s)}} \eZ_{2,c_1},
\end{equation}
an example of which is shown in Fig.~\ref{fig:r2tcsym4} for ${N=3}$. It is straightforward to check that for all $\ga^{(s)}$, $\t U_2$ commutes with $\mathfrak{a}$, $\mathfrak{b}^x$, and $\mathfrak{b}^y$, and therefore $[\t U_2,H] = 0$.

Both unitary operators $\t U_1(\h\ga^{(s)})$ and $\t U_1(\h\ga^{(s)})$ correspond to symmetry operators. When acting on contractible loops, they can be written as
\begin{align}
    \t{U}_1(\h\ga^{(s)} = \pp \h M) &= \prod_{c_1^{(h)}\in \h M}(\mathfrak{a}_{c_1^{(h)}})^{(c_1^{(h)})_x},\\
    \t{U}_2(\ga^{(s)} = \pp M) &= \prod_{c_1^{(h)}\in M}(\mathfrak{a}_{c_1^{(h)}})^{(c_1^{(h)})_y},
\end{align}
where $(c^{(h)}_1)_x$ is the distance of $c^{(h)}_1$ from $\h\ga^{(s)}$ in the $-x$-direction and $(c^{(h)}_1)_y$ is the distance of $c^{(h)}_1$ from $\ga^{(s)}$ in the $-y$-direction. Evidently, both the symmetries generated by $\t U_1$ and $\t U_2$ are sublattice 1-form symmetries, defined on their respective sublattices. Like for the sublattice 1-form symmetry $U_3$, the details of the symmetry are sensitive to both the geometry and topology of the lattice. Indeed, with periodic boundary conditions the ${N\times 1}$ (${1\times N}$) unit cell defined for the $\t U_1$ ($\t U_2$) symmetry operator becomes a ${ \operatorname{gcd}(L_x,N)\times 1}$ (${1\times \operatorname{gcd}(L_y,N)}$) unit cell ($\ga^{(s)}$ and $\h\ga^{(s)}$ are still made of length $N$ segments in the $y$ and $x$ directions, respectively). Note that when $\t U_1$ ($\t U_2$) is supported on a non-contractible loop in the $y$ ($x$) direction, it is related to $\t W_3$ ($\t W_2$) in Eq.~\eqref{eq:post-projection-Z-holonomy}. Similarly, when $\t U_1$ ($\t U_2$) is supported on a non-contractible loop in the $x$ ($y$) direction, it becomes $\t W_4$ ($\t W_1$) in Eq.~\eqref{eq:post-projection-Z-holonomy}

\begin{figure}[t!]
\centering
    \includegraphics[width=.48\textwidth]{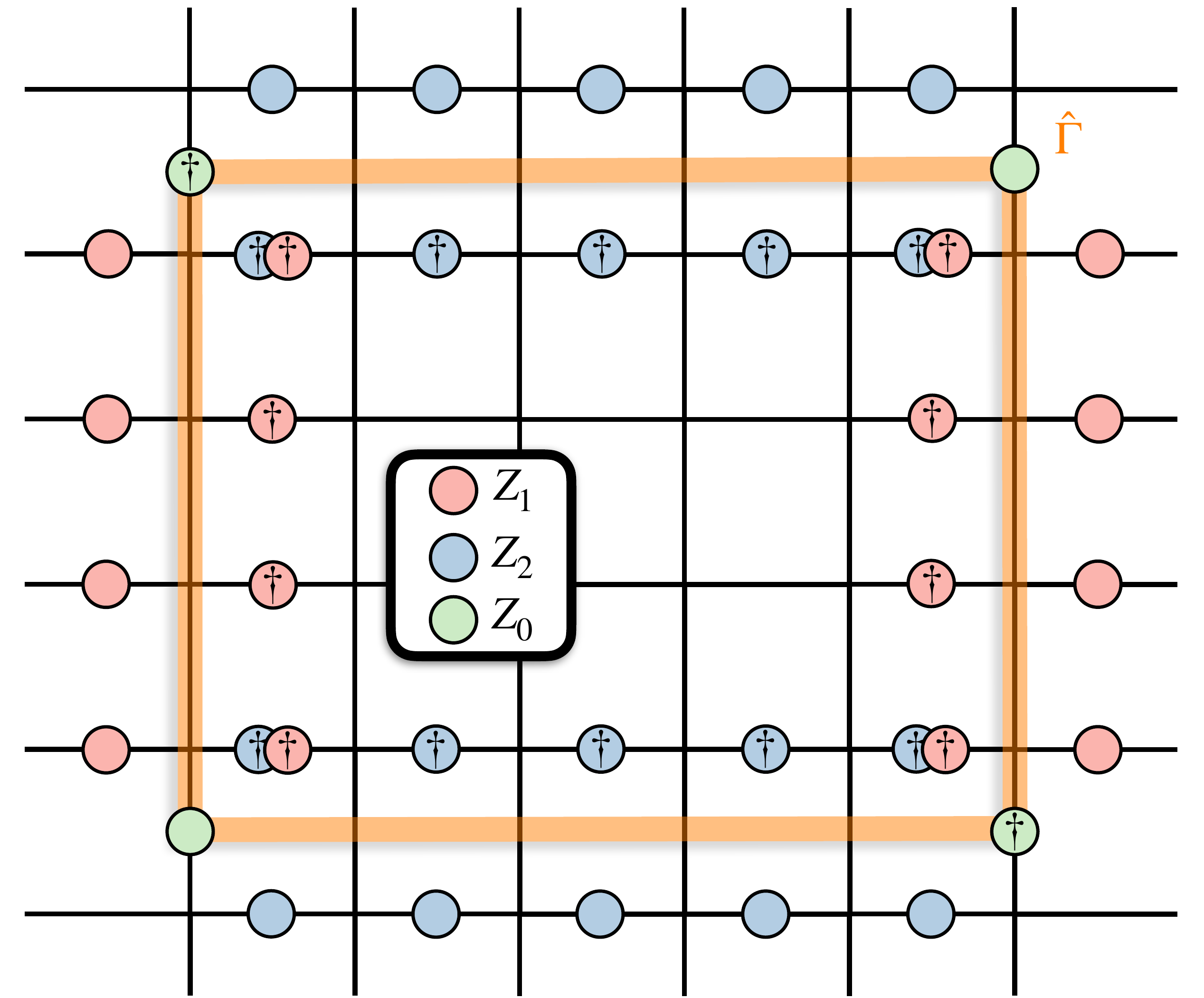}
    \caption{The R2TC symmetry operator $\t U_3(\h\Ga)$ defined by Eq.~\eqref{eq:r2tcSymOp6} acts on closed loops $\Ga$ of the dual $\cV_{\txt{vh}}$ lattice. Here we show a graphical representation of $\t U_3(\Ga)$ acting on a particular loop $\h\Ga$ drawn in orange.}
    \label{fig:r2tcsym1}
\end{figure}

Let us now construct the final symmetry operator. We define the operator $\eZ_3$ which acts only on the vertical links $c_1^{(v)}$ of the lattice. $\eZ_3$ is interpreted as a lattice vector field whose $x$ component acts on the plaquette $c_2$ but whose $y$ component acts on the horizontal links of the dual lattice ${\h c_1^{(h)} = \hstar c_1^{(v)}}$. It turns out it is most natural to formulate this symmetry operator on the previously mentioned $\cV_{\txt{vh}}$ lattice. We will denote the sites of the dual $\cV_{\txt{vh}}$ as $\h{v}$ and note that they are shown in Fig.~\ref{fig:quadrupole} as discs. $\eZ_3$ is related to $Z_0$, $Z_1$, and $Z_2$ by
\begin{equation}
\begin{aligned}
    \eZ_{3,\hat{v}} &= (
    Z_{0,\h v}
    Z_{0,\h v+\h x}^\da 
    Z^\da_{2,\h v + \h x/2 + \h y/2}
    Z_{2,\h v + \h x/2 - \h y/2},\\
    &\hspace{30pt}Z_{2,\h v  -\h x/2 + \h y /2}^\da
    Z_{2,\h v + \h x/2 + \h y /2}
    ).
\end{aligned}
\end{equation}
With $\eZ_3$ defined, we can then consider the unitary operator
\begin{equation}\label{eq:r2tcSymOp6}
    \t{U}_3(\h\Ga) = \prod_{\h v\in \h \Ga}\eZ_{3,\h v}
\end{equation}
where $\h\Ga$ is an oriented closed loop on the dual $\cV_{\txt{vh}}$ lattice (see Fig.~\ref{fig:r2tcsym1} for an example).

It is straight forward to check that $\t U_3$ commutes with the Hamiltonian for all $\h\Ga$ and therefore corresponds to a symmetry operator. When $\t U_3$ is supported on a contractible loop, it can be written as
\begin{equation}
    \t{U}_3(\h\Ga= \pp M) = \prod_{c_1^{(h)}\in M}\mathfrak{a}_{c_1^{(h)}}.
\end{equation}
Since ${\mathfrak{a}_{c_1^{(h)}} = 1}$ in the ground state subspace, $\t{U}_3$ is a topological operator. Therefore, in the IR, $\t U_3$ is the symmetry operator of a $\ZN{1}$ symmetry. In fact, when supported on a non-contractible loop winding around the system in the $x$ ($y$) direction, $\t U_3$ becomes $\t W_6$ ($\t W_5$) from Eq.~\eqref{eq:post-projection-Z-holonomy}.

\subsubsection{Analysis and discussion of R2TC symmetries}\label{r2TCsymDisc}

Having identified the generalized symmetries of the R2TC, let us now use them to interpret the model's interesting properties from a symmetry point of view. Recall that the six symmetry operators are supported on loops and commute with the R2TC Hamiltonian for all respective loops. Their expectation values with respect to excited states depend on more than just the topology of the loops, so in this sense these microscopic (UV) symmetries are non-topological 1-form symmetries. Their existence reflects the lack of dynamics for $e$ and $\vec{m}$ anyons. Throughout the rest of the deconfined phase of $\Z_N$ rank-2 gauge theory, away from the R2TC point, these non-topological 1-form symmetries are explicitly broken.

The symmetry operators of the R2TC are much richer and more complex than those in the R1TC, which were reviewed in section~\ref{R1TCsymReview}. As demonstrated in Fig.~\ref{fig:loops}, the R1TC symmetry operators are nicely defined on 1-cycles of the direct and dual lattice (so, they admit a straightforward description using cellular homology). Furthermore, for a given symmetry operator, each edge of the 1-cycle was acted on by the same $X$ or $Z$ operator (up to taking the hermitian conjugate, which arises from the 1-cycles orientation and lattice's branching structure). The R2TC symmetry operators, examples of which are shown in Figs.~\ref{fig:r2tcsym2}--\ref{fig:r2tcsym1}, go beyond all of these convenient simplicities. For example, they include the following features, absent from the R1TC's symmetries:
\begin{enumerate}
    \item The symmetry operators $\t U_1(\h\ga^{(s)})$, $\t U_2(\ga^{(s)})$, and $\t U_3(\h\Ga)$ act on both the spins on the loops $\h\ga^{(s)}$, $\ga^{(s)}$, and $\h\Ga$, respectively, and the spins near the loops.
    \item For all symmetry operators, the operators acting on/near the loop's edges depend on whether the loop is parallel to the $x$ or $y$ direction. For example, as shown in Fig.~\ref{fig:r2tcsym2}, $U_1(\ga)$ has $X_0$ act on edges when $\ga$ is parallel to the $y$-direction but has $X_1$ act on edges when $\ga$ is parallel to the $x$-direction.
    \item The operators acting on the spins for symmetry operators $U_3(\Ga^{(s)})$, $\t U_1(\h\ga^{(s)})$, and $\t U_2(\ga^{(s)})$, depend on the position of those spins.
    \item The symmetry operators $U_3$ and $\t U_3$ act on loops of the direct/dual $\cV_{\txt{vh}}$ lattice instead of the direct/dual ($\La_2$) lattice. In terms of the $\La_2$ lattice, these operators act on loops defined on both the direct/dual lattice and therefore require additional framing structure, which makes them framed 1-form symmetries~\cite{QRH201002254}.
    \item The symmetry operator $\t U_3(\h \Ga)$ has operators which only act on the corners of the loop $\h\Ga$ while absent from other parts of the loop (i.e., $Z_0$ in Fig.~\ref{fig:r2tcsym1}). From the $\La_2$ lattice point of view, these corners coincides with where the framing structure connects the direct and dual lattices' loops to create $\h\Ga$.
\end{enumerate}

Unlike the expectation values with respect to excited states mentioned previously, the vacuum expectation values of the symmetry operators depend only on the topology of these loops. Thus, in the ground state sub-Hilbert space---the IR---of the R2TC, all six of the generalized symmetries are 1-form symmetries. Three of these ($U_1$, $U_2$, and $\t U_3$) were conventional 1-form symmetries. However, the other three ($U_3$, $\t U_1$, and $\t U_2$) were not conventional 1-form symmetries since their symmetry operators relied on an underlying sublattice structure. These nonconventional 1-form symmetries were called sublattice 1-form symmetries in the previous section to emphasize this additional structure.

The lattice symmetries generally mix these sublattices and act nontrivially on sublattice 1-form symmetries. Therefore, the total symmetry group of the R2TC is ${\txt{(1-form symmetries)$\rtimes$(lattice symmetries)}}$. This interplay between the sublattice 1-form and spatial symmetries can also be noticed by the R2TC's symmetry-enriched topological order, where position-dependent excitations~\cite{pace-wen} reflect the existence of sublattice 1-form symmetries. From a generalized symmetry point of view, this interplay is reflected by the total symmetry group being described by a 2-group, a type of monoidal category (see Refs.~\onlinecite{BH180309336, BCH221111764}). It would be interesting to investigate 't Hooft anomalies of such 2-group symmetries, where mixed anomalies between lattice translations and sublattice 1-form symmetries would lead to LSM theorems.

Throughout the rest of the deconfined phase of $\Z_N$ rank-2 gauge theory, away from the R2TC point, we expect that all of these generalized symmetries are exact emergent IR symmetries~\cite{PW230105261}. This means that despite being emergent symmetries, explicitly broken in the microscopic Hamiltonian, they constrain the IR in the thermodynamic limit as if they were exact microscopic symmetries.

Since all of the R2TC's generalized symmetries are 1-form symmetries, they are sensitive to the topology of the spatial lattice. The sublattice 1-form symmetries, however, also depend on the geometry of the lattice. Indeed, as we discussed in the previous section, with periodic boundary conditions the size of their underlying sublattices depends on the system size. Furthermore, the sublattices in $U_3$'s, $\t U_1$'s, and $\t U_2$'s respective definitions are unique to the square lattice, so the R2TC on a different lattice would generally have different sublattice 1-form symmetries. Therefore, the sublattice 1-form symmetries give rise to UV/IR mixing in the R2TC~\cite{pace-wen}. The emergent IR symmetries depending on the UV lattice is a general diagnosis for UV/IR mixing and, in fact, may be a unified mechanism for UV/IR mixing in all topological and fracton phases.

The R2TC's symmetry operators satisfy the algebra
\begin{align}
    U_1(\ga)~\t U_1(\h\ga^{(s_1)}) 
    &= 
    \om^{\#(\ga,\h\ga^{(s_1)})}~ \t U_1(\h\ga^{(s_1)})~U_1(\ga),\label{r2tcalg1}\\
    U_2(\h \ga) ~\t U_2(\ga^{(s_2)}) 
    &= 
    \om^{\#(\h \ga,\ga^{(s_2)})}~\t U_2(\ga^{(s_2)})~U_2(\h \ga) ,\\
    U_3(\Ga^{(s_3)})~\t U_3(\h\Ga) 
    &=
    \om^{\#(\Ga^{(s_3)}, \h\Ga)} ~\t U_3(\h\Ga)~U_3(\Ga^{(s_3)}),\label{r2tcalg2}
\end{align}
where ${\om \equiv \ee^{\ii 2\pi/N}}$ and ${\#(~~,~)}$ is the signed intersection number. We thus see that the $U_i$ ($\t U_i$) symmetry operator transforms nontrivially under the $\t U_i$ ($U_i$) symmetry transformation---$U_i$ ($\t U_i$) is a charged operator of the $\t U_i$ ($U_i$) symmetry. Recall from the previous section that when supported on contractible loops, ${\langle U_i\rangle_{\txt{gs}} = \langle \t U_i\rangle_{\txt{gs}} = 1}$. The fact that ${\langle U_i\rangle_{\txt{gs}} = 1}$ (${\langle \t U_i\rangle_{\txt{gs}} = 1}$) for contractible loops means that the $\t U_i$ ($U_i$) symmetry charges are condensed in the R2TC ground state, and the $\t U_i$ ($U_i$) symmetry is spontaneously broken. Therefore, the R2TC ground state spontaneously breaks all six of the 1-form symmetries. 

Discrete symmetries spontaneously breaking always gives rise to a ground state degeneracy. The GSD is computed by finding the smallest faithful representation of the spontaneously broken symmetry operators. This exact calculation was done in section~\ref{subsec:pph}, where the holonomies $W_i$ and $\t W_i$ are the generators of the spontaneously broken symmetries, and yields the correct GSD Eq.~\eqref{eq:GSD}. Therefore, the GSD is system size dependent because some of the spontaneously broken symmetries are sublattice 1-form symmetries that encode geometrical information of the lattice.

The algebra Eqs.~\eqref{r2tcalg1}-\eqref{r2tcalg2} also reveals that the R2TC realizes these generalized symmetries in a projective representation. This prevents the 1-form symmetries from being gauged, and is thus a manifestation of an 't Hooft anomaly\footnote{See footnote~\ref{anomalyFootnoteExplain}.}. In particular, there is a mixed 't Hooft anomaly between the $U_i$ and $\t U_i$ symmetries. Like in the R1TC discussed in Sec.~\ref{R1TCsymReview}, mixed 't Hooft anomalies for 1-form symmetries realized through projective representations are physically reflected through the nontrivial braiding statistics of anyons. Since some of the R2TC's anomalous symmetries are sublattice 1-form symmetries, the braiding statistics will generally depend on the sublattice the anyon resides on. However, this is precisely the position dependent-braiding discussed in section~\ref{posDepBraidSubSec}.


\section{Summary and Outlook}
\label{sec:discussion}
We have applied the idea of coupled-layer construction, previously invented to understand the emergence of fracton models out of toric codes in three dimensions~\cite{hermele17,vijay}, to shed light on the appearance of symmetric rank-2 gauge fields in two dimensions and from there the rank-2 toric code through Higgsing. Condensation of gauge fields can take place in either one of the two conjugate gauge fields $A$ and $E$, and leads to theories with either vector-electric or vector-magnetic charges that are ultimately dual to each other. 

Construction of holonomy (Wilson line) operators for the rank-2 toric code follows rather naturally in this approach, as one can start by identifying the holonomy operators in the Hilbert space before the condensation took place. We thus arrive at the picture of holonomies as the creation/annihilation of magnetic and electric charge-anti-charge pairs, and of their dipole-anti-dipole pairs. The dependence of the ground state degeneracy on the system size (the UV/IR mixing) can be thoroughly understood from analysis of the Wilson loop operators thus obtained. We further suggest an easy-to-implement, heuristic derivation of the holonomies based on the rank-2 gauge theory. This may well have applications in the holonomy construction of other, rank-2 gauge theories and the corresponding stabilizer models. 

Furthermore the exact tensor network expression of the ground state of the rank-2 toric code is derived starting from two copies of the rank-1 toric code's ground state wave functions, by sewing them together with an isometry operation that faithfully reflects the condensation of the gauge fields. This, too, may have application in the construction of other rank-2 based stabilizer ground state wave functions. For one thing, analyzing entanglement entropy becomes easy with the tensor network wave function at hand. Additionally, the tensor network projection of R2TC provides a clear picture of how coupled toric code layers engender higher-rank gauge theory in terms of anyon condensation. This also sheds light on exploring phase transitions between conventional gauge theory and R2TC, where one can replace the tensor projection procedure with an additional parameter in the tensor element. We will explore these issues in a future study~\cite{oh-forthcoming}.

The anyon condensation idea can lead to a number of powerful applications. As an example we showed how the Levin-Gu semionic topological model~\cite{levin05} can undergo a similar condensation procedure to result in a new model. The notion of generalized symmetry is a new and powerful description of the topological order in the toric code, and we have discussed how the notion applies to the rank-2 toric code. We believe the `generalization' of the generalized symmetry idea to other rank-2 based models can find interesting applications in the future. Futhermore, it would be interesting to investigate if general rank-$N$ symmetric tensor gauge theories, with ${N>2}$, can be constructed from many copies of rank-1 theories in a particular condensed phase.

\acknowledgments
Y.-T.O. was supported by National Research Foundation\,(NRF) of Korea under Grant NRF-2022R1I1A1A01065149. 
S.D.P. is supported by the National Science Foundation Graduate Research Fellowship under Grant No. 2141064 and by the Henry W. Kendall Fellowship. J.H.H. was supported by Grant No. NRF-2019R1A6A1A10073079. He also acknowledges financial support from EPIQS Moore theory centers at MIT and Harvard. Y.Y. is supported by Northeastern University COS start-up grant. H.-Y.L. was supported by NRF of Korea under Grant No. NRF-2020R1I1A3074769. H.-Y.L. and Y.-T.O. were supported by the Basic Science Research Program funded by the Ministry of Education (2014R1A6A1030732). J.H.H. acknowledges informative discussion with T. Hughes, B. Kang, H. T. Lam, Z. X. Luo, N. Tantivasadakarn, and X.-G. Wen.

\appendix

\section{Review of discrete differential geometry for ${d}$-dimensional cubic lattices}\label{sec:diffgeoLat}

\begin{figure}[t!]
\centering
    \includegraphics[width=.48\textwidth]{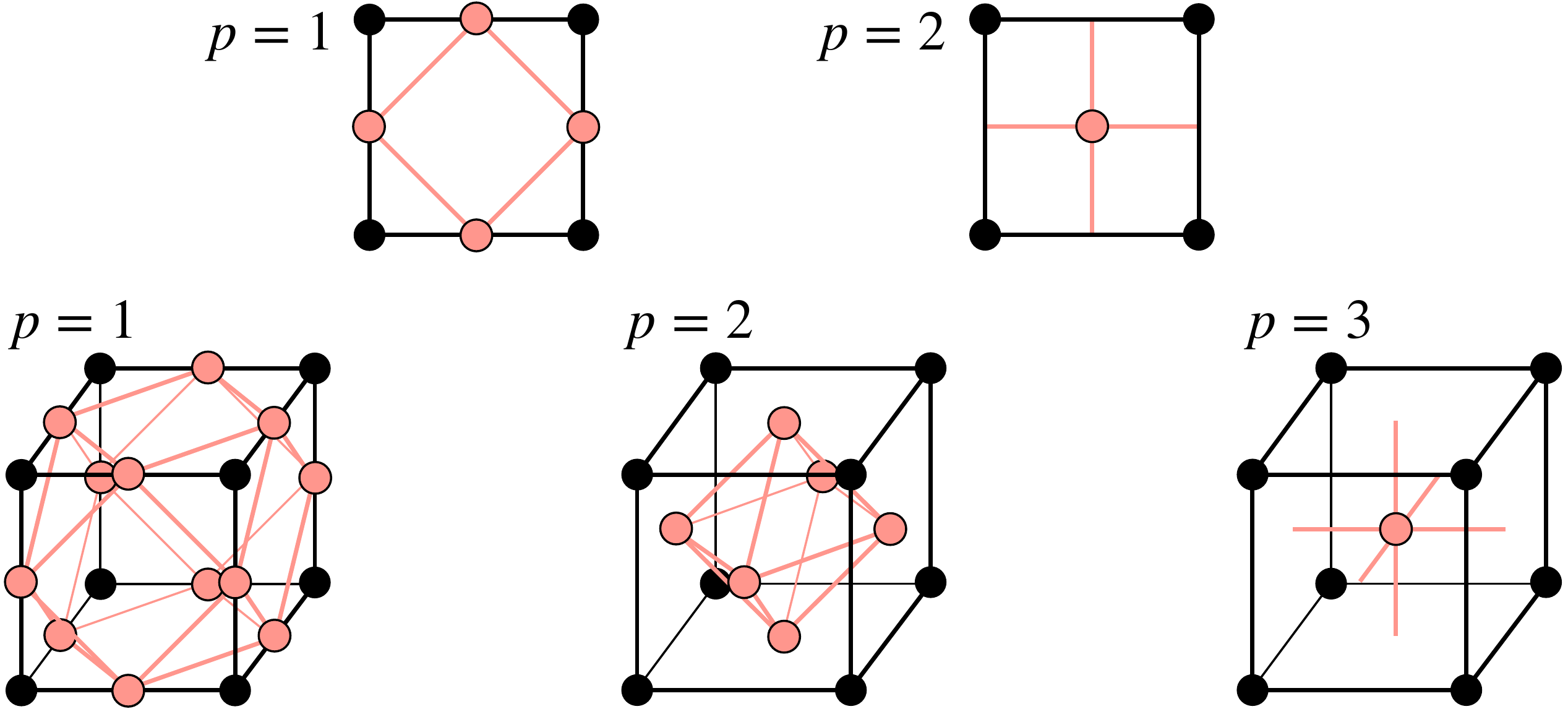}
    \caption{%
    The ${p}$-cells of the ${d}$-dimensional cubic lattice are equivalently the ${0}$-cells---the sites---of some other ${d}$-dimensional lattice. Shown here are examples of this equivalent lattice (drawn in pink) embedded in the conventional unit cell of the cubic lattice (drawn in black). (First row) In ${2}$ dimensions, the ${1}$-cells form another square lattice, rotated by 45 degrees, whose lattice constant is ${1/\sqrt{2}}$ times that of the original square lattice. The ${2}$-cells also form another square lattice, which is the original shifted by the vector ${(\bm{\h\mu}_1+\bm{\h\mu}_2)/2}$. (Second row) In ${3}$ dimensions, both the ${1}$-cells and also the ${2}$-cells form a lattice of corner-sharing octahedra with a lattice constant that is ${1/\sqrt{2}}$ times the cubic lattice's. When ${p=1}$, the octagons are centered at the cubic lattice's ${0}$-cells. When ${p=2}$, the octagons are centered at the cubic lattices ${3}$-cells. Lastly, the ${3}$-cells form another cubic lattice of the same size, but shifted by the vector ${(\bm{\h\mu}_1+\bm{\h\mu}_2+\bm{\h\mu}_3)/2}$.
    }
    \label{fig:lattices}
\end{figure}

In this appendix section, we review relevant parts of discrete differential geometry (in a non-rigorous fashion) used in Sec.~\ref{sec:generalized-symmetry} of the main text. Consider a cubic lattice in ${d}$-dimensional space with periodic boundary conditions, denoted by ${M_d}$. While a Bravais lattice is a collection of lattice sites ${\bm x \in \Z^d}$, it is useful to view it as also formed by higher-dimensional objects, like links, plaquettes, cubes, \etc. We call a ${p}$-dimensional object a ${p}$-cell, with ${0\leq p \leq d}$. So, a ${0}$-cell is a lattice site, a ${1}$-cell is a link, a ${2}$-cell is a plaquette, \etc. This does not add additional structures to the lattice, but instead is just a useful way of organizing the lattice sites. Indeed, denoting a ${p}$-cell associated with site ${\bm x}$ as ${c_p(\bm x)_{\mu_1\mu_2\cdots\mu_p}}$, where ${\mu_1 < \mu_2 < \cdots < \mu_p}$ and ${\mu_i\in\{1,2,\cdots,d\}}$, a ${p}$-cell of the cubic lattice is the set of ${2^p}$ lattice sites\footnote{We adopt the discrete differential geometry and exterior calculus notations and conventions used in \Rf{SG190102637}.}
\begin{equation}
\begin{aligned}
\hspace{-4pt}c_p(\bm x)_{\mu_1\mu_2\cdots\mu_p} \hspace{-4pt}= 
\{\bm x\} 
&\cup \{\bm x+\bm{\h\mu}_{i} ~|~ 1 \leq i \leq p\} \\
&\cup\{\bm x+\bm{\h\mu}_{i}+\bm{\h\mu}_{j} ~|~ 1 \leq i<j \leq p\} \\
&\cup \cdots\cup \{\bm x+\bm{\h\mu}_{1}+\ldots+\bm{\h\mu}_{p}\},
\end{aligned}
\end{equation}
where ${\bm{\h\mu}_i}$ is the unit vector in the ${\mu_i}$-direction. It is often convenient to drop the requirement that the indices are canonically ordered (i.e., that they satisfy ${\mu_1<\mu_2<\cdots<\mu_p<\nu}$) and instead let ${c_p(\bm x)_{\mu_1\mu_2\cdots\mu_p}}$ obey the relation ${c_p(\bm x)_{\cdots\mu_1\mu_2\cdots} = -c_p(\bm x)_{\cdots\mu_2\mu_1\cdots}}$. The ${p}$-cells of the ${d}$-dimensional cubic lattice are equivalently viewed as the ${0}$-cells of some other lattice in ${d}$-dimensions, as demonstrated for ${d=2}$ and ${3}$ in Fig.~\ref{fig:lattices}.

\begin{figure}[t!]
\centering
    \includegraphics[width=.48\textwidth]{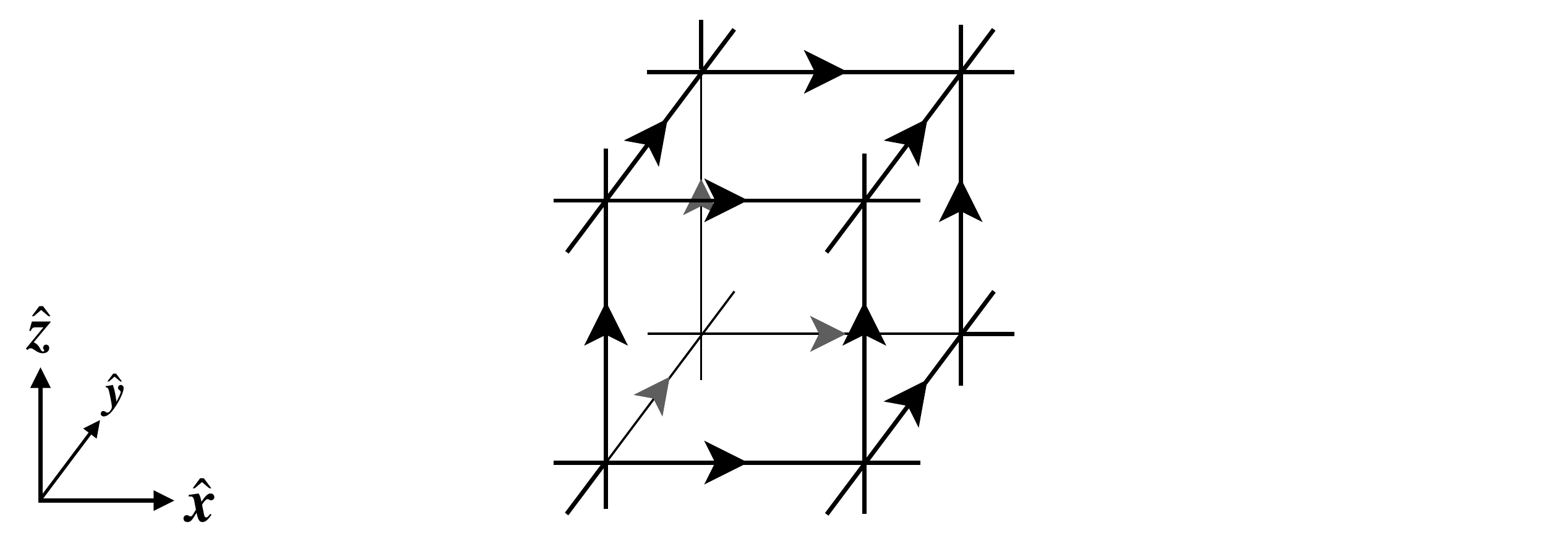}
    \caption{%
    Example of the branching structure used for a chunk of the cubic lattice in three-dimensional space.
    }
    \label{fig:branchingStructure}
\end{figure}

Introducing the concept of ${p}$-cells is strictly unnecessary but very convenient because ``sewing'' ${p}$-cells together gives a natural way to form ${p}$-dimensional subspaces of the lattice. Furthermore these subspaces can also be given an orientation by defining an orientation structure to the lattice. A nice local scheme for the lattice orientation is a branching structure, where the orientation on each ${1}$-cell is chosen such that a collection of ${1}$-cells cannot form an oriented closed loop. A canonical orientation on all other ${p}$-cells then follows from the branching structure. We use the branching structure where each ${1}$-cell ${c_1(\bm x)_\mu}$ has an arrow pointing in the ${\h{\bm{\mu}}}$ direction (see Fig.~\ref{fig:branchingStructure}). However, it is important to note that the choice of lattice orientation is a formal convention, and choosing different branching structures does not affect the physics\footnote{However, according to a conjecture from \Rf{WWW211212148}, observables are independent of the branching structure only if the continuum effective field theory is free of a framing anomaly~\cite{W8951}.}.

A ${p}$-cell can be related to ${(p-1)}$ cells using the boundary operator ${\pp}$. The boundary operator acting on a ${p}$-cell---${\pp c_{p}}$---is the \txti{oriented} sum of $(p-1)$-cells on the boundary of ${c_{p}}$. For the branching structure we use, it is given by
\begin{equation}\label{boundaryDef}
\begin{aligned}
\partial c_p(\bm x)_{\mu_1 \cdots \mu_p}&\hspace{-4pt}=\hspace{-2pt}\sum_{k=1}^{p}(-1)^{k+1}\hspace{-2pt}\left[c_{p-1}\hspace{-2pt}\left(\bm x+\bm{\h\mu}_{k}\right)_{\mu_1 \cdots \stackrel{\txt{o}}{\mu}_k \cdots \mu_p }\right.\\
&\left.\hspace{70pt}-c_{p-1}(\bm x)_{\mu_1 \cdots \stackrel{\txt{o}}{\mu}_k \cdots \mu_p }\right],
\end{aligned}
\end{equation}
where the notation ${\stackrel{\txt{o}}{\mu}_k}$ indicates that the ${\mu_k}$ index is omitted. From its definition, the boundary operator satisfies ${\pp^2 c_p = 0}$ for any ${p}$-cell. Furthermore, as there are no ${(-1)}$-cells, the boundary operator acting on a ${0}$-cell is defined to be zero.

On the other hand, a ${p}$-cell can be related to ${(p+1)}$-cells using the coboundary operator ${\del}$. The coboundary operator acting on a ${p}$-cell---${\del c_{p}}$---is an \txti{oriented} sum of all ${(p+1)}$-cells whose boundary includes ${c_{p}}$. For the branching structure we use, it is given by
\begin{equation}\label{coboundaryDef}
\del c_p(\bm x)_{\mu_1 \cdots \mu_p}\hspace{-2pt}=\sum_{\nu}c_{p+1}(\bm x)_{\nu \mu_{1} \ldots \mu_{p} }-c_{p+1}(\bm x-\bm{\h\nu})_{\nu \mu_{1} \ldots \mu_{p} }.
\end{equation}
From its definition, the coboundary operator satisfies ${\del^2 c_p = 0}$ for any ${p}$-cell. Furthermore, as there are no ${(d+1)}$-cells, the coboundary operator acting on a ${d}$-cell is defined to be zero.

Lastly, the lattice has an associated dual lattice. The dual lattice has its lattice sites centered at the ${d}$-cells of the direct lattice. For the cubic lattice, one choice of framing that relates a dual lattice site ${\bm{\h x}}$ to a direct lattice site ${\bm x}$ is by ${\bm{\h x} = \bm x + \frac{1}{2}\bm{\h r}}$ with ${\bm{\h r} = \sum_i \bm{\h\mu}}_i$.

Each ${p}$-cell ${c_p}$ on the direct lattice is associated with a ${(d-p)}$-cell ${\h{c}_{d-p}}$ on the dual lattice. This is implemented by the dual operator ${\hstar}$. For this choice of framing, a ${p}$-cell ${c_{p}(\bm x)_{\mu_{1} \cdots \mu_{p}}}$ (with canonical ordering ${\mu_1<\cdots<\mu_p}$) and a ${(d-p)}$-cell of the dual lattice ${\h c_{d-p}(\bm{\h x})_{\mu_{1} \cdots \mu_{d-p}}}$ (with canonical ordering ${\mu_{1}<\cdots<\mu_{d-p}}$) are related to one another by
\begin{align}
\hstar c_{p}(\bm x)_{\mu_{1} \cdots \mu_{p}}&= \eps_{\mu_1 \cdots \mu_p \mu_{p+1} \cdots \mu_{d}}\\
&\quad\quad\times\h{c}_{d-p}(\bm{\h x}-\bm{\h\mu}_{p+1} - \ldots-\bm{\h\mu}_{d})_{\mu_{p+1} \cdots \mu_{d}},\nonumber\\
\hstar \h{c}_{p}(\bm{\h x})_{\mu_1 \ldots \mu_p}&=\eps_{\mu_{1} \cdots \mu_{p} \mu_{p+1} \cdots \mu_{d}}\label{stardualcell}\\
&\quad\quad\times c_{d-p}(\bm x+\bm{\h\mu}_{1} + \ldots+\bm{\h\mu}_p)_{\mu_{p+1} \cdots \mu_{d}} \nonumber,
\end{align}
where summation is \txti{not} implied on the right hand side. Here ${\eps}$ is the Levi-Civita symbol, which takes into account the lattice's and dual lattice's relative orientations. From the definition of ${\hstar}$, acting ${\hstar}$ twice on a ${p}$-cell of the direct (dual) lattice yields ${\hstar\hstar c_p = (-1)^{p(d-p)}c_p}$ (${\hstar\hstar \h{c}_p = (-1)^{p(d-p)}\h{c}_p}$). Furthermore, from the definitions of the boundary, coboundary, and dual operators, they are related to one another by
\begin{equation}\label{cobdyBdy}
\del c_p = (-1)^{d(p+1)+1}\hstar\pp\hstar c_p,
\end{equation}
which, equivalently, is ${\hstar\del c_p = (-1)^{p}\pp\hstar c_p}$.

\bibliography{reference}
\end{document}